\documentclass[useAMS,usenatbib,twocolumn]{aastex62}
\usepackage{times}
\usepackage{natbib}
\usepackage{amsmath}
\usepackage{amssymb}
\usepackage{morefloats}
\usepackage{hyperref}
\usepackage[toc,page]{appendix}

\newcommand{\arepo}{\texttt{\small AREPO }}

\newcommand{\wombat}{\texttt{\small{WOMBAT}} }
\newcommand{\athena}{\texttt{\small{ATHENA}} }

\newcommand{\Matrix}[1]{\ensuremath{{\bf \sf #1}}}%
\renewcommand{\vec}[1]{\ensuremath{{\bf #1}}}

\begin{document}

\title[WENO-Wombat]{WENO-Wombat: Scalable  Fifth-Order Constrained-Transport Magnetohydrodynamics for Astrophysical Applications}

\author{J. M.F. Donnert}
\affiliation{Leiden Observatory, PO Box 9513, {NL-2300 RA} Leiden, The Netherlands }
\affiliation{INAF Istituto di Radioastronomia, via P. Gobetti 101, I-40129 Bologna, Italy}
\affiliation{School of Physics and Astronomy, University of Minnesota, Minneapolis, MN 55455, USA}
\email{jdonnert@strw.leidenuniv.nl}

\author{H. Jang}
\affiliation{Department of Physics, School of Natural Sciences, Ulsan National Institute of Science and Technology, {UNIST-gil} 50, Ulsan, 44919, Korea}
\email{hanbyul@sirius.unist.ac.kr}

\author{P. Mendygral}
\affiliation{Cray Inc., Bloomington, MN, USA}
\email{pjm@cray.com}

\author{G. Brunetti}
\affiliation{INAF Istituto di Radioastronomia, via P. Gobetti 101, I-40129 Bologna, Italy}
\email{brunetti@ira.inaf.it}

\author{D. Ryu}
\affiliation{Department of Physics, School of Natural Sciences, Ulsan National Institute of Science and Technology, {UNIST-gil} 50, Ulsan, 44919, Korea}
\email{ryu@sirius.unist.ac.kr}

\author{T.W. Jones}
\affiliation{School of Physics and Astronomy, University of Minnesota, Minneapolis, MN 55455, USA}
\affiliation{Minnesota Supercomputing Institute for Advanced Computational Research}
\email{twj@umn.edu}

\date{Accepted ???. Received ???; in original form ???}



\label{firstpage}
 
\begin{abstract}
    Due to increase in computing power, high-order Eulerian schemes will likely become instrumental for the simulations of turbulence and magnetic field amplification in astrophysical fluids in the next years. We present the implementation of a fifth order weighted essentially non-oscillatory scheme for constrained-transport magnetohydrodynamics into the code \wombat. We establish the correctness of our implementation with an extensive number tests. We find that the fifth order scheme performs as accurately as a common second order scheme at half the resolution. We argue that for a given solution quality the new scheme is more computationally efficient than lower order schemes in three dimensions. We also establish the performance characteristics of the solver in the \wombat framework. Our implementation fully vectorizes using flattened arrays in thread-local memory. It performs at about 0.6 Million zones per second per node on Intel Broadwell. We present scaling tests of the code up to 98 thousand cores on the Cray XC40 machine 'Hazel Hen', with a sustained performance of about 5\% of peak at scale.
\end{abstract}

\keywords{general - methods: numerical - MHD}

\section{Introduction}\label{sect.intro}

Most of the Baryonic matter in the Universe is in the form of a thin ionized plasma. Hence many numerical models of astrophysical interest require the solution of the magneto-hydrodynamic (MHD) equations \citep[e.g.][]{2006PhT....59a..58K}, from the solar corona, the intergalactic medium to the jets of active galactic nuclei, the hot atmosphere of galaxy clusters and the filaments of the cosmic web. Finite difference methods for hydrodynamics pre-date the earliest computers and of course transcend the field of Astrophysics \citep{doi:10.1002/qj.49704820311,L2002}. After seminal contributions by \citet{Lax1954,God59,doi:10.1137/0710071,1979JCoPh..32..101V,Roe81} and nearly a century of development, major progress has been made in ever more accurate and efficient finite volume and finite difference schemes to minimize spurious errors and capture shocks accurately \citep{1983JCoPh..49..357H,1984JCoPh..54..174C,1987JCoPh..71..231H,Shu1988,Einfeldt:1988:GTM,LIU1994200}. For MHD, the development of constrained transport schemes (CT) marks the era of numerical magnetic fields with vanishing divergence to machine precision \citet{1988ApJ...332..659E,2000ApJ...530..508L}. \par
The plethora of available algorithms is only dwarfed by the vast amount of codes that use finite volume or finite difference methods today. In astrophysics, implementations would include \texttt{\small AREPO} \citep{2010MNRAS.401..791S}, \texttt{\small ART} \citep{1997ApJS..111...73K}, \texttt{\small ATHENA} \citep{2008ApJS..178..137S}, \texttt{\small CHOLLA} \citep{2015ApJS..217...24S}, \texttt{\small DISPATCH} \citep{2018MNRAS.477..624N}, \texttt{\small ENZO} \citep{2014ApJS..211...19B}, \texttt{\small FLASH} \citep{2000ApJS..131..273F}, \texttt{\small GAMER} \citep{2018MNRAS.481.4815S}, \texttt{\small GIZMO} \citep{2016MNRAS.455...51H},  \texttt{\small Nyx} \citep{2013ApJ...765...39A}, \texttt{\small RAMSES} \citep{2002A&A...385..337T}, \texttt{\small PLUTO} \citep{2007ApJS..170..228M},  \texttt{\small ZEUS} \citep{1992ApJS...80..753S}, and many more. \par
With super computers now approaching the regime of $10^{18}$ floating point operations per second (Flops), algorithms beyond the most commonly used 2nd and 3rd order for MHD are well feasible \citep[see][ for a recent review]{2017LRCA....3....2B}. These methods offer increased accuracy over common codes, which is advantageous in the simulation of turbulence \citep{2018arXiv180602343G} and when problems include supersonic advection of the fluid relative to the mesh. \par

However, improved accuracy comes at the expense of additional computational costs.  It has been argued by \citet[][]{Greenough} that second order codes are computationally more efficient in one dimension, {in the sense of solution quality per computational cost}. In this contribution we will argue that this not necessarily true in three dimensions for a fifth order finite difference weighted essentially non oscillatory (WENO) scheme \citep{Jiang_Shu__1996__WENO-scheme,1999JCoPh.150..561J,Shu__1998__CIME__WENO,2000JCoPh.160..405B,2004ApJ...612....1F}. WENO schemes use a weighted average of several stencils around a zone to achieve high accuracy and low truncation error in spatial interpolation, while avoiding spurious oscillations near discontinuities.  Modern WENO schemes come in too many flavors to list them all. Particularly relevant to this work are contributions by \citet{2005JCoPh.207..542H}, who have shown that the classical scheme is only third order accurate near critical points. \citet{2008JCoPh.227.3191B} have proposed a simple set of weights to restore full order (WENO-Z). Subsequent work has focused further improvements of the scheme and extended it to very high order, for reviews see \citet{doi:10.1137/070679065,2016JCoPh.326..780B}.  \par
We will show that the classical finite difference method doubles the effective resolution of the grid compared to popular PPM or TVD schemes in all dimensions. This is in-line with prior results finding that WENO3, WENO5 and WENO9 double the effective resolution compared to the next lower order scheme \citep{2003PhRvE..68d6709Z}. The excellent fidelity of these schemes allows in principle to simulate a problem at half the resolution with WENO5 compared to PPM or WENO3, which more than makes up for added computational cost by the fifth order scheme. WENO5 also reduces the accumulated round-off error over many time steps and reduces the data size by a factor eight, which might be even more important than the reduction in computational cost. \par
However, due to the wide use of multi-core computers, most fluid simulations today are not compute bound. Thus a WENO implementation will likely run at the same resolution as the lower order scheme to increase the fidelity of the simulation.  Given the wide use of accelerators and { the increasing diversity of} CPU architectures, it is highly desirable to write a performance aware implementation of an efficient scheme using open standards that can use many different architectures. The high compute intensity of the WENO finite difference scheme, i.e. the large byte per flop ratio in the main WENO loop, makes it very attractive for such an approach.  As always in high performance computing (HPC), SIMD\footnote{single instruction multiple data} vectorization is key to achieving good performance on CPUs, and eventually GPUs by exposing instruction level parallelism. The regular data layout of an Eulerian grid makes this significantly easier to achieve than in particle based schemes. \par
Here we present the implementation of a classical fifth order finite difference WENO scheme \citep{1999JCoPh.150..561J} in the \wombat\footnote{wombatcode.org} code framework \citep{2017ApJS..228...23M}. We use constrained transport to evolve the magnetic field with minimal divergence error with  the ''transport-flux'' formulation by \citet{1998ApJ...509..244R}. Formally, spatial interpolation is fifth order, time interpolation is fourth order and CT is second order. We deliberately chose a second order CT scheme to keep the implementation computationally cheap. High order schemes are much more computationally expensive at fifth order \citep[e.g.][]{2004JCoPh.195...17L,2010JCoPh.229.5896M,2019MNRAS.482..416V}. In contrast, we will show that the our simple second order CT scheme affects only magnetic field dispersion, diffusion remains fifth order accurate, at virtually no additional computational cost. \par
This work is structured as follows: We outline the scheme in section \ref{sect.scheme}. The implementation is heavily optimized to expose parallelism in the code at all levels, details are given in section \ref{sect.code}. Code tests with a special emphasis on errors relevant in cosmological simulations (advection error and angular momentum conservation) are presented in section \ref{sect.tests}. Aside from \texttt{\small WOMBAT}'s own TVD+CTU implementation, we frequently compare the code with published results from the order CTU+CT code \texttt{\small ATHENA} \citep{2008ApJS..178..137S}. We test the code performance on modern HPC hardware in section \ref{sect.perf}. Conclusions are drawn in \ref{sect.conclusions}. For reference purposes, we once again present the eigenvectors used in the calculation in appendix \ref{app.ev}.
 
\section{Numerical Method} \label{sect.scheme}

The equations of ideal magneto-hydrodynamics read \citep[e.g.][]{1995ApJ...442..228R}:

\begin{align}  
    \frac{\partial \rho}{\partial t} + \nabla \cdot \left( \rho \vec{v} \right) &= 0, \label{eq.mhdA}\\
    \frac{\partial \vec{v}}{\partial t} + \vec{v} \cdot \nabla\vec{v} + \frac{1}{\rho} \nabla P - \frac{1}{\rho} \left( \nabla \times \vec{B} \right) \times \vec{B} &= 0, \\
    \frac{\partial P}{\partial t} + \vec{v} \cdot \nabla P + \gamma P \nabla \cdot \vec{v} &= 0 \\
    \frac{\partial B}{\partial t} - \nabla \times \left(\vec{v} \times \vec{B} \right) &= 0 \label{eq.mhdB}\\
    \nabla \cdot \vec{B} = 0 \label{eq.divb}
\end{align}
where $\rho$ is the density, $\vec{v} = (v_x, v_y, v_z)^\mathrm{T}$ is the velocity, $\vec{B} = (B_x, B_y, B_z)^\mathrm{T}$ is the magnetic field. We have chosen a rationalized system of units, so factors of $4\pi$ do not appear. Further we define $B = |\vec{B}|$, $v = |\vec{v}|$, the adiabatic index $\gamma$ and the total pressure $P^*$, the total energy $E$ and Entropy $S$ \citep{1993ApJ...414....1R} of the flow:

\begin{align}
    P^* &= P + \frac{1}{2} B_x^2 + B_y^2 + B_z^2 \\
    E &= \frac{P}{\gamma - 1} + \frac{1}{2} \left( \rho v^2 + B^2 \right), \label{eq.P}\\
    S &= \frac{P}{\rho^{\gamma-1}}.
\end{align}

A state vector of \emph{conserved} quantities $\vec{q}$ can be defined: 

\begin{align}
    \vec{q} &= \left(\rho, \rho v_{x}, \rho v_{y}, \rho v_{z}, B_{x},B_{y},B_{z}, E\right)^\mathrm{T}
\end{align}

and a vector of fluxes $\vec{F}$ in x-direction:
 
\begin{align}
    \vec{F}_x &=  \begin{pmatrix}
                    \rho v_{x}                  \\
                    \rho v_{x}^2 + P^* - B_x^2  \\
                    \rho v_x v_y - B_x B_y      \\
                    \rho v_x v_z - B_x B_z      \\
                    0                           \\
                    B_y v_x - B_x v_y           \\
                    B_z v_x - B_x v_z           \\
                    (E+P^*) v_x - B_x ( B_x v_x + B_y v_y + B_z v_z)
                \end{pmatrix} \label{eq.fluxes}
\end{align}

Then the equations \ref{eq.mhdA}-\ref{eq.mhdB} can be written as a conservation law for\footnote{We do not consider source terms like gravity, cosmic-rays, radiative cooling etc.} $\vec{q}$:
\begin{align}
    \frac{\partial \vec{q}}{\partial t} + \frac{\partial \vec{F}_x}{\partial x} = 0 \label{eq.conform}
\end{align}
This equation can be approximated in quasi-linear form \citep{Roe81,ROE1997250,1991JCoPh..92..273E,1979JCoPh..32..101V}:
\begin{align}
    \frac{\partial \vec{q}}{\partial t} + \Matrix{A}_x \frac{\partial \vec{q}}{\partial x}  &= 0 \label{eq.linear}
\end{align}
where we define the \emph{Jacobian} matrix $\Matrix{A}_x = \partial \vec{F} / \partial \vec{q}$. \par
A complete set of left and right eigenvectors (see appendix \ref{app.ev}) and real, albeit possibly degenerate, eigenvalues can be found for this Jacobian, thus the system \ref{eq.linear} is hyperbolic \citep{BrioWu1988}. The seven eigenvalues can be identified with the three MHD waves: the slow, fast and Alfv\`{e}n mode, as well as an entropy mode. With the definition of the (hydrodynamic) sound speed $c_s = \sqrt{\gamma P /\rho}$, the MHD \emph{wave speeds} are \citep[][ Jang et al. in prep.]{BrioWu1988}:

\begin{align}
    c_\mathrm{slow} &= \left( \frac{1}{2} c_s^2 + \frac{B^2}{\rho} - \sqrt{ \left(\frac{B^2}{\rho} + c_s^2  \right)^2 - 4 \frac{B_x^2}{\rho}c_s^2 } \right)^{1/2} \label{eq.slow} \\
    c_\mathrm{fast} &= \left( \frac{1}{2} c_s^2 + \frac{B^2}{\rho} + \sqrt{ \left(\frac{B^2}{\rho} + c_s^2  \right)^2 - 4 \frac{B_x^2}{\rho}c_s^2 } \right)^{1/2} \label{eq.fast}\\
    c_\mathrm{A}    &= \frac{B_x}{\sqrt{ \rho}} \label{eq.ca}
\end{align} 

Using the left ($\vec{L}(m)$) and right hand ($\vec{R}(m)$) eigenvectors of the linearised equations, the system can be decoupled into seven scalar advection equations, by multiplying it with $\vec{L}$ from the left and using  $ \vec{L}\Matrix{A}_x \vec{R} =  \Matrix{\Lambda}$, where $\Matrix{\Lambda}$ is the diagonal matrix containing the eigenvalues $\lambda  = (\lambda_\mathrm{fast}^-,\lambda_\mathrm{A}^-,\lambda_\mathrm{slow}^-,\lambda_\mathrm{e},\lambda_\mathrm{slow}^+,\lambda_\mathrm{A}^+,\lambda_\mathrm{fast}^+)$, where $\lambda^\pm = v_x \pm c_\mathrm{wave}$ and $c_\mathrm{wave}$ is the corresponding wave speed. Thus the linearised MHD equations encode the propagation of seven MHD waves in time along characteristics \citep[e.g.][]{L2002}. The solution can then be obtained component wise from the linearised scalar problem, and be transformed back by multiplying it with $\vec{R}$. \par
In \wombat, we assume that space is discretized as $x_i = i \Delta x -  L_\mathrm{Box}/2$, with $\Delta x = L_\mathrm{Box}/N_x$, $L_\mathrm{Box}$ the size of the computational domain in x-direction and $N_x$ the number of zones in x-direction. One can enforce numerical conservation by writing eq. \ref{eq.linear} with fluxes across the zone boundaries at $i\pm\frac{1}{2}$ in \emph{conservative form}  \citep{God59,L2002}:
\begin{align}
    \frac{\mathrm{d}\vec{q}}{\mathrm{d} t} + \frac{1}{\Delta x} \left (\vec{F}^n_{i+\frac{1}{2}} - \vec{F}^n_{i-\frac{1}{2}} \right) &= 0 , \label{eq.num}
\end{align}
Given suitable initial conditions, the solution of equation \ref{eq.conform} then becomes an initial value problem and can be solved using the method of lines \citep[e.g.][]{zbMATH03302893}. \par
Depending on the scheme, one chooses a suitable discretization in time to integrate forward and in space to interpolate the fluxes to the cell boundaries.

\subsection{Time Discretization} \label{sect.RK}

Following \citet{Jiang_Shu__1996__WENO-scheme}, we use a 4th-order, 4-stage Runge-Kutta (RK4) time integrator. It has been shown that such a scheme cannot have the strong stability preserving property \citep[e.g.][]{doi:10.1137/S0036142901389025}. In fact the scheme used here is not even total variation diminishing (TVD) \citep{Shu1988,1988JCoPh..77..439S}. Nonetheless, it allows us, to use $\mathrm{CFL} = 0.8$ everywhere and is sufficiently robust in practice. Thus, in $N_\mathrm{d}$ dimensions the timestep is set as:

\begin{align}
    \Delta t &< \frac{ \mathrm{CFL} \Delta x}{  \sum_{d=1}^{N_\mathrm{d}} \mathrm{max}(|v_d| + c_\mathrm{fast}^d))   } \label{eq.cfl}
\end{align}
 Omitting CT for now, the RK4 scheme from \citet{Jiang_Shu__1996__WENO-scheme,1999JCoPh.150..561J} can be written as:
\begin{align}
    \vec{q}              &= \vec{q_0} =\vec{q}_{i,j,k}^n \nonumber\\
    \vec{q}_\mathrm{save} &= -\frac{4}{3} \vec{q_0} \nonumber\\
    \vec{q}               &= \vec{q_0} - \frac{1}{2} \frac{\Delta t}{\Delta x} \left( \vec{F}_{i+1/2,j,k}(\vec{q}) - \vec{F}_{i-1/2,j,k}(\vec{q})  \right)  \nonumber\\
    \vec{q}_\mathrm{save} &= \vec{q}_\mathrm{save} + \frac{1}{3} \vec{q} \nonumber\\
    \vec{q}               &= \vec{q}_0 - \frac{1}{2} \frac{\Delta t}{\Delta x} \left( \vec{F}_{i+1/2,j,k}(\vec{q}) - \vec{F}_{i-1/2,j,k}(\vec{q})  \right)  \nonumber\\
    \vec{q}_\mathrm{save} &= \vec{q}_\mathrm{save} + \frac{2}{3} \vec{q} \nonumber\\
    \vec{q}               &= \vec{q}_0 - \frac{\Delta t}{\Delta x} \left( \vec{F}_{i+1/2,j,k}(\vec{q}) - \vec{F}_{i-1/2,j,k}(\vec{q})  \right)  \nonumber\\
    \vec{q}_\mathrm{save} &= \vec{q}_\mathrm{save} + \frac{1}{3} \vec{q} \nonumber\\
    \vec{q}               &= \vec{q}_0 -  \frac{1}{6} \frac{\Delta t}{\Delta x} \left( \vec{F}_{i+1/2,j,k}(\vec{q}) - \vec{F}_{i-1/2,j,k}(\vec{q}) \right)  \nonumber\\
    \vec{q}_\mathrm{save} &= \vec{q}_\mathrm{save} + \vec{q} \nonumber\\
    \vec{q}_{i,j,k}^{n+1}       &= \vec{q}_\mathrm{save}\nonumber
\end{align}

The CT time integration scheme is the same, but updates the face centered magnetic fields after the WENO step (see section \ref{sect.CT}). In 3 dimensions it adds the 3 component magnetic field on the boundary and the 3 component corner flux to global storage.

\subsection{WENO Spatial Discretization}

For reference, we outline the WENO5 spatial discretization to obtain fluxes at the zone boundaries in equation \ref{eq.num}. The classical scheme from \citet{1999JCoPh.150..561J} is a finite difference approach that efficiently computes boundary fluxes from point valued cell centered states and fluxes. This is in contrast to modern schemes that often use a more flexible, but also more computationally expensive finite volume approach. For more details, we kindly ask the reader to refer to the extensive literature on the subject \citep[e.g.][ Jang et al. in prep.]{Jiang_Shu__1996__WENO-scheme,Shu__1998__CIME__WENO,1999JCoPh.150..561J,doi:10.1137/070679065}. \par
The fifth order weighted essentially non-oscillatory scheme uses a weighted average of three third order stencils to approximate the solution at the boundary. Thus the WENO averaging itself requires two boundary zones. Including an additional zone from the flux difference in equation \ref{eq.num}, gives a total of three boundary zones per RK substep. To avoid Gibbs phenomena (ringing) near shocks, the weights ''switch-off'' one of the three polynomials adjacent to discontinuities, and inside a discontinuity, where the method becomes first order. We note that the scheme is formally split, as we always integrate along the x-direction and rotate the grid to integrate along the other directions successively. However, the time integration scheme averages over all directions four times. Thus effectively the scheme is un-split at fourth order. \par
Following \citet{1999JCoPh.150..561J}, we denote quantities in the decoupled system with superscript $s$, denote the individual components with $m \in [1,7]$ and drop the vector notation. Thus the fluxes, state vectors and their differences in the decoupled system are:
\begin{align}
    F_k^s(m) &= \vec{L}_{i+\frac{1}{2}}(m) \cdot \vec{F}_k, \\
    q^s_k(m) &= \vec{L}_{i+\frac{1}{2}}(m) \cdot \vec{q}_k, \\
    \Delta F^s_{k+\frac{1}{2}}(m) &= F^s_{k+1}(m) - F^s_k(m), \\
    \Delta q^s_{k+\frac{1}{2}}(m) &= q^s_{k+1}(m) - q^s_k(m),
\end{align}
respectivey. Note that the left hand eigenvector is taken at the zone boundary using simple arithmetic averaging. In the decoupled system, WENO uses Lax-Friedrichs type flux splitting to upwind the fluxes \citep{Lax1954}: 
\begin{align}
    F_i^{s\pm}(m) & = \frac{1}{2} \left( F^s_i(m) \pm \alpha(m)q_i^s(m) \right) \\
    \Delta F_{k+\frac{1}{2}}^{s\pm}(m) & = \frac{1}{2} \left(\Delta F^s_{k+\frac{1}{2}}(m) \pm \alpha(m)\Delta q_{k+\frac{1}{2}}^s(m) \right),
\end{align}
where $\alpha(m) = \mathrm{max}(|\lambda_i(m)|,|\lambda_{i+1}(m)|)$ is the larger of the two eigenvalues $\lambda(m)$ between zone $i$ and $i+1$.\par
The WENO interpolated fluxes at the zone boundary for equation \ref{eq.num} are given by $\vec{F}_{i+\frac{1}{2}} = F^s_{i+\frac{1}{2}}(m) \cdot \vec{R}_{i+\frac{1}{2}}(m)$, where again the right hand eigenvector is taken at the zone boundary. In the decoupled system, the flux of each component $m$ is \citep{1999JCoPh.150..561J}:
\begin{align}
    {F}^s_{i+\frac{1}{2}} &= \frac{1}{12} \left( -{F}^s_{i-1}+7{F}^s_{i}+7{F}^s_{i+1}-{F}^s_{i+2} \right)\nonumber\\
    &-\varphi_N\left( \Delta{F}^{s+}_{i-\frac{3}{2}},\Delta{F}^{s+}_{i-\frac{1}{2}},\Delta{F}^{s+}_{i+\frac{1}{2}},\Delta{F}^{s+}_{i+\frac{3}{2}} \right) \nonumber\\
    &+\varphi_N\left( \Delta{F}^{s-}_{i+\frac{5}{2}},\Delta{F}^{s-}_{i+\frac{3}{2}},\Delta{F}^{s-}_{i+\frac{1}{2}},\Delta{F}^{s-}_{i-\frac{1}{2}}\right), \label{eq.FWENO}
\end{align}
where we have dropped the $(m)$ notation for clarity. The WENO interpolant $\varphi_N(a,b,c,d)$ is defined as
\begin{align}
    \varphi_N &= \frac{1}{3} \omega_0 \left(a-2b+c\right) + \frac{1}{6} \left( \omega_2 - \frac{1}{2} \right) \left( b - 2c + d \right).
\end{align}
The non-linear weights are:
\begin{align}
    \omega_0 &= \frac{\alpha_0}{\alpha_0 + \alpha_1 + \alpha_2}, & \omega_2 &= \frac{\alpha_2}{\alpha_0 + \alpha_1 + \alpha_2} \\
    \alpha_0 &= \frac{1}{\left(\epsilon + IC_0\right)^2}, & \alpha_1 &= \frac{6}{\left(\epsilon + IC_1\right)^2}, \\
    \alpha_2 &= \frac{3}{\left(\epsilon + IC_2\right)^2},
\end{align}
with $\epsilon = 10^{-6}$ and
\begin{align}
    IS_0 &= 13(a-b)^2 + 3 (a-3b)^2,\\
    IS_1 &= 13(b-c)^2 + 3 (b+c)^2,\\
    IS_2 &= 13(c-d)^2 + 3 (3c-d)^2.
\end{align}

This description of the weights is generally fifth-order accurate in smooth flows, but only third order accurate near critical points (extrema \& saddle points). \citet{2008JCoPh.227.3191B} proposed the WENO-Z weights to make the scheme truly fifth order everywhere:

\begin{align}
    \tau_5    &= |IS_0 - IS_2| \\
    \alpha_0 &= 1 + \frac{\tau_5}{\left(\epsilon_Z + IS_0\right)^2}, \\
    \alpha_1 &= 6 \left( 1 + \frac{\tau_5}{\left(\epsilon_Z + IS_1\right)^2}\right), \\
    \alpha_2 &= 3 \left( 1 + \frac{\tau_5}{\left(\epsilon_Z + IS_2\right)^2} \right),
\end{align}
and $\epsilon_Z = 10^{-40}$. As we will see this is useful in advection dominated problems, but reduces the general robustness of the scheme. For complex problems, WENO-Z will fall back into protection fluxes more often, which might not aid the solution for all problems.

\subsection{Constrained-Transport MHD} \label{sect.CT}

Equation \ref{eq.divb} requires maintainance of a zero magnetic field divergence at all times. In constrained transport (CT), the magnetic field is defined on a staggered mesh, i.e. on the faces of the computational zones \citep{1988ApJ...332..659E,1998ApJ...494..317D,1999JCoPh.149..270B,2000JCoPh.161..605T}. This way magnetic field divergence can be maintained to machine precision. For computational efficiency, we follow \citep[][]{1998ApJ...509..244R} in our implementation of CT, which uses second order accurate averages to obtain the corner fluxes. As we will see this choice introduces additional dispersion to the solution, but not additional diffusion. \par
We denote the magnetic field on the forward zone faces (i.e. $i+\frac{1}{2}$ for the x-component) as $\vec{b} = (b_x, b_y, b_z)^T$. From the Riemann solver, we obtain the fifth-order accurate magnetic field fluxes across the zone faces $f_y,f_z$ during the sweep in x-direction (component five and six), $g_z, g_x$ for the y-direction and $h_x, h_y$ for the z-direction. The scheme includes a correction for the convection of the magnetic field (\citet[][]{1998ApJ...509..244R}, compare to e.g. \citet{2019MNRAS.482..416V}):  
\begin{align}
    f^*_y &= f_y + \frac{1}{2} \left( B_{x,i,j,k} v_{y,i,j,k} + B_{x,i+1,j,k} v_{y,i+1,j,k} \right) \label{eq.fct1} \\
    f^*_z &= f_z + \frac{1}{2} \left( B_{x,i,j,k} v_{z,i,j,k} + B_{x,i+1,j,k} v_{z,i+1,j,k} \right) \label{eq.fct2}\\
    g^*_x &= g_x + \frac{1}{2} \left( B_{y,i,j,k} v_{z,i,j,k} + B_{y,i,j+1,k} v_{z,i,j+1,k} \right) \label{eq.fct3}\\
    g^*_z &= g_z + \frac{1}{2} \left( B_{y,i,j,k} v_{x,i,j,k} + B_{y,i,j+1,k} v_{x,i,j+1,k} \right) \label{eq.fct4}\\
    h^*_x &= h_x + \frac{1}{2} \left( B_{z,i,j,k} v_{x,i,j,k} + B_{z,i,j,k+1} v_{x,i,j,k+1} \right) \label{eq.fct5}\\
    h^*_y &= h_y + \frac{1}{2} \left( B_{z,i,j,k} v_{y,i,j,k} + B_{z,i,j,k+1} v_{u,i,j,k+1} \right) \label{eq.fct6}
\end{align}
Due to our implementation involving the rotation of the grid, the components actually remain the same for every direction. However, the fluxes have to be rotated backward ($g_x,g_z$) or forward ($h_x,h_y$) into the original frame. The corner fluxes $\Omega_x, \Omega_y, \Omega_z$ are then:
\begin{align}
    \Omega_x &= \frac{1}{2} \left( g^*_{x,i,j,k} + g^*_{x,i+1,j,k}\right) - \frac{1}{2} \left( f^*_{y,i,j,k} + f^*_{y,i,j+1,k} \right)  \label{eq.ox1}\\
    \Omega_y &= \frac{1}{2} \left( h^*_{y,i,j,k} + h^*_{y,i,j+1,k}\right) - \frac{1}{2} \left( g^*_{z,i,j,k} + g^*_{z,i,j,k+1} \right) \label{eq.ox2}\\
    \Omega_z &= \frac{1}{2} \left( f^*_{z,i,j,k} + f^*_{z,i,j,k+1}\right) - \frac{1}{2} \left( h^*_{x,i,j,k} + h^*_{x,i+1,j,k} \right)\label{eq.ox3}
\end{align}
The face centered magnetic fields are then updated analogous to equation \ref{eq.num}:
\begin{align}
    \frac{\mathrm{d}b_x}{\mathrm{d}t} &+ \frac{1}{\Delta x} \left( \Omega_{x,i,j,k} - \Omega_{x,i,j-1,k} \right) \nonumber\\ 
                                      &-  \frac{1}{\Delta x} \left( \Omega_{z,i,j,k} - \Omega_{z,i,j,k-1} \right) = 0 \\
    \frac{\mathrm{d}b_y}{\mathrm{d}t} &+ \frac{1}{\Delta x} \left( \Omega_{y,i,j,k} - \Omega_{y,i,j,k-1} \right) \nonumber\\ 
                                      &-  \frac{1}{\Delta x} \left( \Omega_{x,i,j,k} - \Omega_{x,i-1,j,k} \right) = 0 \\
    \frac{\mathrm{d}b_z}{\mathrm{d}t} &+ \frac{1}{\Delta x} \left( \Omega_{z,i,j,k} - \Omega_{z,i-1,j,k} \right) \nonumber\\ 
                                      &-  \frac{1}{\Delta x} \left( \Omega_{y,i,j,k} - \Omega_{y,i,j-1,k} \right) = 0
\end{align}
The zone centered magnetic field is interpolated at fourth order from face values:
\begin{align}
    B_{x,i,j,k} &= \frac{1}{16} \left( -b_{x,i-2,j,k} + 9 b_{x,i-1,j,k} + 9 b_{x,i,j,k} - b_{x,i+1,j,k}  \right),
\end{align}
the other component follow accordingly along the $j$ and $k$ index.\par
We use the same fourth-order Runge Kutta scheme presented in section \ref{sect.RK}. The CT update is interleaved with the WENO update, with boundary communication of $\Omega$ between a WENO and a CT step. As we will see, the CT scheme converges to second order. However, it can be shown that magnetic field dissipation converges to fifth order, so the additional error is in the shape of the field, not its energy (see also Jang et al. in prep.). 

 
\section{Implementation} \label{sect.code}

We implement the scheme in the numerical code \wombat\footnote{\url{wombatcode.org}}. For a detailed description of the code infrastructure and its performance, readers may consider \citet{2017ApJS..228...23M}. The code is written in object-oriented Fortran 2008 and hybrid parallelized with MPI and OpenMP. It decomposes the computational domain into \emph{patches}, blocks of independent work of variable size that can be cache blocked (16-32 zones per dimension depending on architecture). This way, communication is reduced to a boundary problem, i.e. the boundary (ghost) zones between patches need to be communicated (resolved). Patches are implemented as Fortran objects and store boundary zones alongside all necessary information about neighbouring patches and MPI ranks. Thus there is no global data structure keeping track of patch distribution that may grow with increasing number of MPI ranks, and communication and computation are intrinsically separated, i.e. the program is highly modular. \par
Initially, each MPI rank carries a cubic portion of the world grid (in 3D), a \emph{domain}, containing a large number of patches. Patches are load balanced via tunable virtual domains, where neighbouring MPI ranks are included into a ranks domain. Patches are off-loaded to other MPI ranks by exporting them into the virtual domain region and are then communicated analogous to the boundary exchange. \par
OpenMP threads are implemented using a \emph{single} parallel region and combined with \mbox{MPI\_THREAD\_MULTIPLE}, so that threads independently and asynchronously carry out MPI communication of boundaries and resolution of patches (computational work). \wombat currently uses one-sided MPI-RMA with calls to \mbox{MPI\_Put} to place each of the 27 patch boundary regions in a neighbours mailbox of user-definable size. A signal per mailbox of 8 bytes (the ''heartbeat'') is used to notify the neighbouring MPI rank of completed communication. The signal is updated at every loop iteration even if no data was communicated, to achieve a form of weak synchronization among ranks.  Once every boundary is communicated (i.e. all the signals are set as completed), a thread on the other rank unpacks all boundaries from the mailbox into the corresponding patch object. The patch is then ''resolved'' by any OpenMP thread on the rank.  This way communication can react to load-imbalance and also network contention on large multi-user machines. This represents fine-grained communication-computation overlap on the thread level. This implementation shifts the communication pattern from few large MPI messages requiring large buffers, to many small messages with accordingly smaller buffers. Thus this approach requires the MPI library to support lock-free OpenMP and lowest overhead, which is currently the case for Cray's MPI library and also OpenMPI, where \wombat is part of the regression testing. The communication scheme is actively researched for exa-scale systems by performance engineers at Cray Inc and thus subject to continuous improvements.

\subsection{The WENO Solver}

In \wombat, every solver works on a single boundary communicated patch, where the grid data resides in rank global memory. At this point, no communication is required anymore, work and communication are completely separate in the implementation. This eases maintenance of the code considerably.\par
In the WENO solver, we first copy the grid data into OpenMP thread private buffers, to minimize Non Uniform Memory Access (NUMA) effects. This introduces memory overhead of about 25 MB per thread for patches with $18^3$ zones. In multiple dimensions, the grid is explicitly flattened as well, i.e. the number of indices of all arrays is reduced to one along spatial dimensions and all loops run over the flattened array, ignoring the boundaries between dimensions. Thus memory and code-level layout of the data are identical. This increases vector length and decreases cache blocked patch size, which in-turn eases MPI load-balancing. It also improves code readability, as all lower level routines (fluxes, eigenvectors) remain inherently one dimensional. This layout naturally leads to an explicit separation of computation and data movement in the code. In two and three dimensions, the sweeps in y-direction and z-direction are implemented as rotations of the plane/cube in flattened memory. The resulting WENO fluxes are later rotated back into the original coordinate system to compute the state-vector updates (eq. \ref{eq.num} and section \ref{sect.RK}).  Using the Fortran {\small CONTIGUOUS} keyword and implied array sizes, \emph{all} relevant loops in the implementation auto-vectorize with current Cray and Intel compilers. This is a necessary pre-requisite to achieve a significant fraction of peak performance on current CPUs. We anticipate that this implementation will be very convenient to port to accelerators (GPUs) as it naturally exposes long vectors and makes memory movement explicit. \par
Due to \wombat's parallelization strategy the RK4 scheme presented in section \ref{sect.RK} is implemented using a running state vector $\vec{q}$,  a state vector buffer $\vec{q}_\mathrm{save}$ to accumulate intermediate results and a copy of the initial state $\vec{q}_0 = \vec{q}^n_i$. Note that we communicate boundaries twice (once for WENO, once for CT update) per RK4 sub-step. This allows us to retain 3 boundary zones on the patch, which is optimal regarding Flops and memory consumption given that communication in \wombat is local, asynchronous and thus extremely cheap.

\subsubsection{WENO-WOMBAT in one Dimension }

The program proceeds as follows:

\begin{enumerate}
    \item Copy-in $\vec{q}$. If executing the first substep, initialize $\vec{q}_0$ and set $\vec{q}_\mathrm{save}$.
    \item Compute cell centered pressure $P$ eq. \ref{eq.P},  eigenvalues $c_\mathrm{A}, c_\mathrm{fast}, c_\mathrm{slow}$  eq. \ref{eq.slow} to \ref{eq.ca} and fluxes $\vec{F}$ eq. \ref{eq.fluxes}.
    \item Compute eigenvectors $\vec{L}$ and $\vec{R}$ on the cell boundaries, compute WENO fluxes, eq. \ref{eq.FWENO}.
    \item Update $\vec{q}$ using WENO fluxes, update $\vec{q}_\mathrm{save}$ using new $\vec{q}$.
    \item Move $\vec{q}, \vec{q}_0, \vec{q}_\mathrm{save}$ on the patch, communicate patch boundaries.
\end{enumerate}

Repeat 4 times with corresponding Runge-Kutta factors so that $\vec{q}^{n+1} = \vec{q}_\mathrm{save}$ at the last substep.

\subsubsection{WENO-WOMBAT in two Dimensions }

In two dimensions, we have to interleave the CT step into the WENO5 RK4 update. The program proceeds as follows:

\begin{enumerate}
    \item Copy-in and flatten $\vec{q}$. If executing first substep, initialize $\vec{q}_0$ and set $\vec{q}_\mathrm{save}$, else copy-in and flatten them. Set a buffer $\vec{q}_\mathrm{buf} = \vec{q}_0$.
    \item Compute from $\vec{q}$ cell centered pressure $P$ eq. \ref{eq.P},  eigenvalues $c_\mathrm{A}, c_\mathrm{fast}, c_\mathrm{slow}$  eq. \ref{eq.slow} to \ref{eq.ca} and fluxes $\vec{F}$ eq. \ref{eq.fluxes}.
    \item Compute from $\vec{q}$ eigenvectors $\vec{L}$ and $\vec{R}$ on the cell boundaries, compute WENO fluxes, eq. \ref{eq.FWENO}, compute 2d CT-fluxes eq. \ref{eq.fct1}, \ref{eq.fct2}.
    \item Update $\vec{q}_\mathrm{buf}$ using WENO fluxes
    \item Rotate $\vec{q}$ forward so flattened index is along y-direction.
    \item Compute from $\vec{q}$ the cell centered pressure $P$ eq. \ref{eq.P},  eigenvalues $c_\mathrm{A}, c_\mathrm{fast}, c_\mathrm{slow}$  eq. \ref{eq.slow} to \ref{eq.ca} and fluxes eq. \ref{eq.fluxes}.
    \item Compute from $\vec{q}$, eigenvectors $\vec{L}$ and $\vec{R}$ on the cell boundaries, WENO fluxes, eq. \ref{eq.FWENO}. The compute 2d CT-fluxes \ref{eq.fct3}, \ref{eq.fct4}.
    \item Rotate $\vec{q}$ and CT-fluxes backwards.
    \item Update $\vec{q}_\mathrm{buf}$ using WENO fluxes, copy $\vec{q}_\mathrm{buf}$ into $\vec{q}$, compute CT corner flux $\Omega_x$ eq. \ref{eq.ox1}.
    \item Update $\vec{q}_\mathrm{save}$ using new $\vec{q}$, but not the magnetic field.
    \item Move $\vec{q}, \vec{q}_0, \vec{q}_\mathrm{save}$, $\Omega_x$ on the patch, communicate patch boundaries, face centered magnetic fields and corner flux.
    \item Copy-in and flatten $\vec{q}$,$\vec{b}$, $\Omega_x$.
    \item Update face centered magnetic field $\vec{b}$ from $\Omega_x$.
    \item Interpolate face centered magnetic field $\vec{b}$ to zone centered magnetic field $\vec{B}$ at fourth order. Update $\vec{q}_\mathrm{save}$ with new magnetic field.
    \item Move $\vec{q},\vec{b}, \vec{q}_\mathrm{save}$ to the patch object.
\end{enumerate}
Repeat 4 times with corresponding Runge-Kutta factors so that $\vec{q}^{n+1} = \vec{q}_\mathrm{save}$ at the last substep. Perform another CT update on $\vec{q}_\mathrm{save}$ before moving it into $\vec{q}$.

\subsubsection{WENO-WOMBAT in three Dimensions }

In three dimensions the program proceeds as follows:

\begin{enumerate}
    \item Copy-in and flatten $\vec{q}$. If executing first substep, initialize $\vec{q}_0$ and set $\vec{q}_\mathrm{save}$, else copy-in and flatten them. Set a buffer $\vec{q}_\mathrm{buf} = \vec{q}_0$.
    \item Compute from $\vec{q}$ cell centered pressure $P$ eq. \ref{eq.P},  eigenvalues $c_\mathrm{A}, c_\mathrm{fast}, c_\mathrm{slow}$  eq. \ref{eq.slow} to \ref{eq.ca} and fluxes $\vec{F}$ eq. \ref{eq.fluxes}.
    \item Compute from $\vec{q}$ eigenvectors $\vec{L}$ and $\vec{R}$ on the cell boundaries, compute WENO fluxes, eq. \ref{eq.FWENO}, compute 3d CT-fluxes eq. \ref{eq.fct1} \ref{eq.fct2}.
    \item Update $\vec{q}_\mathrm{buf}$ using WENO fluxes
    \item Rotate $\vec{q}$ and $\vec{q}_\mathrm{buf}$ forward so flattened index is along y-direction.
    \item Compute from $\vec{q}$ cell centered pressure $P$ eq. \ref{eq.P},  eigenvalues $c_\mathrm{A}, c_\mathrm{fast}, c_\mathrm{slow}$  eq. \ref{eq.slow} to \ref{eq.ca} and fluxes eq. \ref{eq.fluxes}.
    \item Compute from $\vec{q}$ eigenvectors $\vec{L}$ and $\vec{R}$ on the cell boundaries, compute WENO fluxes, eq. \ref{eq.FWENO}, compute 3d CT-fluxes  eq. \ref{eq.fct3} \ref{eq.fct4}.
    \item Update $\vec{q}_\mathrm{buf}$ using WENO fluxes,
    \item Rotate $\vec{q}$ and $\vec{q}_\mathrm{buf}$ forward so flattened index is along z-direction. Rotate CT-fluxes backwards.
    \item Compute from $\vec{q}$ the cell centered pressure $P$ eq. \ref{eq.P},  eigenvalues $c_\mathrm{A}, c_\mathrm{fast}, c_\mathrm{slow}$  eq. \ref{eq.slow} to \ref{eq.ca} and fluxes eq. \ref{eq.fluxes}.
    \item Compute from $\vec{q}$ eigenvectors $\vec{L}$ and $\vec{R}$ on the cell boundaries, compute WENO fluxes, eq. \ref{eq.FWENO}, compute 3d CT-fluxes using eq. \ref{eq.fct5} \ref{eq.fct6}.
    \item Update $\vec{q}_\mathrm{buf}$ using WENO fluxes
    \item Rotate forward $\vec{q}$ and $\vec{q}_\mathrm{buf}$ and CT-fluxes, so the first index runs along x-direction again.
    \item Update $\vec{q}_\mathrm{buf}$ using WENO fluxes, copy $\vec{q}_\mathrm{buf}$ into $\vec{q}$, compute CT corner flux  eq. \ref{eq.ox1}-\ref{eq.ox3} from rotated CT fluxes.
    \item Update $\vec{q}_\mathrm{save}$ using new $\vec{q}$, but not the magnetic field.
    \item Move $\vec{q}, \vec{q}_0, \vec{q}_\mathrm{save}$, $\vec{\Omega}$ on the patch, communicate patch boundaries, face centered magnetic fields $\vec{b}$ and corner flux.
    \item Copy-in and flatten $\vec{q}$,$\vec{b}$, $\vec{\Omega}$.
    \item Update face centered magnetic field $\vec{b}$ from $\vec{\Omega}$.
    \item Interpolate face centered magnetic field $\vec{b}$ to zone centered magnetic field $\vec{B}$ at fourth order. Update $\vec{q}_\mathrm{save}$ with new magnetic field.
    \item Move $\vec{q},\vec{b}, \vec{q}_\mathrm{save}$ to the patch object.
\end{enumerate}
Repeat 4 times with corresponding Runge-Kutta factors so that $\vec{q}^{n+1} = \vec{q}_\mathrm{save}$ at the last substep. Perform another CT update on $\vec{q}_\mathrm{save}$ before moving it into $\vec{q}$.

\section{Test Calculations} \label{sect.tests}

All simulations in this section are run with $CFL = 0.8$ in 8 byte precision. We also do not use protection fluxes or density/pressure floors, unless noted otherwise. State vectors are defined as primitive variables $\vec{U} = (\rho, v_\mathrm{x}, v_\mathrm{y}, v_\mathrm{z}, B_\mathrm{x}, B_\mathrm{y}, B_\mathrm{z}, P)^T$, or as conserved variables $\vec{Q} = (\rho, \rho v_\mathrm{x}, \rho v_\mathrm{y}, \rho v_\mathrm{z}, B_\mathrm{x}, B_\mathrm{y}, B_\mathrm{z}, E)^T$, where $E(\vec{U})$ can be obtained from equation \ref{eq.P}. We compare our WENO5 or WENO5-Z results with \wombat's second order TVD+CTU implementation presented in \citet{2017ApJS..228...23M}. The $L_1$ error norm is given by:
\begin{align}
    L_1(\vec{q}) & = \sqrt{ \sum\limits_s \left(\sum\limits^{N_\mathrm{Zones}}_{i=1} \frac{\left| \vec{q}_i(t) - \vec{q}_i(0) \right|}{N} \right)^2},
\end{align}
where $s$ denotes the components of the state vector. We are using the geometrical norm of the $L_1$ error to keep the results comparable to \citet{2005JCoPh.205..509G,2008JCoPh.227.4123G,2008ApJS..178..137S,2018JCoPh.375.1365F}. Other definitions are in use as well, e.g. the mean of $L_1$.
\subsection{Linear Wave Convergence} \label{sect.3dwaves}

\begin{figure*}
    \centering
    \includegraphics[width=0.45\textwidth]{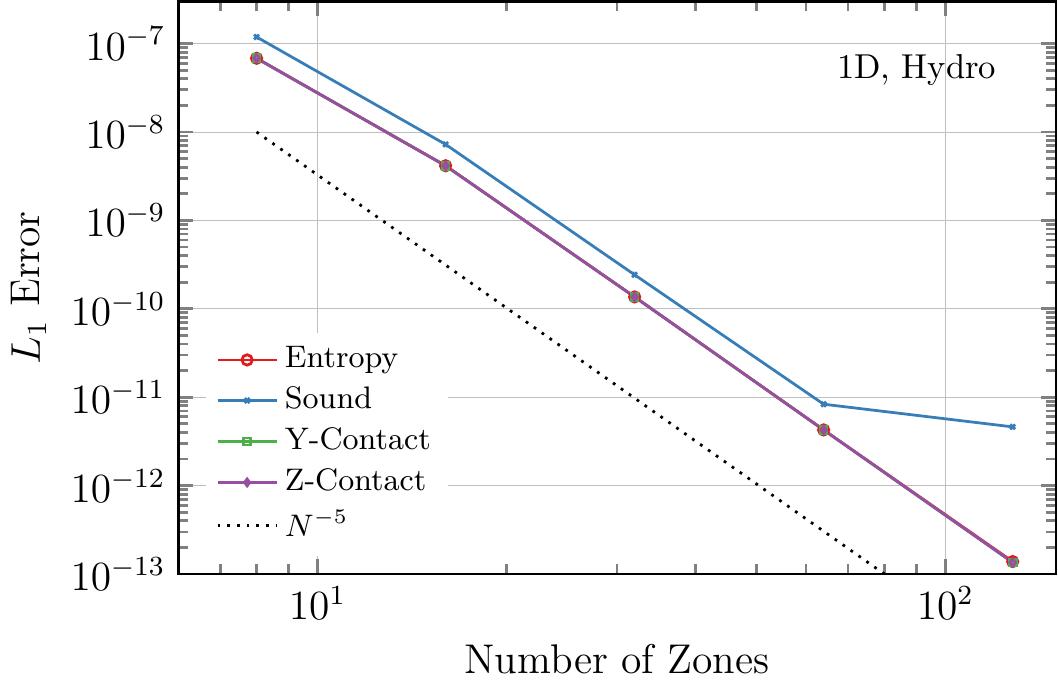}
    \includegraphics[width=0.45\textwidth]{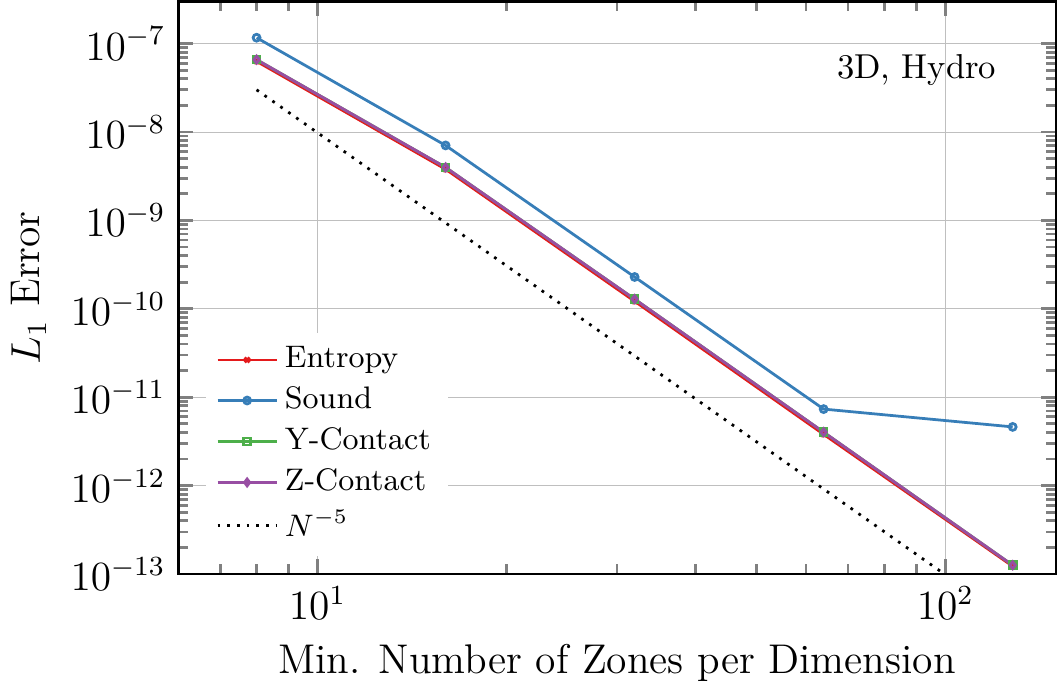}\\
    \includegraphics[width=0.45\textwidth]{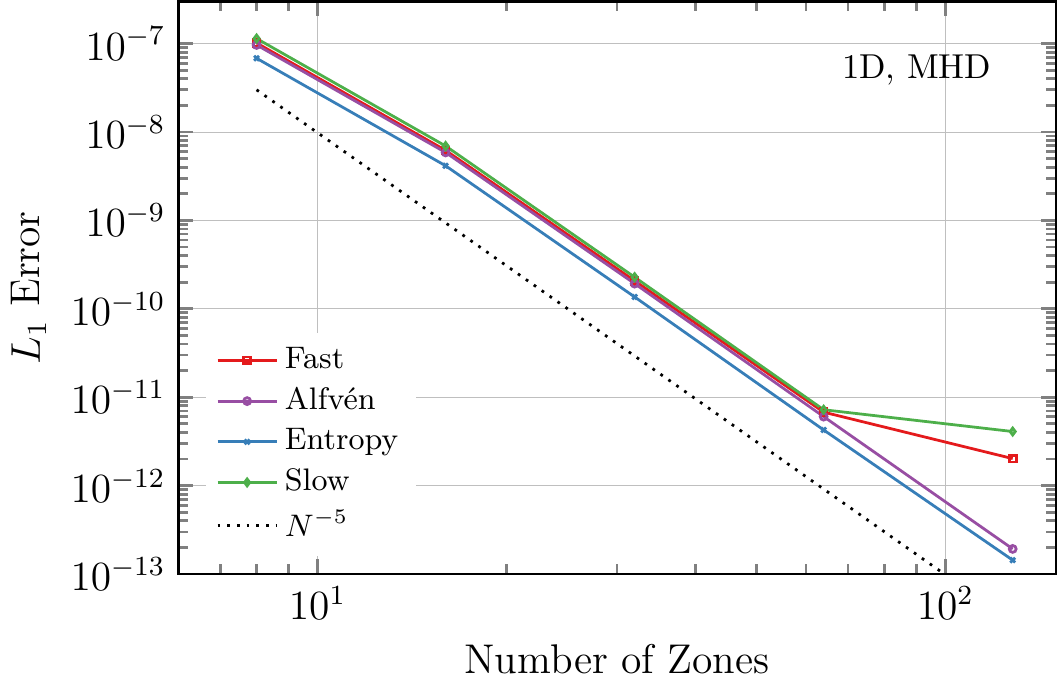}
    \includegraphics[width=0.45\textwidth]{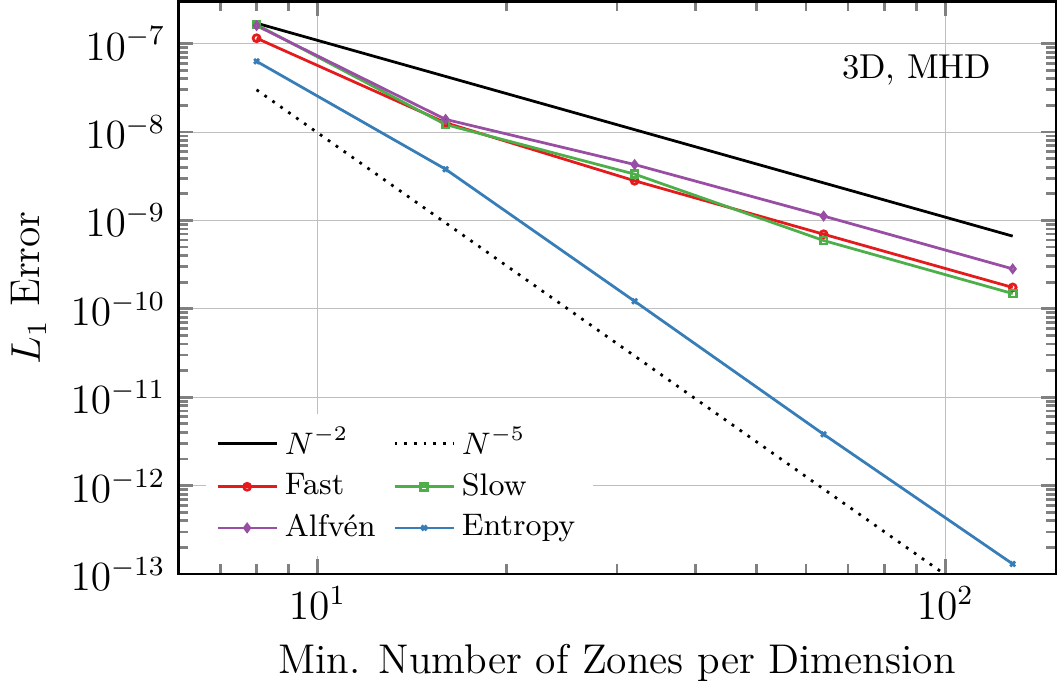}
    \caption{$L_1$ error over resolution for advecting hydrodynamic (top) and MHD waves (bottom) in one dimension (left) and in three dimensions (right) with WENO5-Z across a unit computational domain for one wave length following \citep{2008JCoPh.227.4123G}. Fifth order was marked as dotted black line, second order as black line.}
    \label{fig.waves}
\end{figure*}

\begin{figure}
    \centering
    \includegraphics[width=0.45\textwidth]{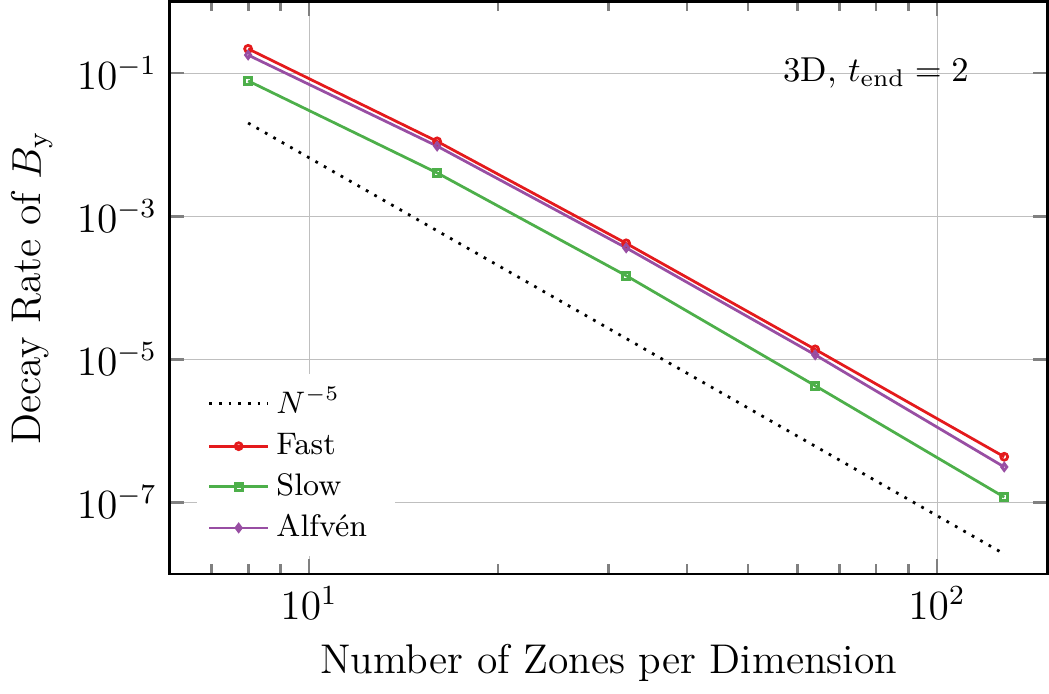}    
    \caption{Decay rate (eq. \ref{eq.decay_rate}) of the y-component of the magnetic field between $t=0$ and $t=2$ for the fast mode (red), the slow mode (green) and the Alfv\'{e}n mode (purple)  with WENO5-Z.}\label{fig.decay}
\end{figure}
 
The convergence order of the scheme can be exposed by advecting a perturbation corresponding to one of the (M)HD eigenvectors through a periodic box with $2N\times N \times N$ zones and a domain of $3\times 1.5 \times  1.5$ in three dimension. This test evaluates the code in the linear regime and is very valuable for debugging. As the setup is not trivial, especially in 3 dimensions, we describe it in greater detail here, but note that the \wombat or \athena source code is likely indispensible to fully reproduce the results.\par
Following \citet{2005JCoPh.205..509G,2008JCoPh.227.4123G,2008ApJS..178..137S}, we set $\gamma = 5/3$ and $\bar{\vec{U}} = (1,v_x,0,0,B_x,B_y,B_z,1/\gamma)^T$, where $v_x = 1$ for shear and entropy modes, and $v_x = 0$ otherwise. For hydro waves, $\vec{B} = 0$, for MHD waves $\vec{B} = (1,\sqrt{2},1/2)^T$. We add a perturbation on the conserved variables: $\vec{Q} = \bar{\vec{Q}}  + A_0 \vec{R}_k\sin(2\pi x)$. Here  $\vec{R}_k$ is the right hand eigenvector of mode $k$ and $A_0 = 10^{-6}$. We note that the perturbation is applied only once per component, i.e. the momentum fluctuation uses the unperturbed background density. The right eigenvectors for the (M)HD waves are:
\begin{align}
    \vec{R}_\mathrm{Alfven} &= \frac{1}{6\sqrt{5}} \left(0,0,1,-2\sqrt{2},0,-1,2\sqrt{2},0\right)^T \\
    \vec{R}_\mathrm{fast}   &= \frac{1}{6\sqrt{5}} \left(6, 12, -4\sqrt{2},-2,0, 8\sqrt{2},4,27 \right)^T \\
    \vec{R}_\mathrm{slow}   &= \frac{1}{6\sqrt{5}} \left(12,6,8\sqrt{2},4,0,-4\sqrt{2},-2,9  \right)^T\\
    \vec{R}_\mathrm{entropy}&= \left(1,1,0,0,0,0,0,\frac{1}{2} \right)^T. \label{eq.entropyEV} \\
    \vec{R}_\mathrm{sound}  &= \left(1,1,0,0,0,0,0,1/(\gamma-1) \right)^T\\
    \vec{R}_\mathrm{shear,y}&= \left(0,0,1,0,0,0,0,0 \right)^T
\end{align}
In three dimensions, we rotate the zone positions parallel to the wave vector following \citet{2008JCoPh.227.4123G}, with the rotation matrix:
\begin{align}
    M(\alpha,\beta) &= \begin{bmatrix} 
                                \cos\alpha \cos\beta & -\sin\beta & \sin\alpha \cos\beta \\
                                \cos\alpha \sin\beta & \cos\beta & -\sin\alpha\sin\beta \\
                                 \sin\alpha & 0 & \cos\alpha \\
                             \end{bmatrix}, 
\end{align}
where $\sin\alpha = 2/3$ and $\cos\beta = 2/\sqrt{5}$. We use its inverse/transpose to rotate back into the lab frame. We set a magnetic vector potential:
\begin{align}
    A_x &= 0 \\
    A_y &= \frac{A_0}{2\pi} R_{k,7} \sin(2\pi x) + B_z x_k \\
    A_z &= -\frac{A_0}{2\pi} R_{k,6} \sin(2\pi x) -  B_y x_k +  B_x y_k,
\end{align}
where $x_k$ and $y_k$ are the x and y components of the zone position rotated with the inverse rotation matrix. We note that the vector potential has to be evaluated on the edges of the zone in three dimension, so that the resulting magnetic field is field centered. E.g. in our case $A_x$ is defined at ($i, j+\frac{1}{2}, k+\frac{1}{2}$). The vector potential is then multiplied with the forward rotation matrix and the magnetic field on the interfaces is obtained using a standard first order finite difference rotation operator. We observe $\nabla \cdot \vec{B} < 10^{-12}$ in all tests at all times. \par
In figure \ref{fig.waves} we plot the resulting $L_1$ error for resolutions between $8$ and $128$ zones run with WENO5-Z. Top left to bottom right  we show 1D hydro waves, 3D hydro waves, 1D MHD waves and 3D MHD waves. In the 1D case, the simulation converges as $N^{-5}$ as expected down to $10^{-11}$, where the effect of wave steepening begins to limit further convergence of the compressive modes. We note that for $N > 128$, the round off error from the 8 byte variables leads to an increase in $L_1$ again \citep[see also][]{2018Galax...6..104D}. This suggests that the scheme needs to be run with at least 8 byte precision to take advantage of the high order convergence properties. \par
In three dimensions, all hydro waves converge to fifth order. The entropy MHD wave converges to fifth order in $L_1$ as well. This mode does not feature a fluctuation in the magnetic field (eq.\ref{eq.entropyEV}), the background magnetic field is only advected. All other MHD waves converge to second order in $L_1$, because our CT scheme is geometrically only second order. However, the fluxes used are of course fifth order accurate, thus this error should be dominated by dispersion, not dissipation \citep{L2002}. The $L_1$ norm measures both errors, dispersion and diffusion. \par
To further investigate this, we consider the decay rate of the magnetic field in the wave that is sensitive only to the dissipation error of the scheme (numerical diffusion), not the dispersion of the wave. The decay rate $\tau_\mathrm{D}$ of $B_y$ is defined as:
\begin{align}
    \tau_\mathrm{D}(B_y) &= -\frac{1}{t_\mathrm{end}}\log\left( \frac{\delta B_{y}(t=t_\mathrm{end})}{\delta B_{y}(t=0)} \right), \label{eq.decay_rate}
\end{align}
where $\delta B_{y}(t)$ is the root-mean-square magnetic field fluctuation in the frame parallel to the wave vector. We plot the decay rate of the fast, slow and Alfv\'{e}n waves in figure \ref{fig.decay} between the initial conditions and time $t = 2$. The time was chosen so that all wave families have been advected through the domain at least once. It is important to evaluate the decay time at a late enough time, as the RMS values fluctuate with time, indicative of additional waves being present in test setup. Fifth order convergence of the decay rate is observed for all three remaining MHD waves.\par
This confirms that the error introduced by the second order CT scheme is indeed in the form of wave dispersion only, not dissipation. I.e. the price paid for the computationally cheap constrained transport method from \citet{1998ApJ...509..244R} is in the form of magnetic field shape, but not magnetic energy conservation. This view is supported by results in the Alfv\'{e}n wave and magnetic field loop advection tests shown later in the paper. Nonetheless, a comparison of the absolute value of the $L_1$ error with \citet{2008ApJS..178..137S} (their figure 32) yields that CT-WENO5 increases the effective resolution compared to \athena in three dimensions by roughly a factor two, even for linear MHD waves  propagating with an oblique angle to the grid in three dimensions. The $L_1$ error is significantly smaller even at $N=16$, so the statement is true despite the second order convergence for MHD waves. 

\subsection{Circularly Polarized Alfv\'{e}n Wave}

\begin{figure*}
    \centering
    \includegraphics[width=0.45\textwidth]{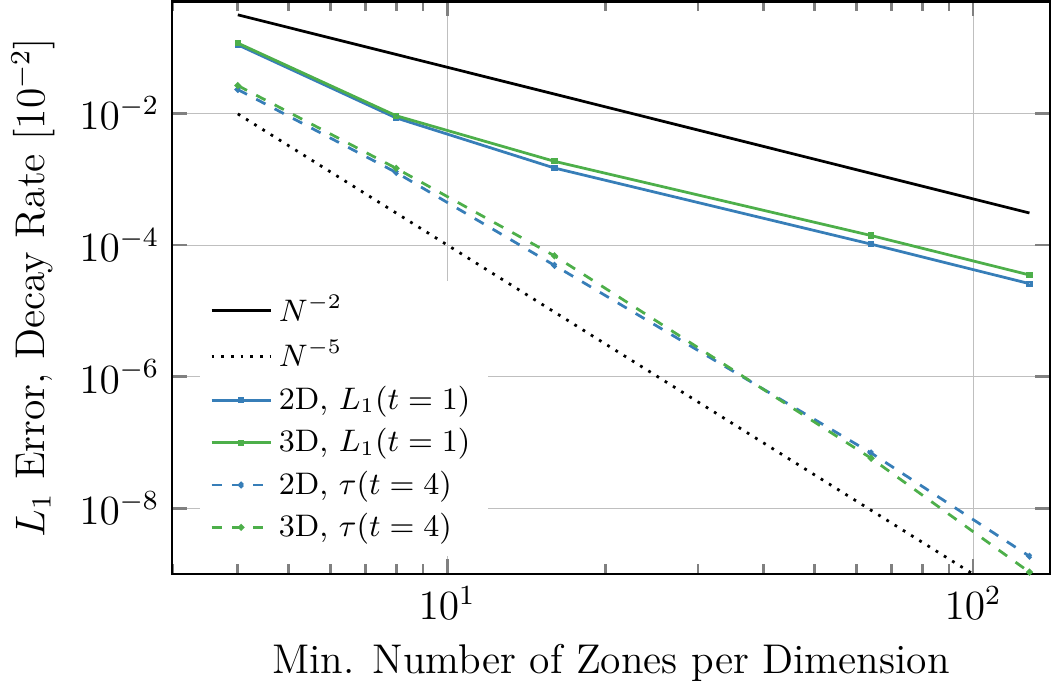}
    \includegraphics[width=0.45\textwidth]{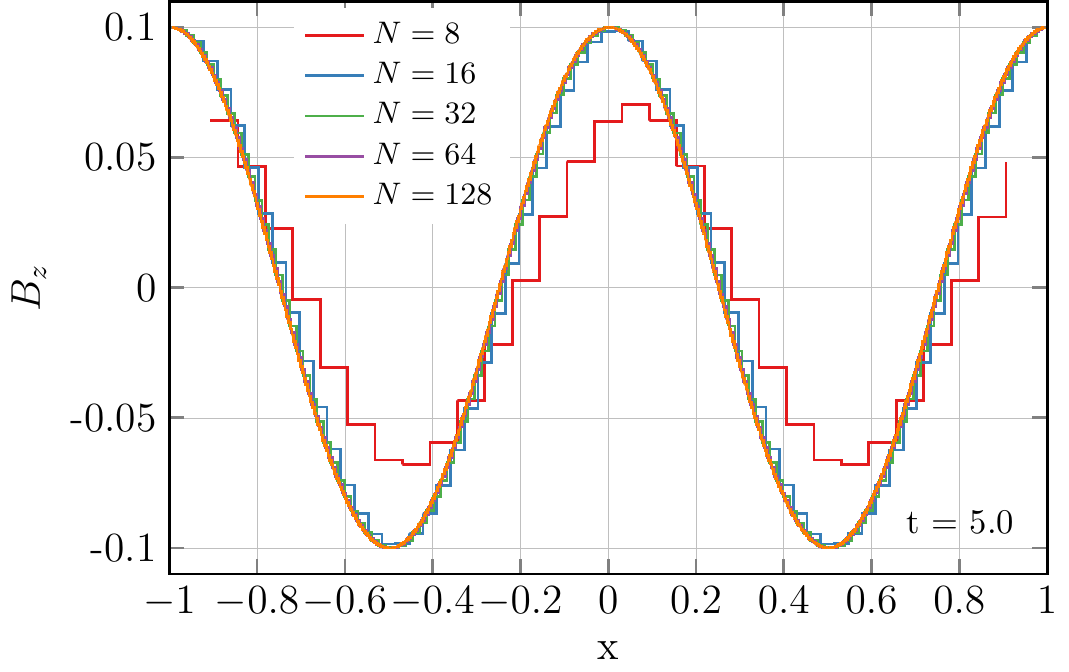}
    \includegraphics[width=0.45\textwidth]{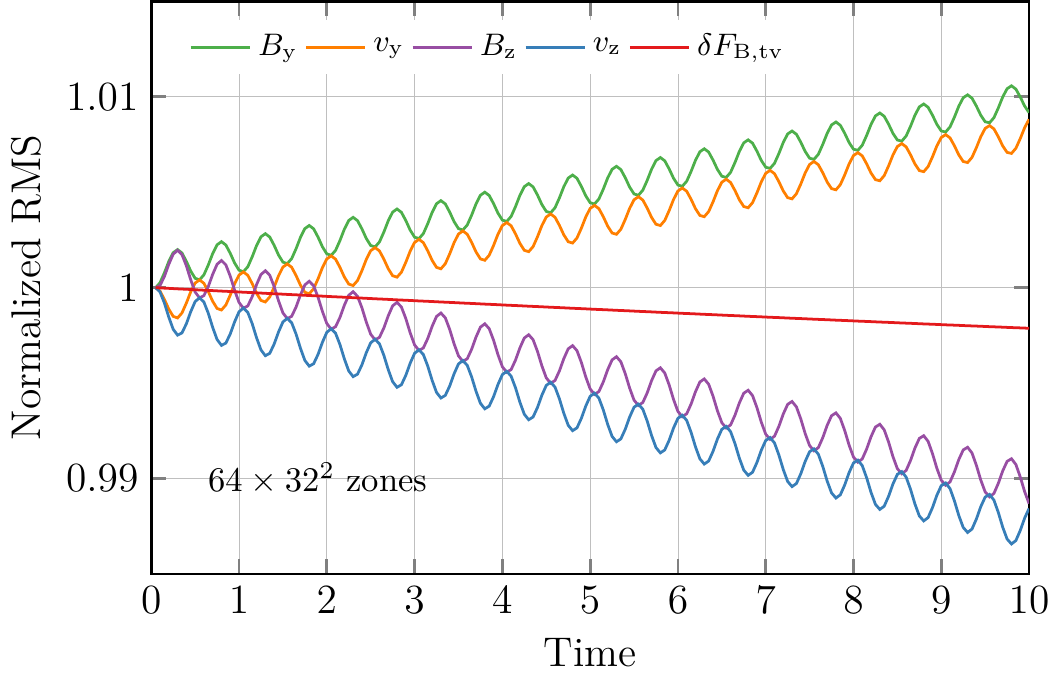}
    \caption{Top left: $L_1$ error of the circularly polarized Afv\'{e}n wave test over resolution in 2d (blue) and 3d (green). Decay rate of transverse magnetic flux as dashed lines in units of 100. We mark second order as solid and fifth order as dotted black lines   Top right: $B_z$ for the circularly polarized Alfv\'{e}n wave test in two dimensions at $t=5$ with $ N \in [8,16,32,64,128]$ in red, blue, green, purple and orange, respectively. Right: Time evolution of root mean square fluctuations around the background state in the Alfv\'{e}n wave test with $ 64\times 32^2$ zones for the state vector components that carry a fluctuation.}\label{fig.cpaw}
\end{figure*}

Polarized Alfv\'{e}n waves are important in heating the solar corona due to instabilities, and have been studied for many decades \citep[e.g.][]{1978ApJ...219..700G,2001A&A...367..705D,2005SSRv..121..287R,2007Sci...318.1574D}. As a numerical test, polarized Alfv\'{e}n waves can be used to evaluate the fidelity of the CT scheme \citep{2000JCoPh.161..605T}.\par
Following \citet{2005JCoPh.205..509G}, we set $\vec{U}=(1,0,v_y,v_z,1,B_y,B_z,0.1)^T$, with $v_y = B_y = 0.1 \sin(2\pi x)$, $v_z = B_z = 0.1 \cos(2\pi x)$. In two dimensions, we rotate $\vec{U}$ by the second Euler angle $\theta=\tan^{-1}(2) \approx 63.5 \,\mathrm{deg}$. The computational domain has $2N \times N $  zones and covers $L_\mathrm{Box,x}, L_\mathrm{Box,y} = \sqrt{5},\sqrt{5}/2$. In three dimensions, the wave is rotated similarly to section \ref{sect.3dwaves}. The computational domain of $2N \times N \times N$ zones covers $(L_x, L_y, L_z) = (3, 3/2, 3/2)$. \citet{2005JCoPh.205..509G} did not find an instability using these parameters. We note that the analytical RMS at $t=0$ is given as $0.1/\sqrt{2}$ for fluctuating quantities, while for all other quantities the initial RMS is zero. Thus the former errors dominate the simulation, while the latter quantities are subject to truncation errors from the rotation operation. \par
On the top left of figure \ref{fig.cpaw}, we show the $L_1$ error norm in 2d (blue) and 3d (green) over number of zones after the wave traveled for five wave period ($c_\mathrm{A} = 1$) in a frame parallel to the wave vector. The scheme converges at second order in $L_1$. This is in-line with the result from the previous linear wave convergence test. \par
We also show the decay rate convergence as dashed lines in figure \ref{fig.cpaw}. However, this time  oscillations in the RMS of the state vector components can lead to negative decay rates even at rather late times. In figure \ref{fig.cpaw} bottom, we show these RMS fluctuations over time for $64\times 32^2$ zones in three dimensions for the components that carry a fluctuation in the initial conditions. Dissipation in the solution is observed as asymmetry in the RMS at late times.  To obtain a robust measure for the wave decay, one may consider the RMS of the transverse magnetic flux in the wave (red line): 
\begin{align}
    \delta F_\mathrm{B,tv} &= \mathrm{RMS}\left(\sqrt{ \left( B_\mathrm{y} v_\mathrm{y}\right)^2 + \left( B_\mathrm{z}  v_\mathrm{z}\right)^2}\right)
\end{align}
For this quantity, fifth order convergence is observed at all times in figure \ref{fig.cpaw}, top left. \par
In figure \ref{fig.cpaw}, top right we show the z-component of the magnetic field in two dimensions after advecting for five wave periods for resolutions of $ N \in [8,16,32,64,128]$ in red, blue, green, purple and orange, respectively. Only the lowest resolution shows some dissipation, dispersion is observed for the two lowest resolution. This is a significant improvement over the results shown in \citet{2000JCoPh.161..605T} for the same CT scheme. It also compares favourably with \wombat's TVD+CTU implementation (see \citet{2017ApJS..228...23M}) and matches results from the \athena solver (\citet{2008JCoPh.227.4123G}, their figure 4), despite the simpler CT scheme.

\subsection{RJ95 2a Shock Tube}
\begin{figure*}
    \centering
    \includegraphics[width=\textwidth]{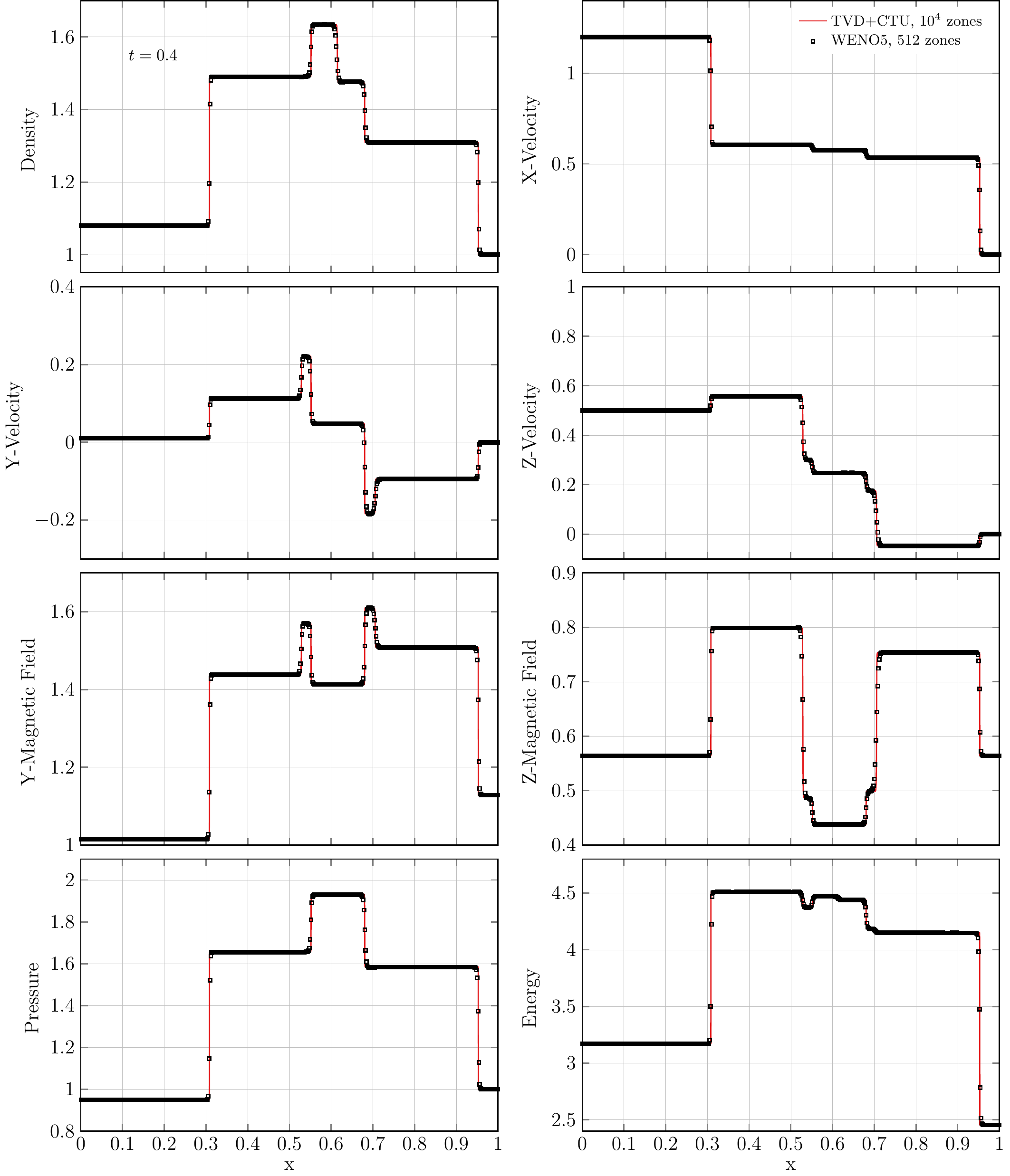}
    \caption{Shock tube test 2a from \citet{1995ApJ...442..228R} at time $t = 0.2$ in one dimension. TVD+CTU result with 10000 zones as red line, WENO5 result with 512 zones as black squares. Density, velocity components, magnetic field components, pressure and energy are shown from top left to bottom right.}
    \label{fig.rj2a_1d}
\end{figure*}

\begin{figure*}
    \centering
    \includegraphics[width=\textwidth]{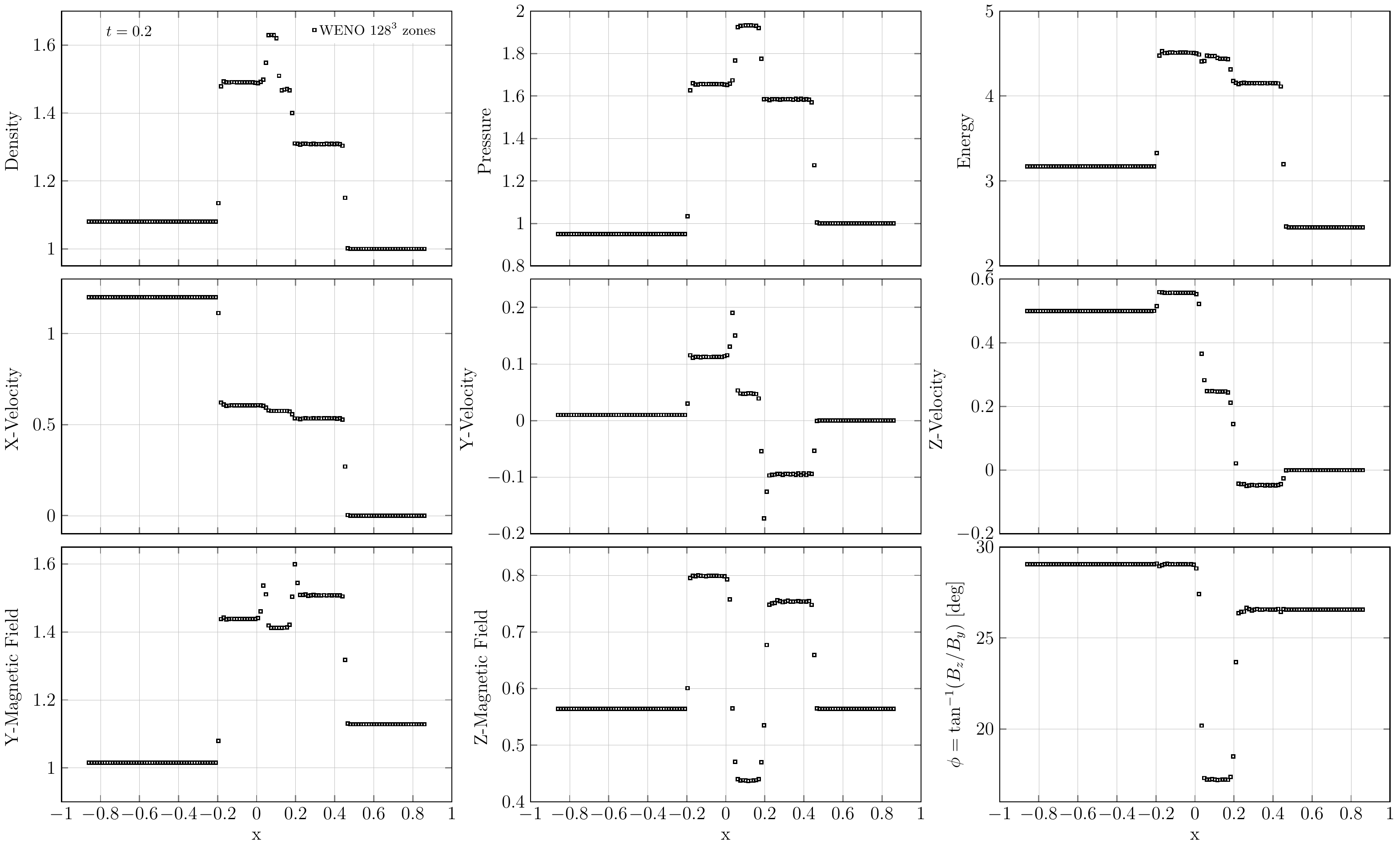}
    \caption{Shock tube test 2a from \citet{1995ApJ...442..228R} at time $t = 0.2$ in three dimensions along the zone diagonal of the computational domain. WENO5 result with $128^3$ zones as black squares. Density, pressure, energy, velocity components, magnetic field components and magnetic field angle $\phi$ in degrees are shown from top left to bottom right.}
    \label{fig.rj2a_3d}
\end{figure*}

We compute shock tube number 2a from \citet{1995ApJ...442..228R} with \wombat's TVD+CTU and WENO5 solver. The shock tube features all four MHD modes. The left hand state vector is $\vec{U}_\mathrm{L} = (1.08,1.2,0.01,2/\sqrt{4\pi},3.6/\sqrt{4\pi},2/\sqrt{4\pi},0.95)^T$. The right hand state vector is $\vec{U}_\mathrm{R} = (1,0,0,0,2/\sqrt{4\pi},4/\sqrt{4\pi},2/\sqrt{4\pi},1)^T$. We plot the TVD+CTU result with $10^4$ zones as red line and the WENO5 result with 256 zones as black circles. The evolved state vector components at $t = 0.2$ for the 1 dimensional computation are found in figure \ref{fig.rj2a_1d}, for 3 dimensions in figure \ref{fig.rj2a_3d}. For 3 dimensions, we rotated the shock normal along $\vec{k} = (1,1,1)^T$ in a $128^3$ zone simulation. The primitive variables alongside the magnetic field angle $\phi = \arctan^{-1}(B_y/B_z)$ in degrees are shown in figure \ref{fig.rj2a_3d}.
  
\subsection{RJ95 4d Shock Tube}

\begin{figure*}
    \centering
    \includegraphics[width=\textwidth]{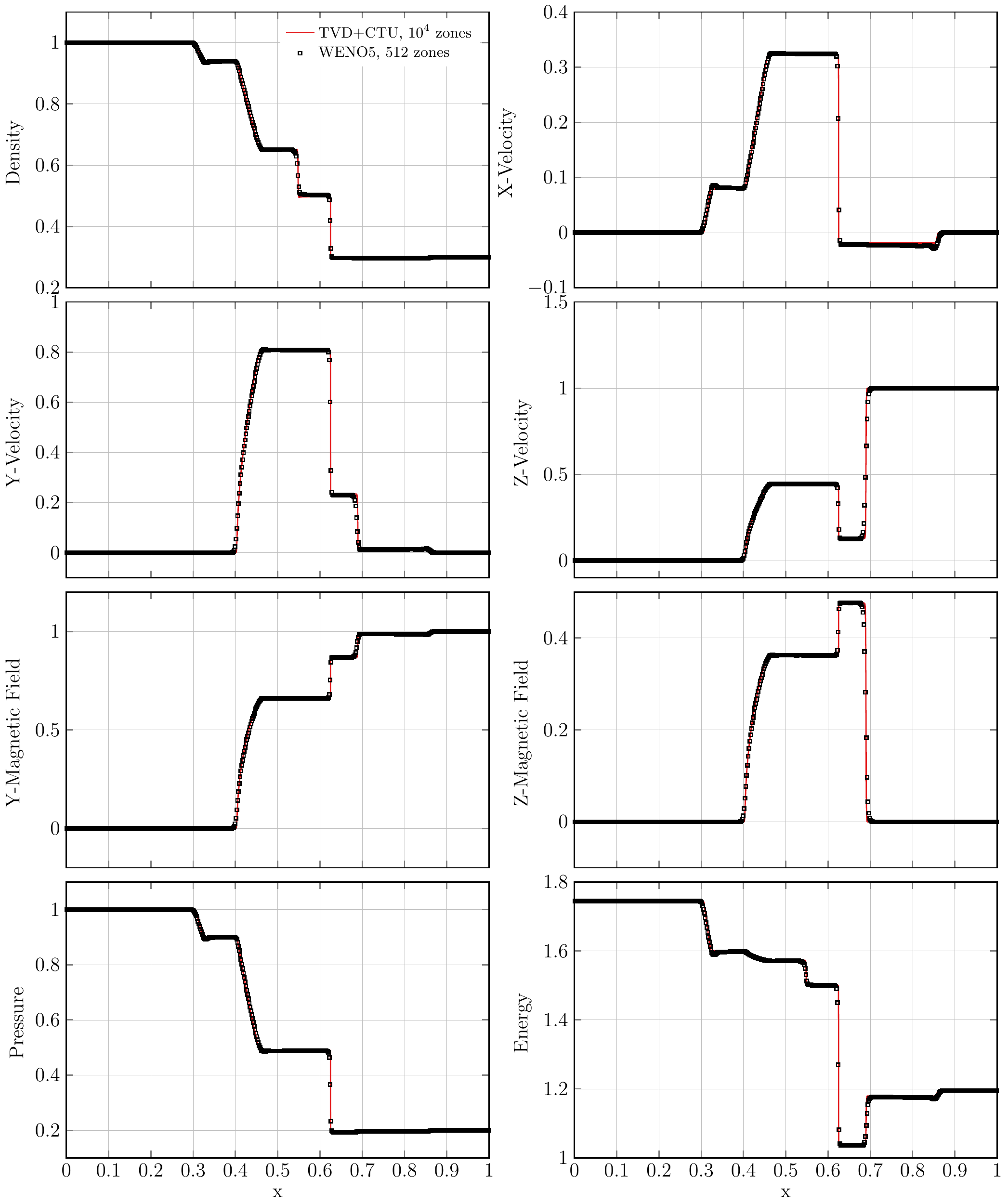}
    \caption{Shock tube test 4d from \citet{1995ApJ...442..228R} at time $t = 0.2$ in one dimension. TVD+CTU result with 10000 zones as red line, WENO5 result with 256 zones as black squares. Density, velocity components, magnetic field components, pressure and energy are shown from top left to bottom right.}
    \label{fig.rj4d_1d}
\end{figure*}

We compute shock tube number 4d from \citet{1995ApJ...442..228R} with \wombat's TVD+CTU and WENO5 solver. The left hand state vector is $\vec{U}_\mathrm{L} = (1,0,0,0,0.7,0,0,1)^T$. The right hand state vector is $\vec{U}_\mathrm{R} = (0.3,0.0.1,0.7,1,0,0.2)^T$. We plot the TVD+CTU result with $10^4$ zones as red line and the WENO5 result with 256 zones as black circles. The evolved state vector components at $t = 0.16$  are shown in figure \ref{fig.rj4d_1d}.

\subsection{Brio \& Wu Shock Tube}

\begin{figure*}
    \centering
    \includegraphics[width=0.9\textwidth]{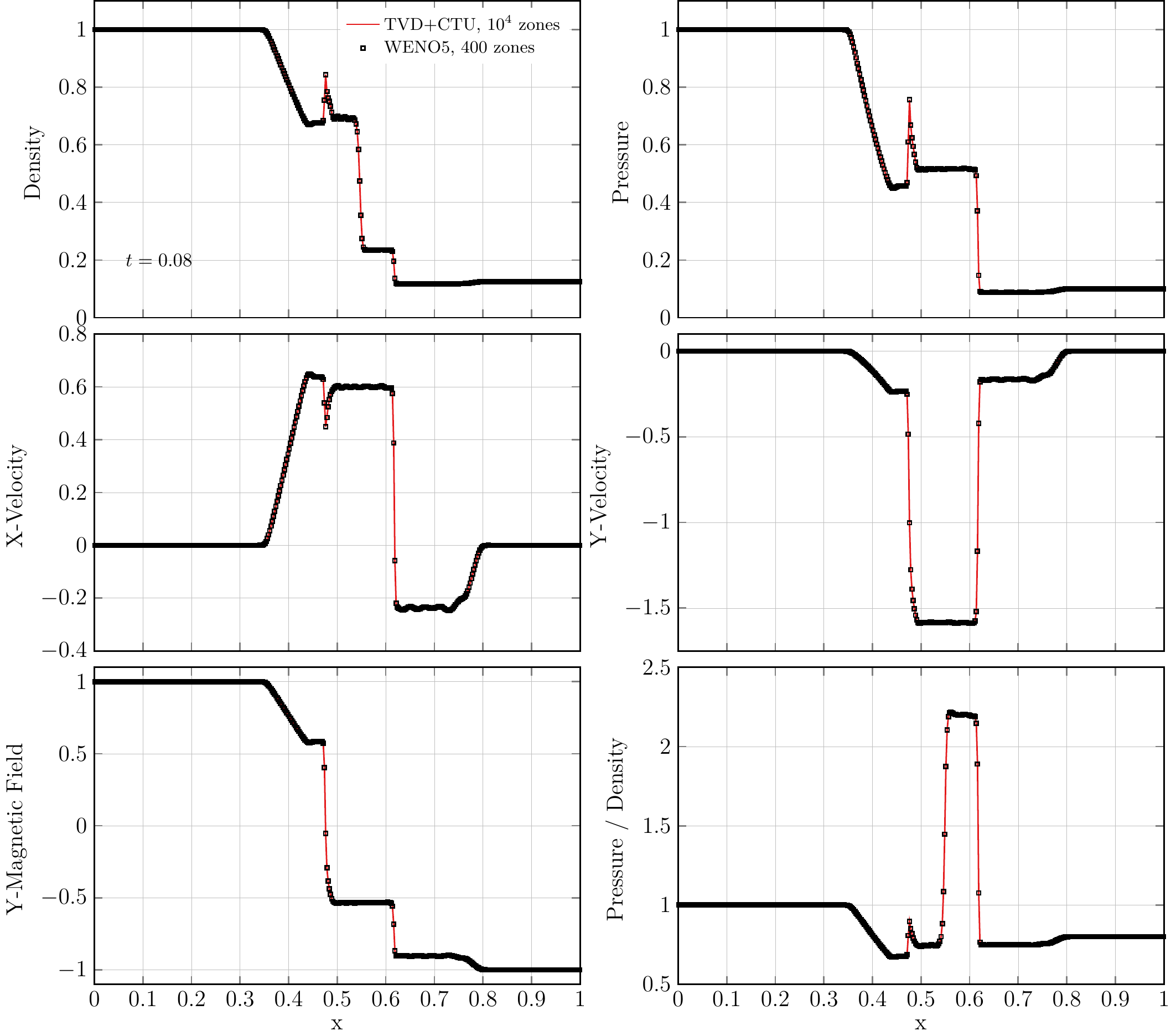}
    \caption{Shock tube test from \citet{1988JCoPh..75..400B} at time $t=0.08$. We show Density, x and y component of velocity, y component of magnetic field, pressure and energy from top left to bottom right. Results from WENO5 with 400 zones as black squares, TVD+CTU with 10000 zones as red line.}
    \label{fig.bw_1d}
\end{figure*}
The MHD shock tube from \citet{BrioWu1988} has the initial conditions: $\vec{U}_\mathrm{L} = (0.125,0,0,0,0.75,1,0,1)^T$, $\vec{U}_\mathrm{R} = (0.125,0,0,0,0.75,-1,0,0.1)^T$, with $\gamma = 2$. In figure \ref{fig.bw_1d}, we show the solution at $t=0.08$ with 400 zones as black squares and the solution from TVD+CTU with $10^4$ zones as red line.\par
We note that this problem starts with a degeneracy in the magnetic field when computing the eigenvectors of the initial conditions in the single zone at $x=0.5$. Following \citet{BrioWu1988}, we resolve this degeneracy by setting $B_\mathrm{y} = B_\mathrm{z} = 1/\sqrt{2}$. However, in this particular problem, $B_\mathrm{z} = 0$ at all times. Thus the test exposes this choice in the eigenvectors. Setting $B_\mathrm{z} = 0$ for this problem only, would remove the oscillations in our figure \ref{fig.bw_1d}. \citet{doi:10.1137/040610246} have shown that the issue can also be fixed with global smoothness indicators.  In real world applications in 2 or more dimension, this is not an issue. { Jang et al. in prep. found that the oscillations disappear with WENO3, i.e. in lower order codes increased dissipation leads to a constant solution.}

\subsection{Shock-Density Wave Interaction}

 \begin{figure}
     \centering
     \includegraphics[width=0.45\textwidth]{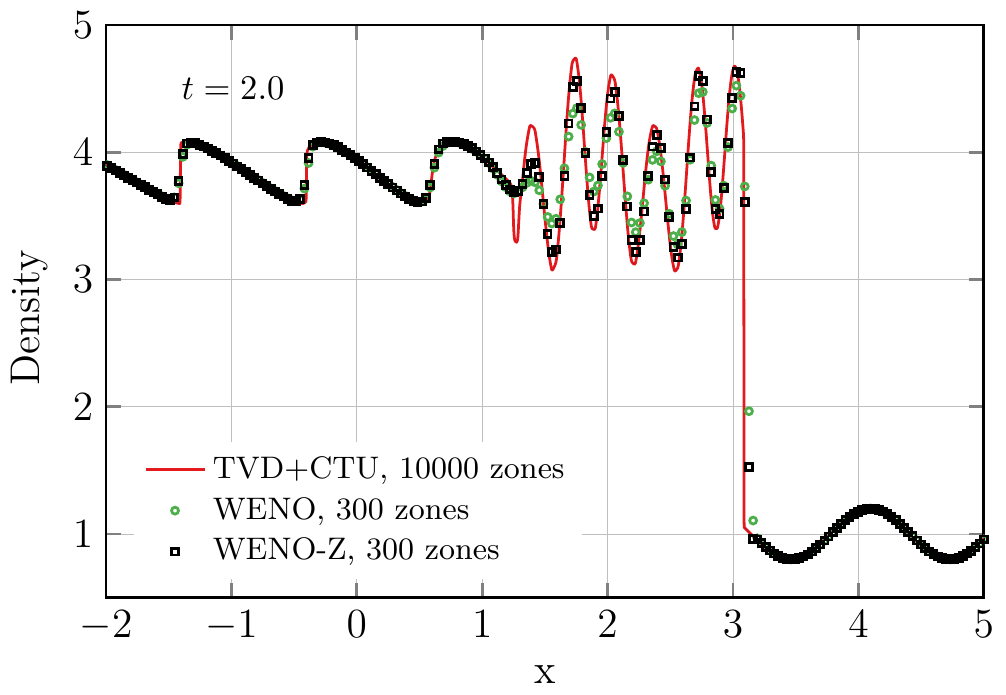}
     \caption{Density of the shock-density wave interaction test at time $t=1$}
     \label{fig.ShockDWave_1d}
 \end{figure}
 
The Shu-Osher shock tube simulates the interaction of a shock with a smooth flow \citep{1989JCoPh..83...32S}. \citet{Greenough} used it to argue that second order PPM methods give results comparable to third order WENO methods and are thus computationally more efficient. It exposes that the standard WENO5 weights are at best third order accurate in critical points \citep{2008JCoPh.227.3191B}. \par
The initial conditions contain a  shock tube with left \& right state $\vec{U}_\mathrm{L} = (3.857143,2.629369,0,0,0,0,0,31/3)^T$, $\vec{U}_\mathrm{R} = (1+0.2 \sin(5x),0,0,0,0,0,0,1)^T$ on a domain of size $L_x = 10$. The shock is located at $x = -4$. The shock tube is evolved until $t=2$. \par
In figure \ref{fig.ShockDWave_1d}, we show the result from WENO5 (green circles) and WENO5-Z simulations (black squares) with 300 zones and TVD+CTU with 10000 zones (red line). Clearly the WENO-Z result traces the complex flow behind the shock more accurately than the WENO5 result.

\subsection{Implosion} \label{sect.implosion}

\begin{figure*}
    \centering
    \includegraphics[width=0.45\textwidth]{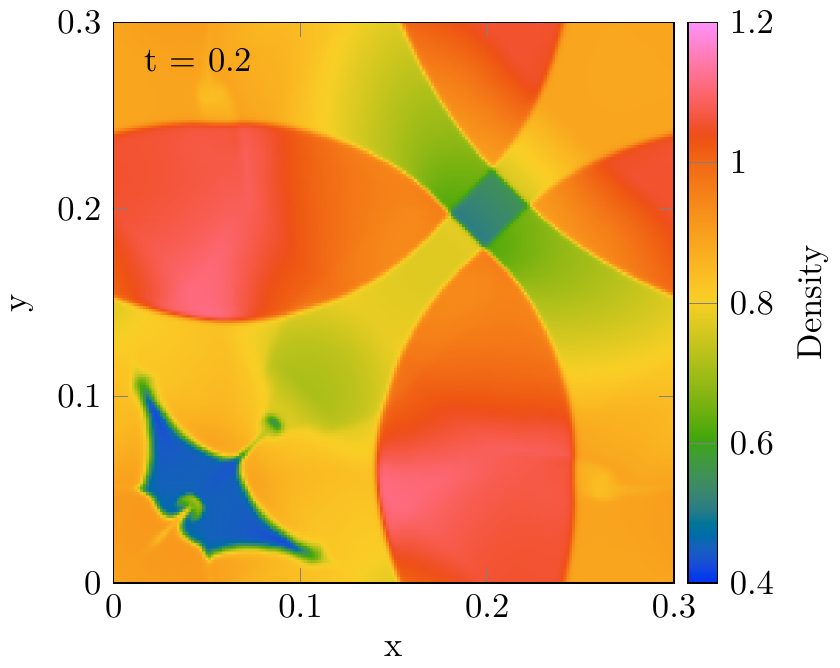}
    \includegraphics[width=0.45\textwidth]{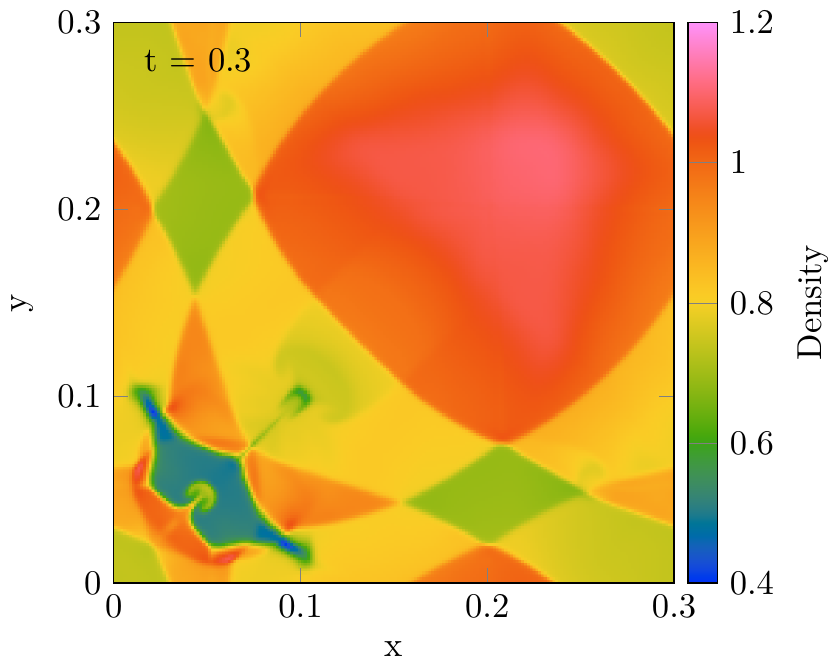}\\
    \includegraphics[width=0.45\textwidth]{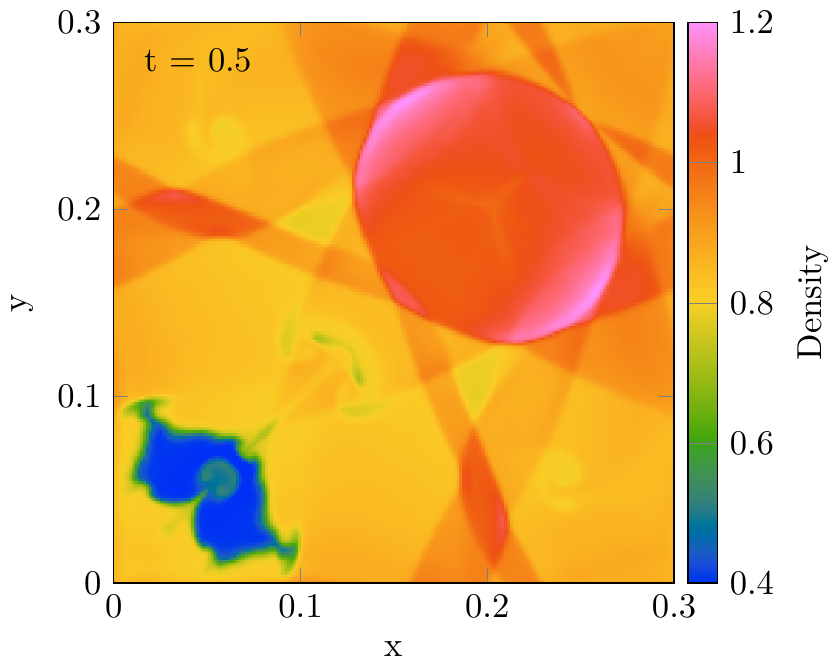}
    \includegraphics[width=0.45\textwidth]{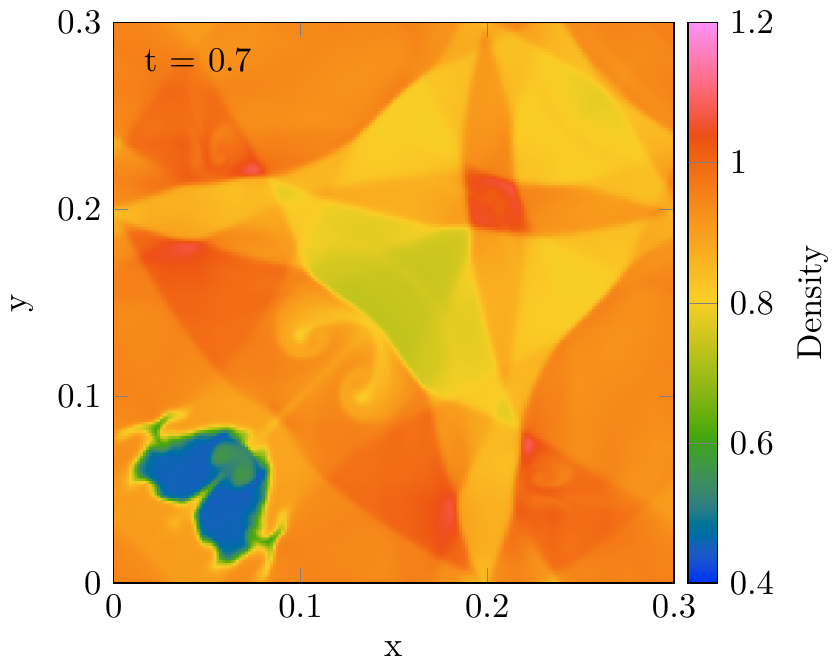}
    \caption{Density of the implosion test  at times $t=0.2$, $t=0.3$, $t=0.5$ and $t=0.7$ (top left to bottom right). The color bar ranges from 0.4 to 1.2, the resolution was $200^2$ zones.}
    \label{fig.impl_2d}
\end{figure*}

\begin{figure}
    \centering
    \includegraphics[width=0.45\textwidth]{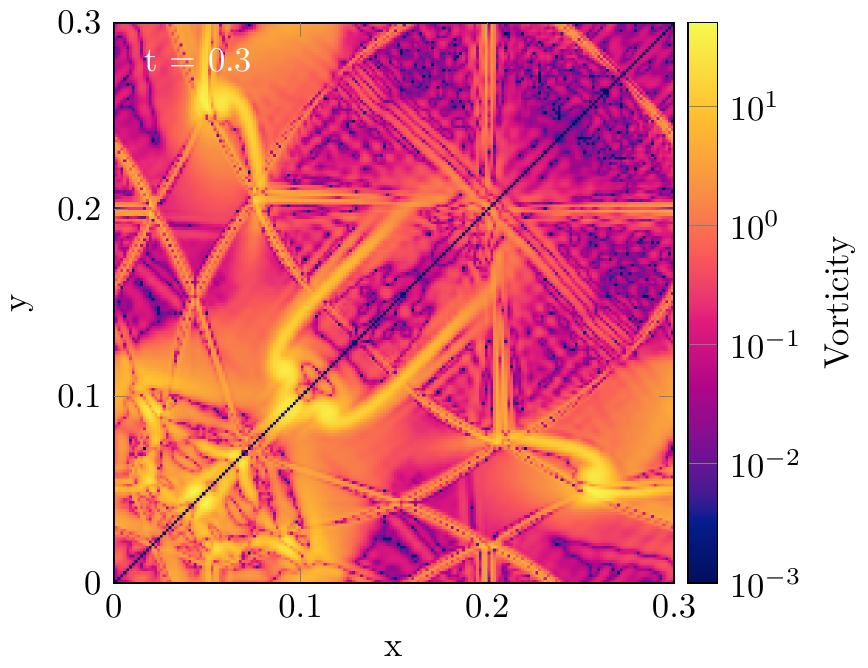}
    \caption{Logarithm of vorticity of the implosion test at time $t=0.3$. The color bar ranges from 0.001 to 50, the resolution was $200^2$ zones.}
    \label{fig.impl_vort_2d}
\end{figure}

The noise inherent in an implementation can be exposed in the two dimensional implosion test \citep{Liska2003}.  The setup leads a series of shock waves, which are intersecting many times in the simulation. The problem is highly non-linear and thus not very useful to compare algorithmic fidelity. There is simply no convergence to a universal solution that results could be compared to. However, the non-linearity ensures that computational noise gets amplified very quickly, and solutions diverge strongly from correct less-noisy results. Using this test we found noise injected by the compiler at the  $\sqrt{10^{-16}}$ level in the eigenvector calculation (see Appendix \ref{app.ev}), by comparing the solution to TVD+CTU. This may be a word of warning for running Eulerian methods with 4 byte floating point precision, where numerical noise itself resides at $10^{-8}$ and thus will influence this test significantly. Many real world applications are strongly non-linear and thus may not be correct with four byte variables. \par
We set $\vec{U} = (1,0,0,0,0,0,0,1)^T$ everywhere in a periodic domain with size $L_x = L_y = 0.3$ and $200^2$ zones. When $x+y < 0.15$, we set $\vec{U} = (0.125,0,0,0,0,0,0,0.14)^T$. We show the resulting density in figure \ref{fig.impl_2d} at $t=0.2,0.3,0.5,0.7$ (top left to bottom right) on a colour scale ranging from 0.4 to 1.2. We notice that the symmetry in the simulation is not visibly broken and the results are reasonably similar to the tests shown in \citet{Liska2003,2012MNRAS.424.2999S}. We show the vorticity of the test at $t=0.3$ in figure \ref{fig.impl_vort_2d}, where no noise is visible. In particular, the result is symmetric along the $x=y$ axis down to single zones. \par
The implosion test was used by \citet{2012MNRAS.424.2999S} to compare results between a traditional SPH and a Lagrangian finite volume method. Both show more computational noise than our Eulerian method due to particle motion. There are sizable differences in the shape of the "bird" in the lower left corner of the simulation. In particular, interfaces in the Lagrangian finite volume result differ from our WENO5 simulation. Numerical noise seems to trigger Raleigh-Taylor instabilities in the Lagrangian simulation that are absent in the Eulerian simulations. This noise is clearly visible in the vorticity maps (their figure 3).  Runs at much higher resolution show that indeed these interfaces do become unstable eventually. However due to the non-linearity of the flow, the high resolution Eulerian result differs significantly from both low resolution runs in that all instabilities grow faster in the high resolution Eulerian runs. As mentioned above it is unclear, which result is closer to the ''true solution'' in this test, and we conclude that numerical noise affects the growth of instabilities in highly non-linear solutions of the hydrodynamic equations, which is of course not a new result at all \citep[e.g.][]{2012ApJS..201...18M}.

\subsection{Orszag–Tang Vortex}

\begin{figure*}
    \centering
    \includegraphics[width=0.45\textwidth]{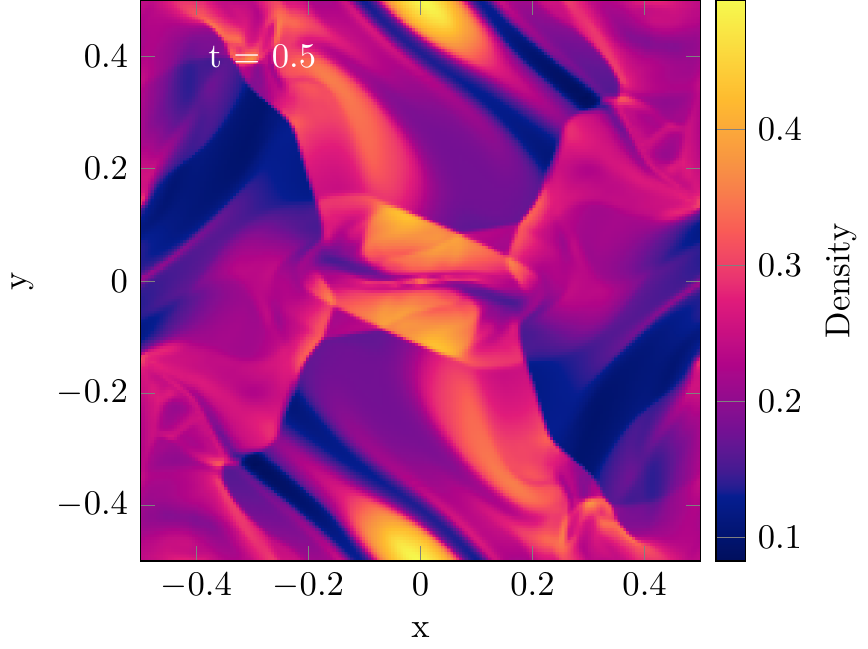}
    \includegraphics[width=0.45\textwidth]{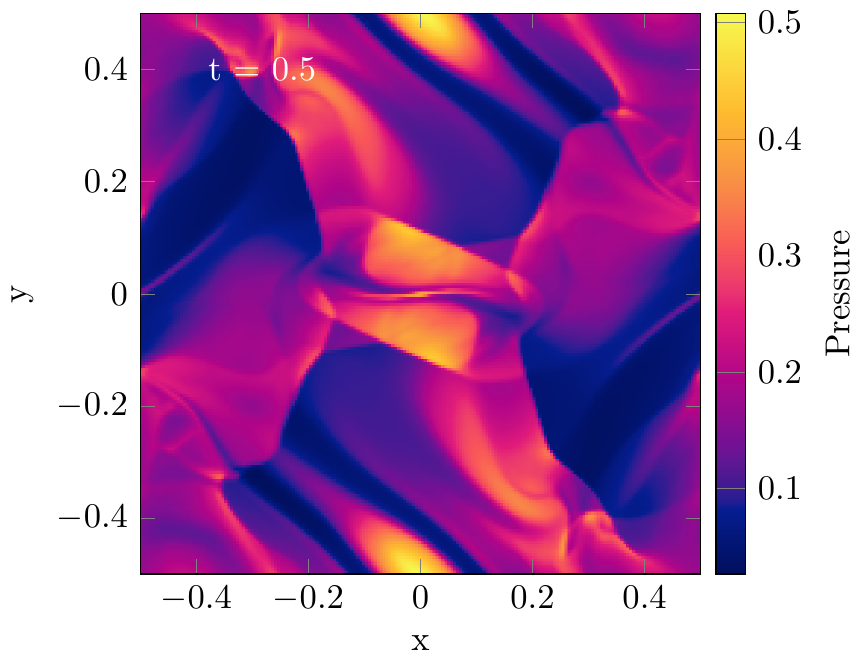}\\
    \includegraphics[width=0.45\textwidth]{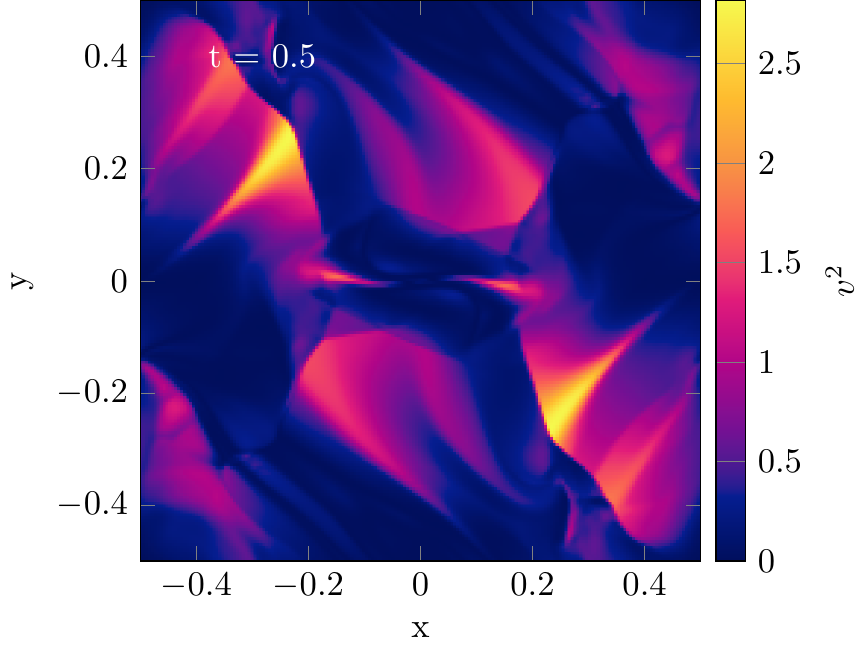}
    \includegraphics[width=0.45\textwidth]{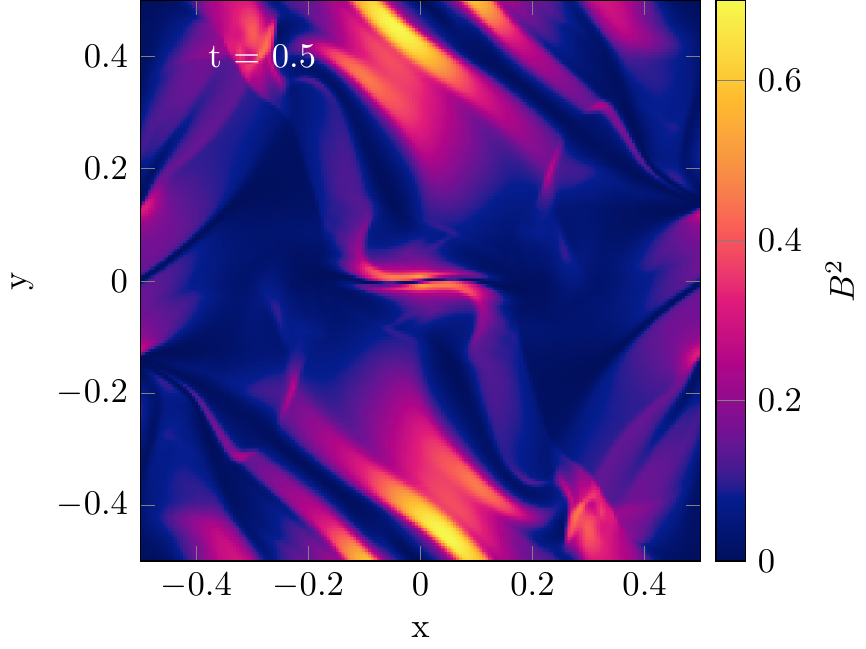}
    \caption{Images of the Orszag-Tang vortex test at time $t=0.5$ with $192^2$ zones. Top-left to bottom right: density, pressure, $v^2$ and $B^2$.}
    \label{fig.OT_2d}
\end{figure*}
 \begin{figure}
    \centering
    \includegraphics[width=0.45\textwidth]{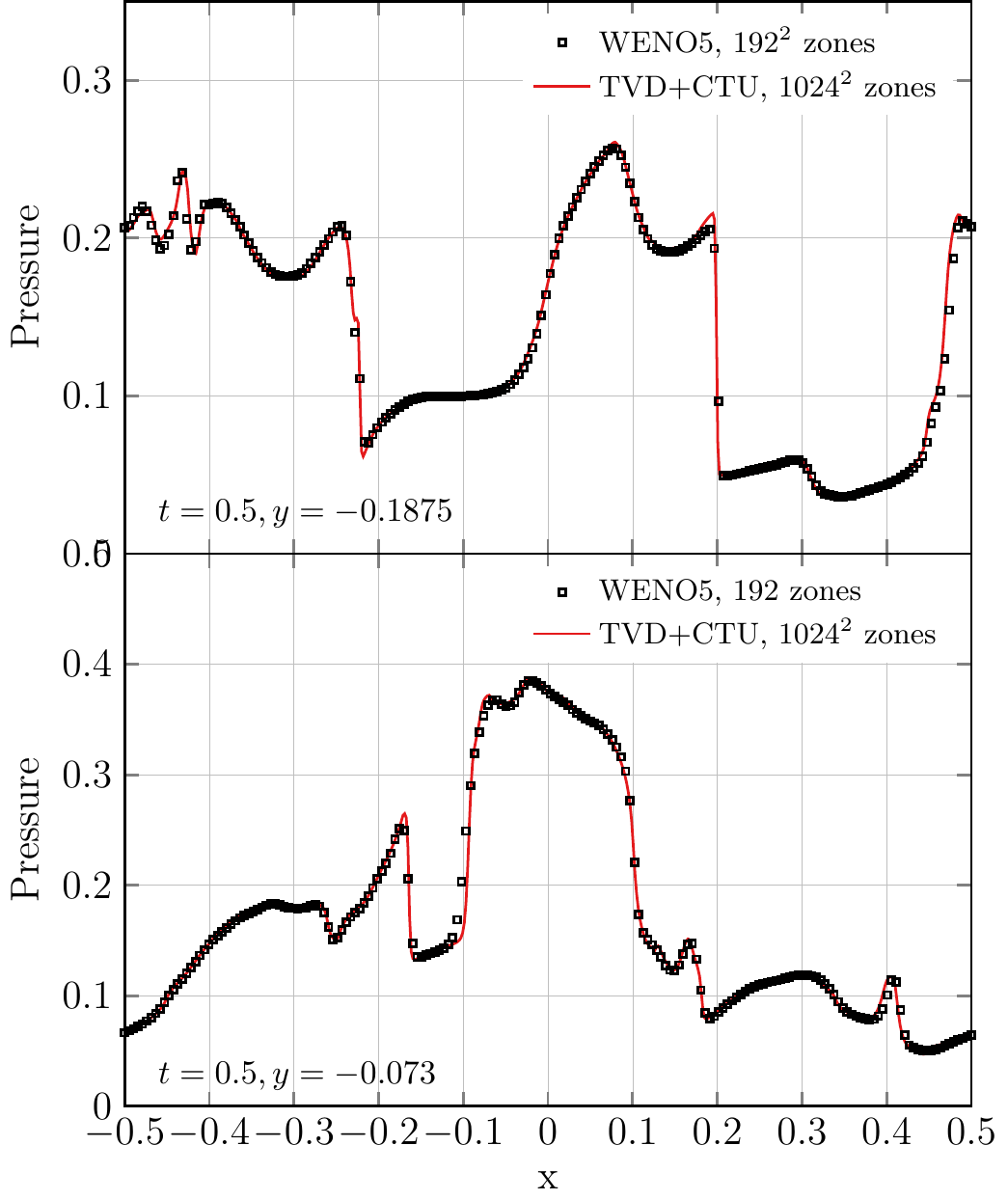}\\
    \caption{Pressure in the Orszag-Tang vortex test at $y = -01875$ (top) and $y = 0.073$ (bottom) and at time $t=0.5$. WENO5-Z with $192^2$ zones as black squares, TVD+CTU with $1024^2$ zones as red line.}
    \label{fig.OT_slice}
\end{figure}
 \begin{figure}
    \centering
    \includegraphics[width=0.45\textwidth]{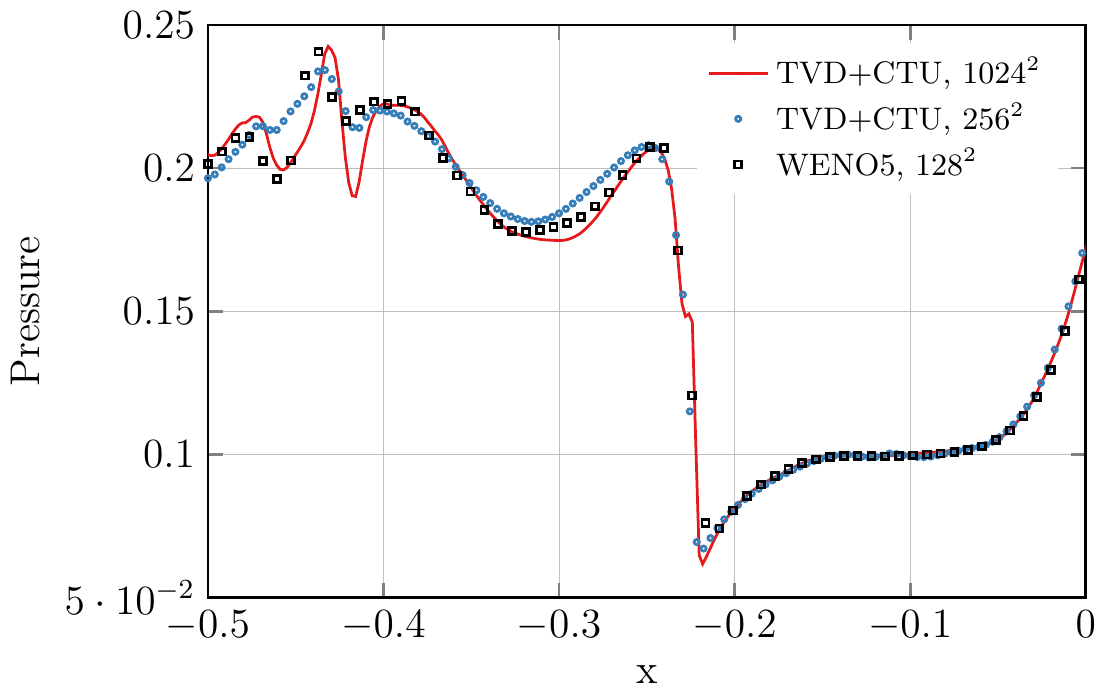}
    \caption{Pressure in the Orszag-Tang vortex test at $y = -01875$, $-0.5 < x < 0$ and at time $t=0.5$. WENO5-Z with $128^2$ zones as black squares, TVD+CTU with $1024^2$ zones as red line, TVD+CTU with $256^2$ zones as blue circles.}
    \label{fig.OT_slice2}
\end{figure}

 \begin{figure*}
    \centering
    \includegraphics[width=1\textwidth]{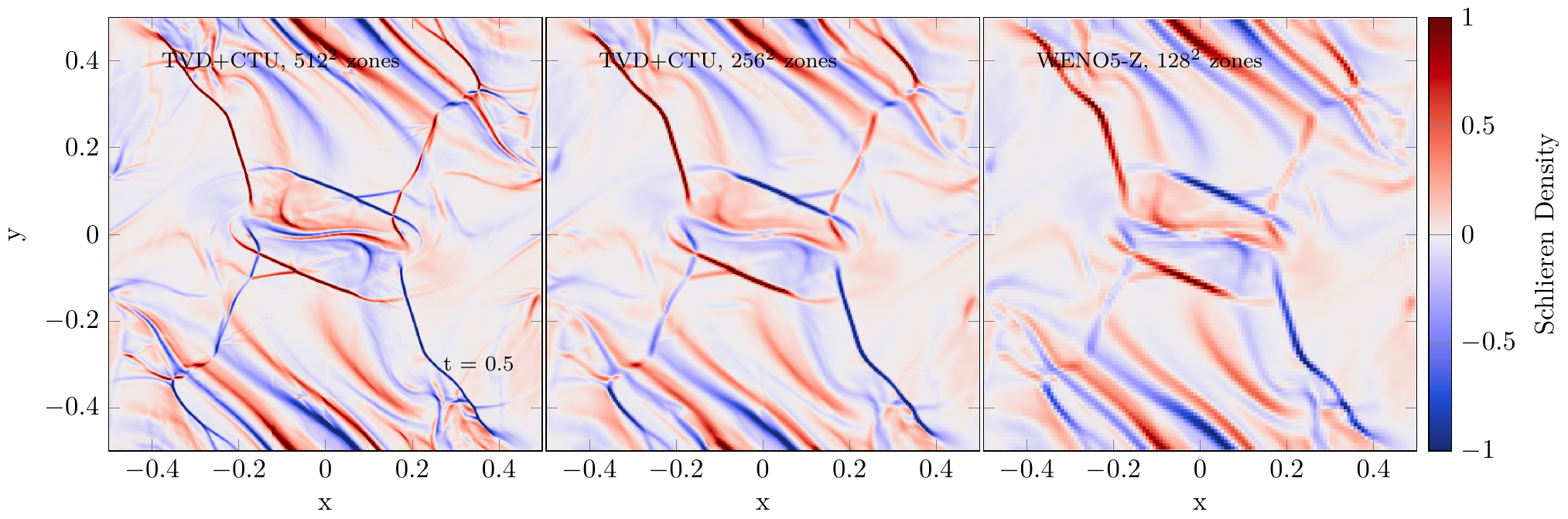}
    \caption{Schlieren image of the density in the Orszag–Tang vortex test at $t=0.5$ for TVD+CTU with $512^2$ zones (left), $256^2$ zones (middle) and WENO5-Z with $128^2$ zones (right).}
    \label{fig.OT_schlieren}
\end{figure*}

The Orszag-Tang vortex is a standard MHD test that mimics the transition of a smooth flow into 2D turbulence \citep{Orzang,2000JCoPh.161..605T,2008ApJS..178..137S}. The initial conditions in a unit domain are: $\vec{U} = (25/(36\pi), -\sin(2\pi y), \sin(2\pi y), 0, B_x, B_y, B_z, 5/(12\pi))^T$, where the magnetic field components are obtained from a  vector potential with the z-component $A_z = (B_0/4\pi) \cos(4\pi x) + (B_0/2\pi) \cos(2\pi y)$, with $B_0 = 1/\sqrt{4 \pi}$. \par
We show in figure \ref{fig.OT_2d}, top left to bottom right: Density, pressure, velocity magnitude and magnetic field strength at time $t=0.5$ at a resolution of $192^2$ zones with WENO5-Z. In figure \ref{fig.OT_slice}, we show the pressure in two slices through the domain at $t = 0.5, y=-0.1875$ (top) and $t=0.5, y=0.073$ (bottom). A high-resolution run with TVD+CTU is included as a reference solution.  \par
In figure \ref{fig.OT_slice2} we compare the pressure at $y = -01875$, $-0.5 < x < 0$ and $t=0.5$ of WENO5-Z with $128^2$ zones (green squares) with TVD+CTU (blue circles) with $256^2$ zones. The reference solution with $1024^2$ zones is plotted as red line. Apparently, WENO5-Z has comparable or better fidelity than TVD+CTU at double the resolution in this test. In particular, the blip at $x=-0.45$ is better resolved in the WENO5-Z solution. A comparison with figure \ref{fig.OT_slice} shows that the blip becomes fully resolved at $192^2$. \par
In figure \ref{fig.OT_schlieren} we compare Schlieren images \citep[][]{Hooke1665,2011JCoPh.230.3803H} of the density in TVD+CTU runs at $512^2$ and $256^2$ zones (left, middle), with the WENO5-Z run at $128^2$. Following \citet{Guo2004}, we obtained a synthetic brightness value by adding the red and yellow channels with a Gladstone-Dale constant of $K_\mathrm{GD} = 1$, $L = 1$ and a contrast parameter of $C = 0.075$. Schlieren images visualise the gradient of the solution, which makes the Orszag-Tang vortex test more discriminative between schemes.  We observe that all main features of the low resolution TVD+CTU simulation are reproduced in the WENO5-Z simulation at half that resolution. We conclude that WENO5-Z doubles the effective resolution compared to TVD+CTU in this test.

\subsection{MHD Rotor}

\begin{figure*}
    \centering
    \includegraphics[width=0.45\textwidth]{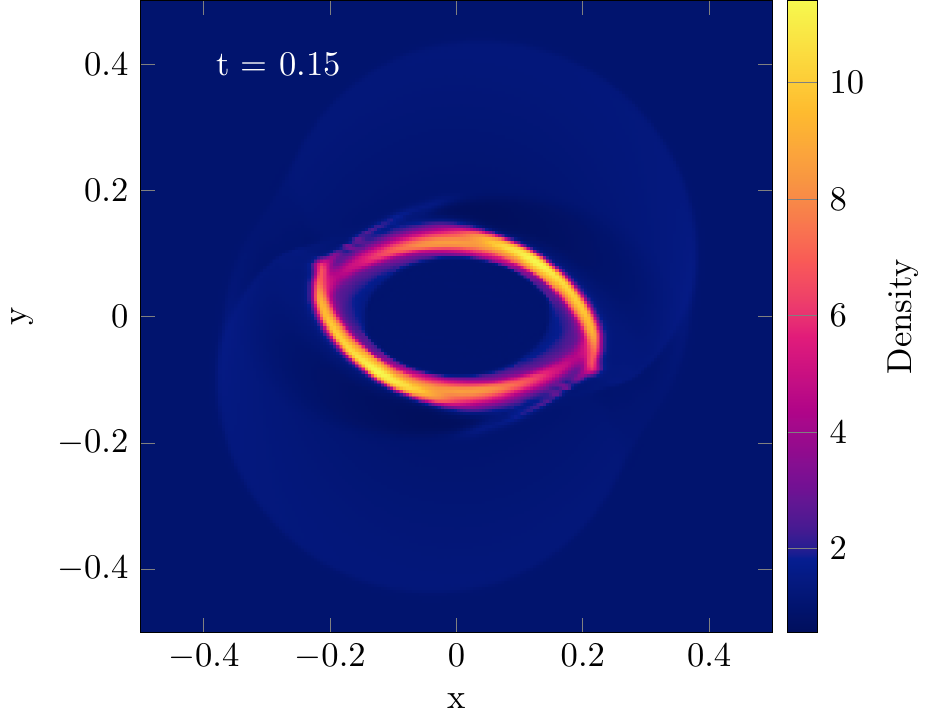}
    \includegraphics[width=0.45\textwidth]{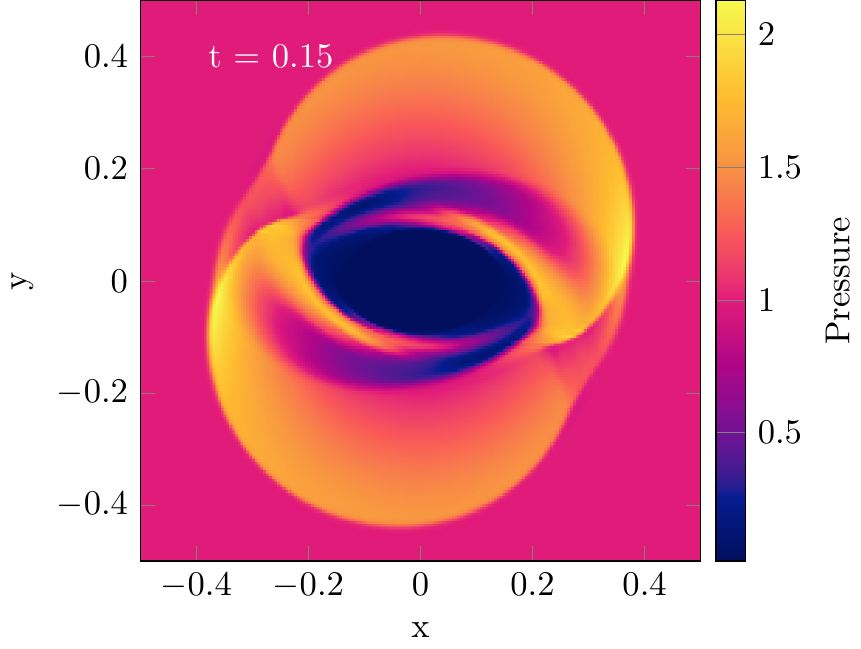}\\
    \includegraphics[width=0.45\textwidth]{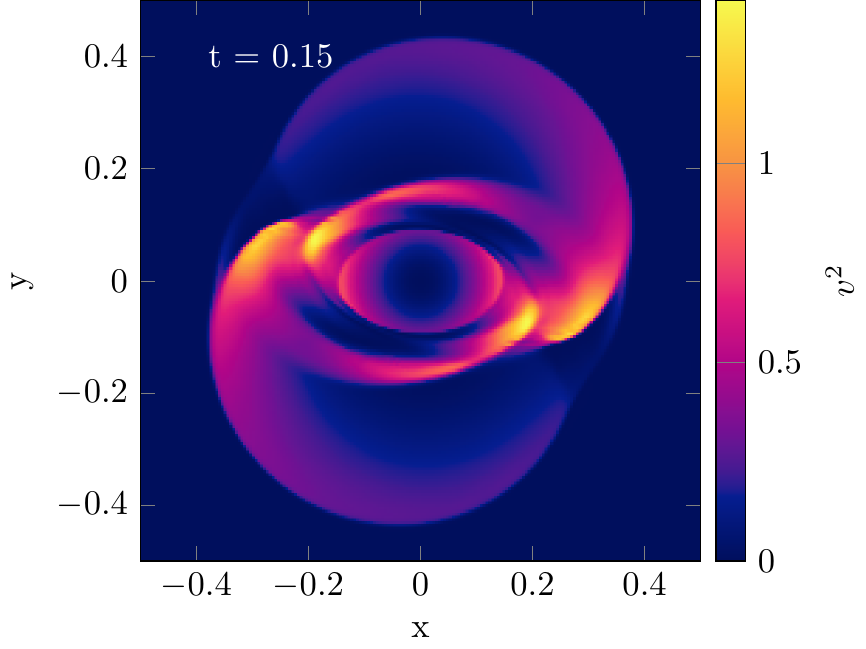}
    \includegraphics[width=0.45\textwidth]{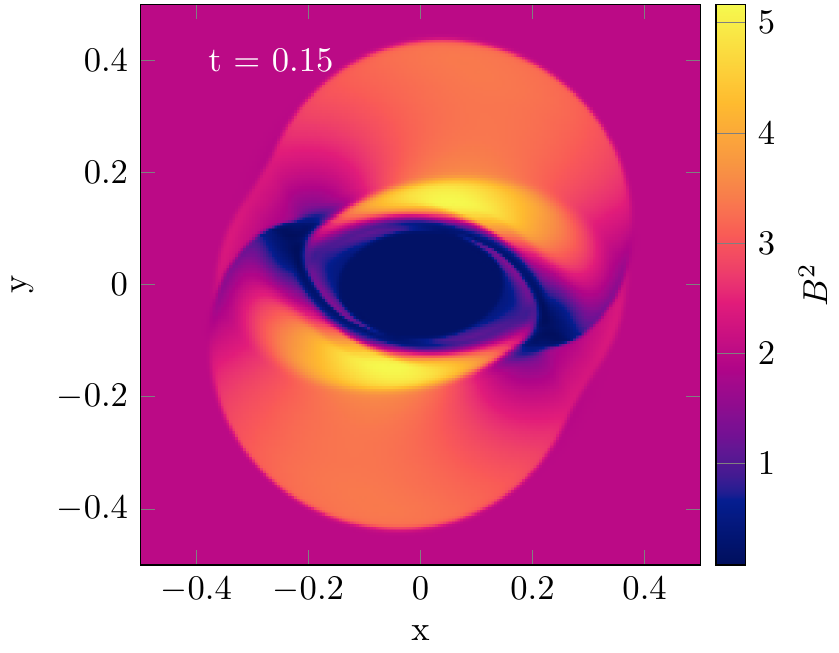}
    \caption{Images of the MHD rotor test at time $t=0.15$ with $400^2$ zones. Top-left to bottom right: density, pressure, $v^2$ and $B^2$.}
    \label{fig.MHDR_2d}
\end{figure*}

\begin{figure}
   \centering
   \includegraphics[width=0.45\textwidth]{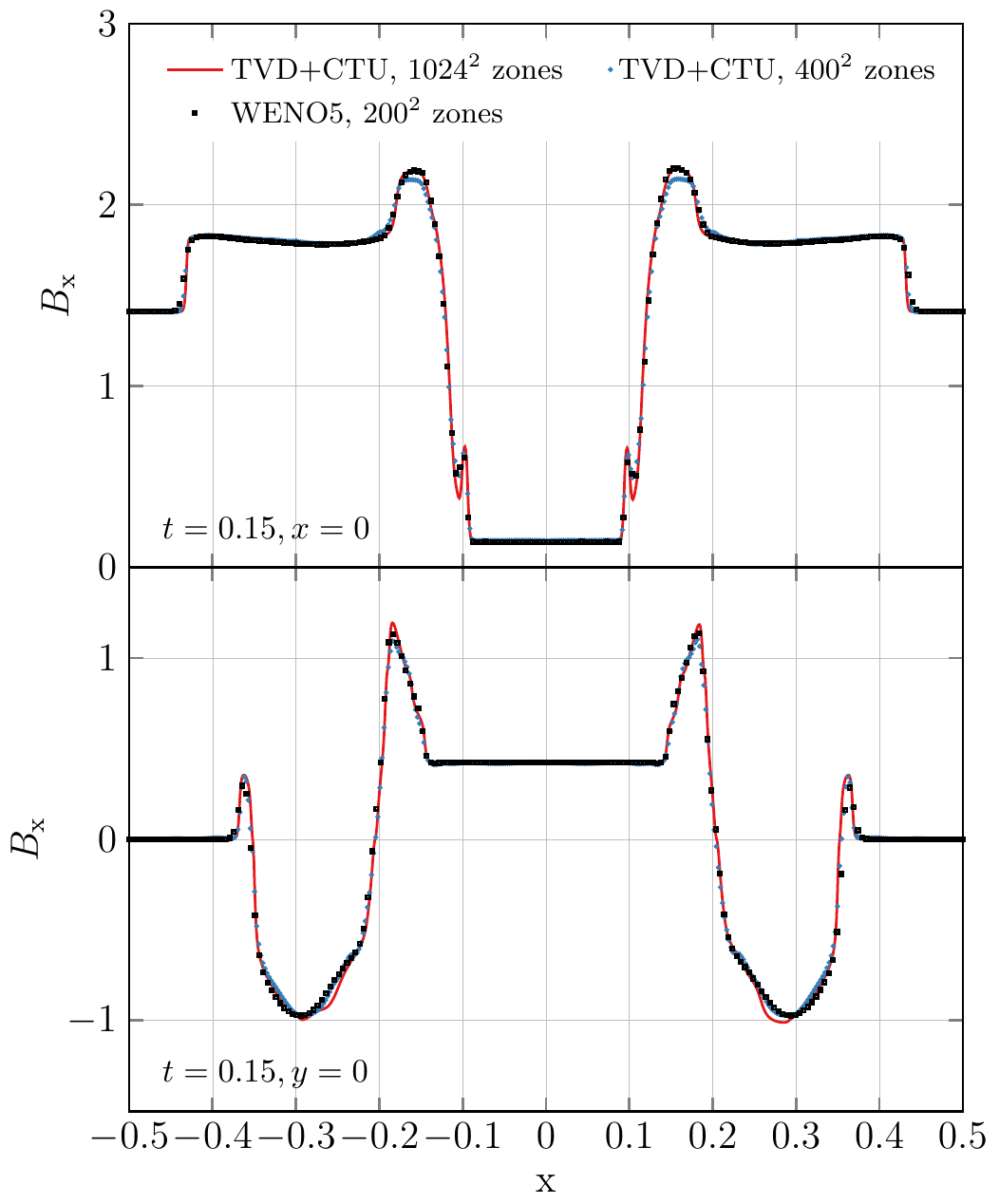}
   \caption{X-component of the magnetic field (top) and y-component of the magnetic field (bottom) in the MHD rotor test at $y = -0.1875$ (top) and $y = 0.073$ (bottom), at time $t=0.5$. WENO with $200^2$ zones as black squares, TVD+CTU with $400^2$ zones as blue dots, TVD+CTU with $1024^2$ zones as red line for reference.}
   \label{fig.MHDR_slice}
\end{figure}

The MHD rotor test simulates a rapidly rotating disk in 2 dimensions, which produces shocks and rare-fraction waves as the disk expands due to the centrifugal forces \citet{1992ApJS...80..753S,Balsara99,2008ApJS..178..137S}. The solution needs to remain symmetric and no spurious oscillations may occur. We set initial conditions with $\gamma=5/3$ in a unit domain of $P=1$, $\vec{B} = (5/\sqrt{4\pi},0)^T$,$v_0 = 2$, $r_0 = 0.1$, $r_1 = 0.115$ and $f(r) = \frac{r_1 - r}{r_1 - r_0}$ and \citep{2000JCoPh.161..605T}:
\begin{align}
    \rho &= \begin{cases}
             10        &  \Leftrightarrow r < r_0 \\
             1 + 9 f(r)&  \Leftrightarrow r_0 \le r \le r_1 \\
             1         &  \mathrm{otherwise}.
            \end{cases} 
\end{align}
the velocity components are:
\begin{align}
    v_x &=  \begin{cases}
             -y \frac{v_0}{r_0}      &  \Leftrightarrow r < r_0 \\
             -f(r) y \frac{v_0}{r_0} &  \Leftrightarrow r_0 \le r \le r_1 \\
             0                       &  \mathrm{otherwise}
            \end{cases} \\
    v_y &=  \begin{cases}
             -x \frac{v_0}{r_0}      &  \Leftrightarrow r < r_0 \\
             -f(r) x \frac{v_0}{r_0} &  \Leftrightarrow r_0 \le r \le r_1 \\
             0                       &  \mathrm{otherwise},
            \end{cases}
\end{align}
We show the test result with WENO5 and $200^2$ zones at time $t=0.15$ in figure \ref{fig.MHDR_2d}. A slice through the rotor at $t=0.5$ and $y=-0.1875$ (top) and $y = 0.073$ are shown in figure \ref{fig.MHDR_slice}. We show the WENO5 solution as black squares, TVD+CTU with $1024^2$ zones as red line and $400^2$ zones (blue dots). No asymmetries or oscillations are visible, the WENO5 solution at $200^2$ resolves the small features comparably or better than TVD+CTU. A by-eye comparison with published results from \athena \citep[][ their figure 26]{2008ApJS..178..137S}, yields that WENO5 at half resolution is not quite as resolved as \athena at full resolution, likely because our CT scheme is only second order in space. There is some ringing visible in the density maps, which could likely be alleviated using global weights. We leave the improvement to future work.

\subsection{Double Mach Reflection Test}

\begin{figure*}
    \centering
    \includegraphics[width=0.9\textwidth]{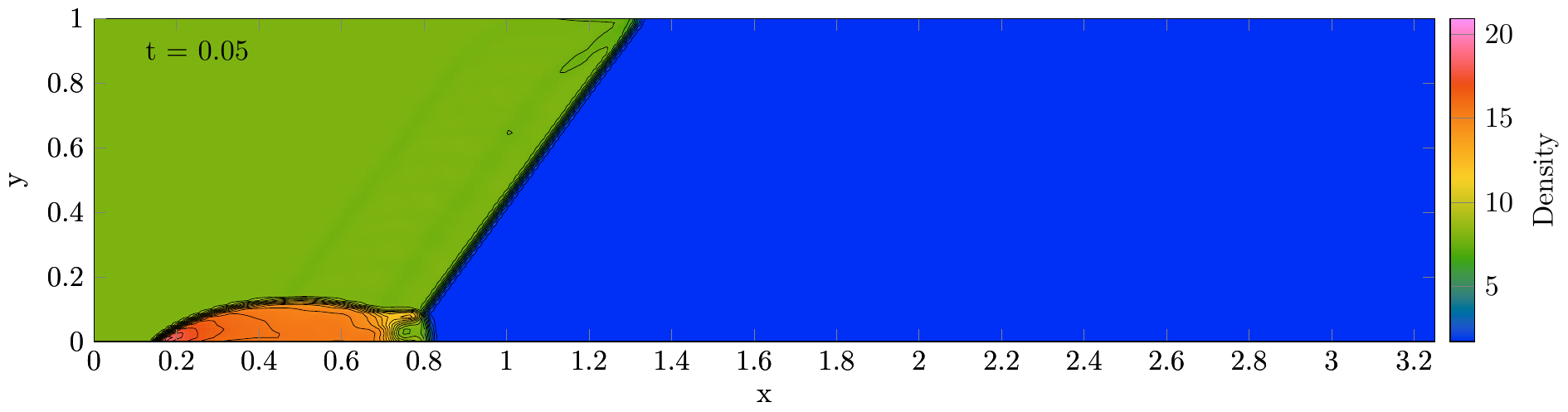}
    \includegraphics[width=0.9\textwidth]{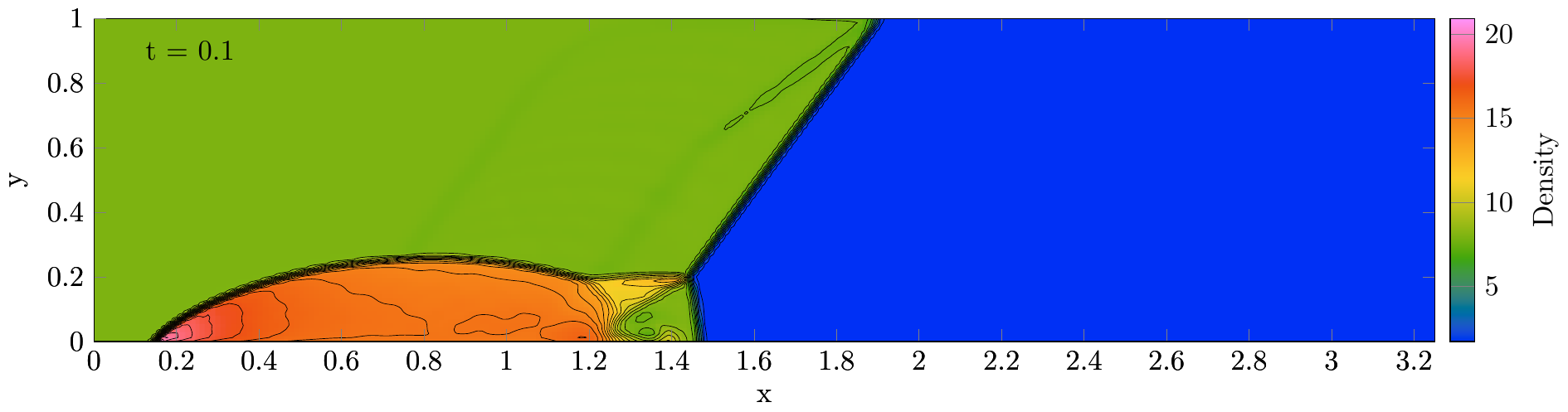}
    \includegraphics[width=0.9\textwidth]{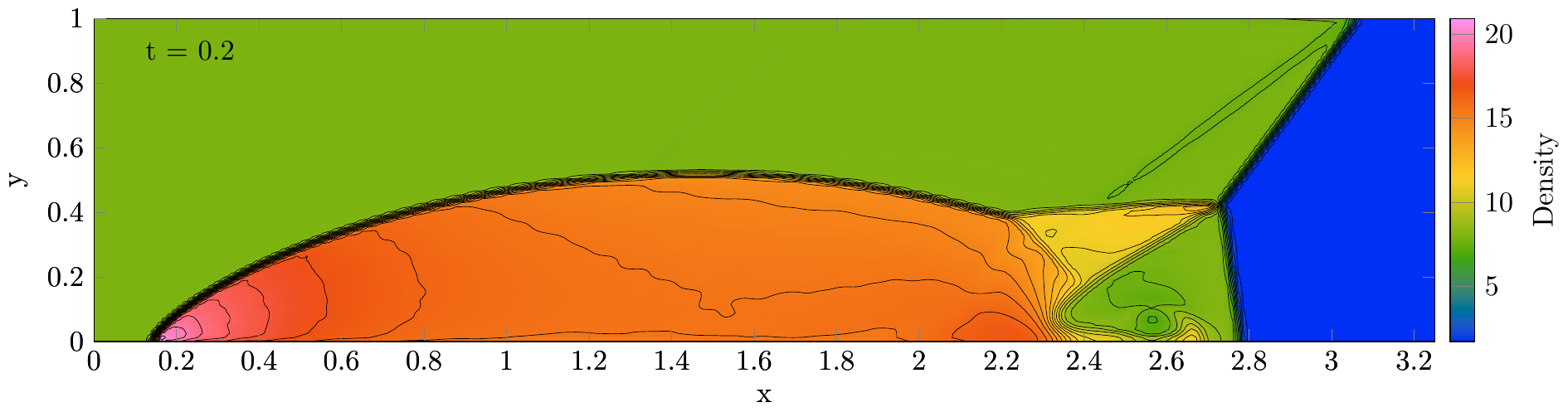}
    \caption{Density of the 2D double Mach reflection test with WENO5 at times 0.05, 0.1, 0.2 (top to bottom) and resolution $260\times80$ zones. We overplot 30 contours between 1.73 and 20.92, following \citet{1984JCoPh..54..115W}.}
    \label{fig.dmr_2d}
\end{figure*}

The double Mach reflection test simulates the evolution of a Mach 10 shock at an oblique $60 \,\mathrm{deg}$ angle to the grid with reflecting boundaries on the bottom x-axis, continuous upper x-boundary and open y-boundaries \citep{1984JCoPh..54..115W,2008ApJS..178..137S}. Here the adiabatic index $\gamma = 1.4$ for air, the state ahead of the shock is $\vec{U} = (1.4, 0,0,0, 0,0,0, 1)^T$, the domain size is $3.25 \times 1$. Because the shock is reflected of the bottom wall, a second weak shock and a jet form. In the upper boundary the evolution of the shock is followed analytically, i.e. $x_\mathrm{shock} = 1/6 + (1 + 20 t)$. As the shock in the setup is only one cell wide, the initial conditions contain an error, which is also continuously injected at the upper boundary. A one cell wide shock is not a ''natural'' solution to the numerical scheme, similar to the common shock tube setup used in SPH code tests. The test foremost exposes the robustness of the numerical scheme that has to handle these errors by injecting numerical diffusion. This differs from SPH shock tube tests, where dissipation and conduction has to be explicitly added. This test produces negative pressures when run with the WENO5-Z scheme without protection floors or protection fluxes. \par
In figure \ref{fig.dmr_2d}, we show the density from the test run with the standard WENO5 scheme at times $t=0.05, 0.1, 0.2$ (top to bottom). We also overlay 30 contours between 1.73 and 20.92, which can be directly compared to \citet{1984JCoPh..54..115W}, their figure 4. In the top panel, the error of the initial shock is visible in the color scheme as two stripes with slightly lower density left of and parallel to the shock. Over time another such dip forms starting at the location of the shock at the upper y-boundary traveling toward the lower left. In some codes, the reflection of the shock off the bottom y-boundary causes a carbuncle instability, because the Mach number is very high here. We do not observe the instability in the standard WENO5 scheme, which points towards good robustness of the scheme in applications that contain complicated sub-grid physics. This is very desirable in cosmological simulations.

\subsection{Kelvin-Helmholtz Instability}

\begin{figure*}
    \centering
    \includegraphics[width=0.9\textwidth]{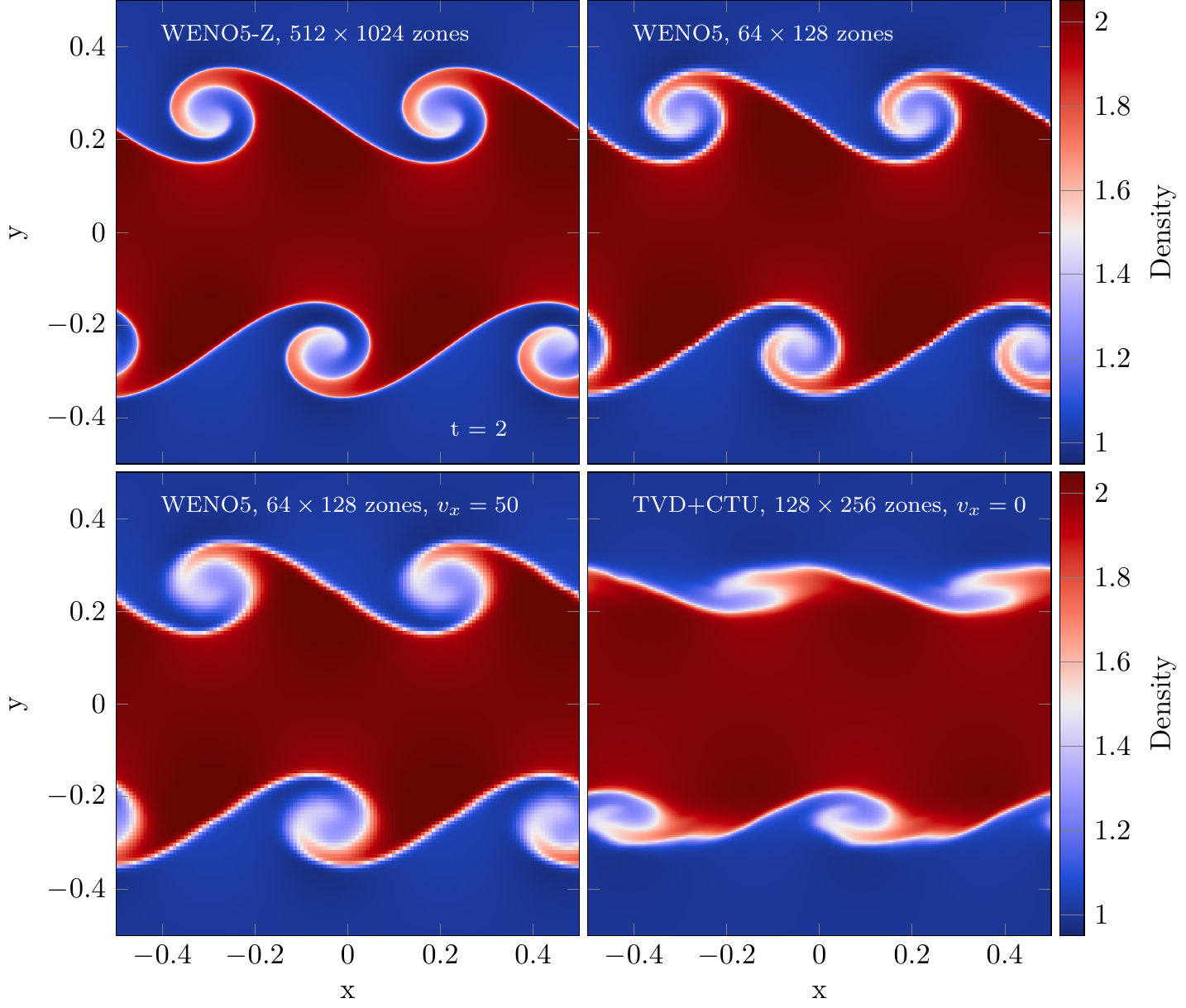}
    \caption{Density in Kelvin-Helmholtz simulations at $t= 2$. The colorbar ranges from 0.95 to 2.05. Top left to bottom right: WENO5-Z with $512 \times 1024$ zones, WENO5 with $32\times 64$ zones, WENO5 with  $32\times 64$ zones and a velocity boost of $v_x = 10$, TVD+CTU with $128\times 256$ zones and $v_x = 0$. We have duplicated the domain once along the x-direction to ease comparison with other publications.}
    \label{fig.kh_2d}
\end{figure*}

The growth of instabilities plays a crucial role for mixing and the injection of vorticity in astrophysical fluid flows \citep[e.g.][]{2018ApJ...865..118S}. We evaluate the performance of the code using the Kelvin-Helmholtz (KH) test, where the kinetic energy of a slightly perturbed shear flow generates vorticity at the interface \citep[e.g.][]{1958ApJ...128..664P,1982JGR....87.7431M,1996ApJ...460..777F}. This test has been used extensively in the literature to test hydrodynamic codes \citep[e.g][]{2010MNRAS.401..791S,2016MNRAS.455.2110B,2015MNRAS.453.4278S,2012ApJS..201...18M,2015MNRAS.450...53H}. In particular, the KH problem was used to expose the artifical surface tension of traditional SPH methods \citep[e.g.][]{2013MNRAS.428.2840H}. It also exposes the velocity dependent truncation error in Eulerian grid methods \citep{2010MNRAS.401.2463R}. Our setup follows \citet{2016MNRAS.455.4274L}:
\begin{align}
    \rho &= 1 + \frac{\Delta\rho}{\rho_0} \frac{1}{2} \left[ \tanh\left( \frac{z-z_1}{a} \right) - \tanh\left( \frac{z-z_2}{a} \right) \right] \\
    v_x &= v_\mathrm{flow} \left[ \tanh\left( \frac{z-z_1}{a} \right) - \tanh\left( \frac{z-z_2}{a} \right) - 1 \right] \\
    v_z &= A \sin(2\pi x) \left[ \exp\left(-\frac{(z-z_1)^2}{\sigma^2}\right) + \exp\left(-\frac{(z-z_2)^2}{\sigma^2}\right) \right] \\
    P &= P_0,
\end{align}
with the parameters, $A=0.01$, $P_0=10$, $a=0.05$, $\sigma=0.4$, $z_1 = 0.5$, $z_2=1.5$, $u_\mathrm{flow}=1$, $\Delta\rho/\rho_0 = \rho_0 = 1$. We note that the instability is not resolved in our runs with this choice of parameters, i.e. the instability grows too slowly due to numerical effects. Simulations are run in a periodic domain with $L_y = 2,L_x = 1$.  \par
In figure \ref{fig.kh_2d}, we show the density of four simulations at $t=2$.  Top left to bottom right panels show WENO5-Z without boost with $512\times1024$ zones resolution, WENO5 with $32\times64$ zones without a velocity boost, WENO5 with $32\times 64$ zones boosted by $v_x = 50$ and TVD+CTU with $128 \times 256$ zones without boost. \par
The TVD+CTU run does not develop a vortex roll due to numerical diffusivity, even at twice the resolution from the WENO5 runs. In contrast, the low resolution WENO5 runs develop a vortex similar to the high resolution simulation \citep[see][]{2016MNRAS.455.4274L}. This shows that WENO5 resolves instabilities much better than the second order TVD+CTU, even at half the grid resolution. WENO5 is also much less susceptible to advection errors. The boost of $v_x = 50$ (!) barely changes the result. Some differences are visible at the tip of the roll, because the truncation error is not Galilean invariant. Nonetheless, we expect significant improvements in the growth of instabilities in cosmological simulations, where advection is commonplace and may be a major source of diffusivity for second order Eulerian methods \citep{2010MNRAS.401..791S}.

\subsection{Gresho-Vortex}

The Gresho-Vortex \citep{1990IJNMF..11..621G,Liska2003,2010MNRAS.401..791S} is a two dimensional rotating structure in hydrodynamic equillibrium that evaluates angular momentum and vorticity conservation of the code. The test becomes very challenging for Eulerian codes, when the vortex is advected to the grid. It exposes how the increase in truncation error affects angular momentum conservation. \par
In a 2 dimensional unit domain we set $\vec{U} = (1,v_x(r),v_y(r),0,0,0,0,P(r))^T$, with the radius $r$, $v_x(r) = -v(r) \cos(\phi) + v_\mathrm{boost}$, $v_y(r) = -v(r) \sin(\phi)$, $\phi = \tan^{-1}(y/x)$\footnote{We use the Fortran \texttt{ATAN2} function here.} and
\begin{align}
    v(r) &= \begin{cases}
             5r         &\Leftrightarrow r \le 0.2 \\
             2 - 5r     &\Leftrightarrow r < 0.4 \\
             0          &\mathrm{otherwise}
            \end{cases} \\
    P(r) &= \begin{cases}
             5 + \frac{25}{2} r^2  &\Leftrightarrow r \le 0.2 \\
             9 + \frac{25}{2} r^2 - 20 r + 4 \log\left( \frac{r}{0.2} \right)    & \Leftrightarrow r < 0.4 \\
             3 + 4 \log(2)          &\mathrm{otherwise}
            \end{cases}
\end{align}

\begin{figure}
    \centering
    \includegraphics[width=0.45\textwidth]{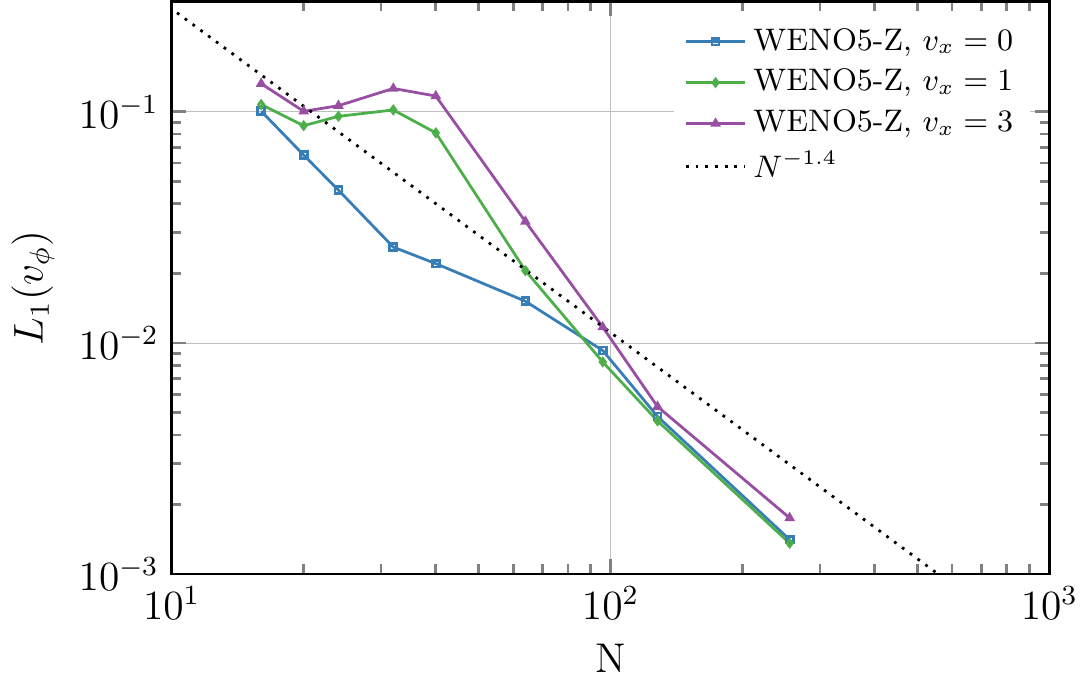}
    \caption{$L_1$ error norm of the angular velocity $v_\phi$ in the Gresho Vortex test at $t=3$ for three WENO5 simulations with $v_x=0$ (blue), $v_x = 1$ (green) and $v_x= 3$ (purple). The dotted line corresponds to a scaling of $N^{-1.4}$ and is normalized to approximately correspond to the results shown in \citet{2010MNRAS.401..791S}, their figure 29.}
    \label{fig.grconv}
\end{figure}
\begin{figure*}
    \centering
    \includegraphics[width=0.9\textwidth]{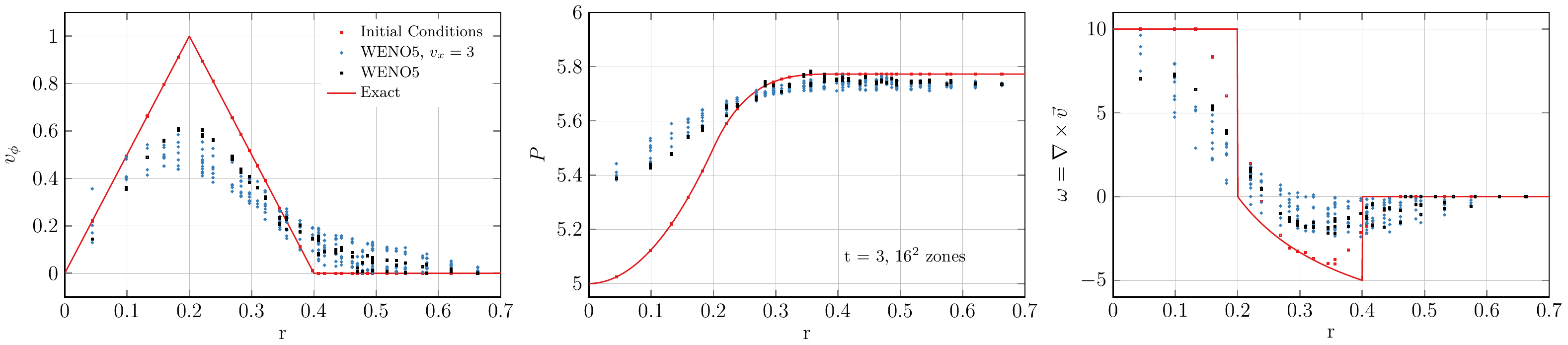}
    \includegraphics[width=0.9\textwidth]{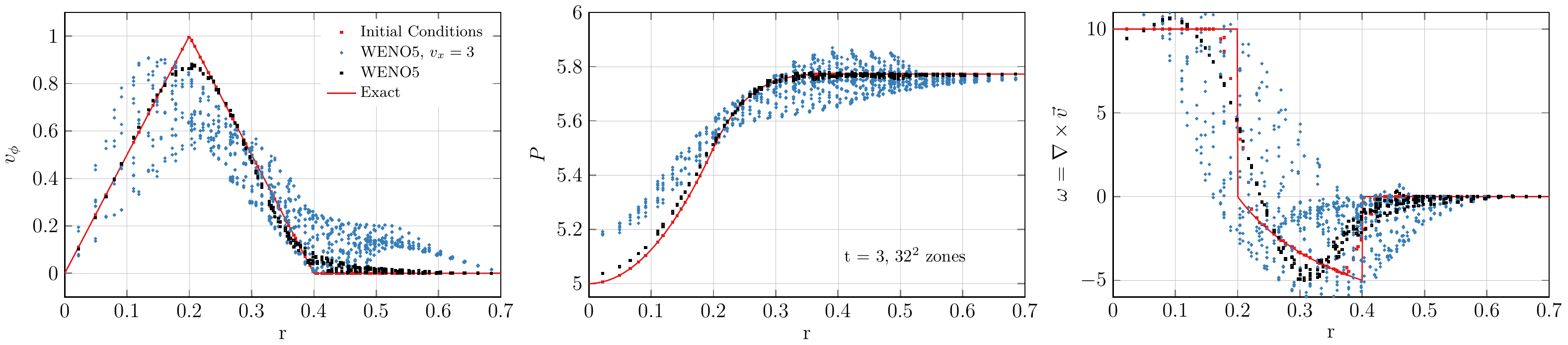}
    \includegraphics[width=0.9\textwidth]{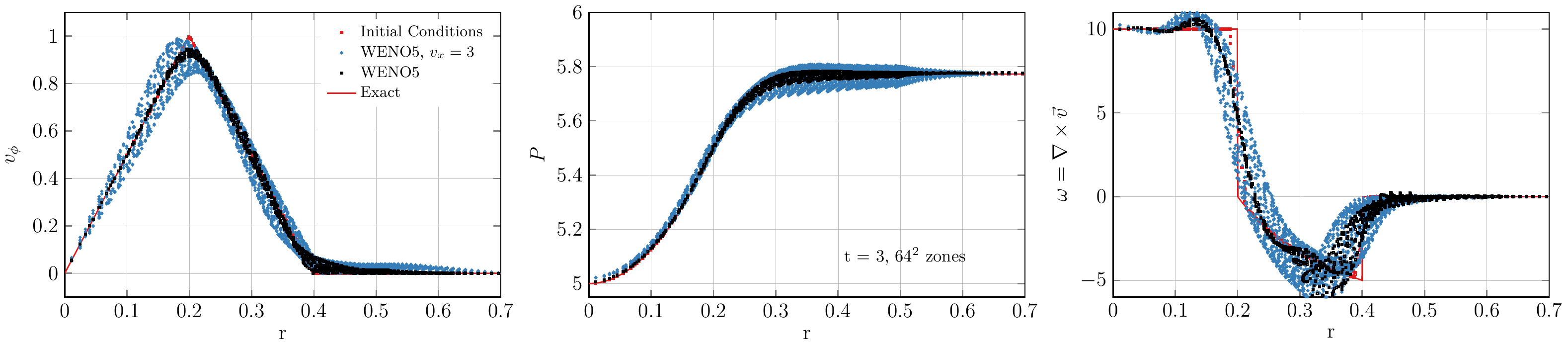}
    \caption{Left to right: Angular velocity, pressure and vorticity profile of the Gresho Vortex simulations at (top to bottom) $16^2, 32^2, 64^2$ zones. In each panel we show the exact solution as red line, the initial conditions as red dots and  at $t=3$ the WENO5 solution without boost as black dots alongside the WENO5 solution with $v_x = 3$ as blue dots.}
    \label{fig.gr}
\end{figure*}
We run a convergence study of the test with the WENO5-Z algorithm using three different boost velocities in x-direction: $0,1,3$. The results are shown in figure \ref{fig.grconv} where we show the $L_1$ error over resolution. Blue, green and purple lines correspond to boost velocities of $0,1,3$, respectively. We add a dotted line scaling with $N^{-1.4}$, which corresponds to the scaling obtained with \texttt{\small AREPO} in \citet{2010MNRAS.401..791S}. Our results compare quite favourably to the results obtained with lower order codes. In particular, the error remains small even with the velocity boost at most resolutions and is smaller than the errors observed with \texttt{\small ATHENA} at  all times. The boost mostly affects resolutions below $100^2$, where the vortex becomes under-resolved. Below $20^2$ the $L_1$ error with velocity boost reduces again. We note that the error changes depending on how the center of the vortex is placed on the grid, likely because the initial conditions contain a discontinuity at $r=0.2$. As discussed in \citet{2013MNRAS.428..254M}, the sampling of the discontinuity affects the convergence properties of the simulations. \par
To illustrate this further,  we show in figure \ref{fig.gr} (left to right) angular velocity $v_\phi$, pressure and vorticity $\omega(r)=\nabla \times \vec{v}$ over radius. We compute :
\begin{align}
    v(\phi) &= r \frac{x v_y - y v_x}{x^2 + y^2},
\end{align}
and the vorticity from the standard finite difference operator. Analytic profiles are shown as red lines, the initial conditions as red dots, the simulation at $t=3$ without boost as black dots and the simulation with velocity boost of $v_x = 3$ as blue dots. From top to bottom we show resolutions of $16^2,32^2$ and $64^2$ zones. The run with $32^2$ zones and velocity boost shows the largest scatter, in-line with the increase in $L_1$ error observed before. At $16^2$ zones the profiles show less scatter, but are only sampled with 15 unique values below $r=0.4$ in the ICs. At this low resolution the problem is under-resolved and it stands to reason that the reduced scatter in the evolved boosted profiles is a result of the poor sampling. In particular, the discontinuous peak of the $v_\phi$ profile is not sampled directly, which should lead to additional error \citet{2010MNRAS.401..791S,2013MNRAS.428..254M}. The pronounced increase in $L_1$ error around $32^3$ zones seems to be a feature of the broad WENO5 reconstruction. Indeed \citet{Liska2003} find a relatively high $L_1$ error in total kinetic energy for their WENO5 scheme at low resolutions. \par
We conclude that the stationary solution with WENO5 is very competitive with the \arepo and \athena solutions presented in \citet{2010MNRAS.401..791S}. The error in the boosted solutions is comparable with second order static mesh solutions at intermediate resolutions and approaches the Lagrangian error at resolutions above $100^2$ zones.

\subsection{Advection of a Density Square}

\begin{figure*}
    \centering
    \includegraphics[width=0.9\textwidth]{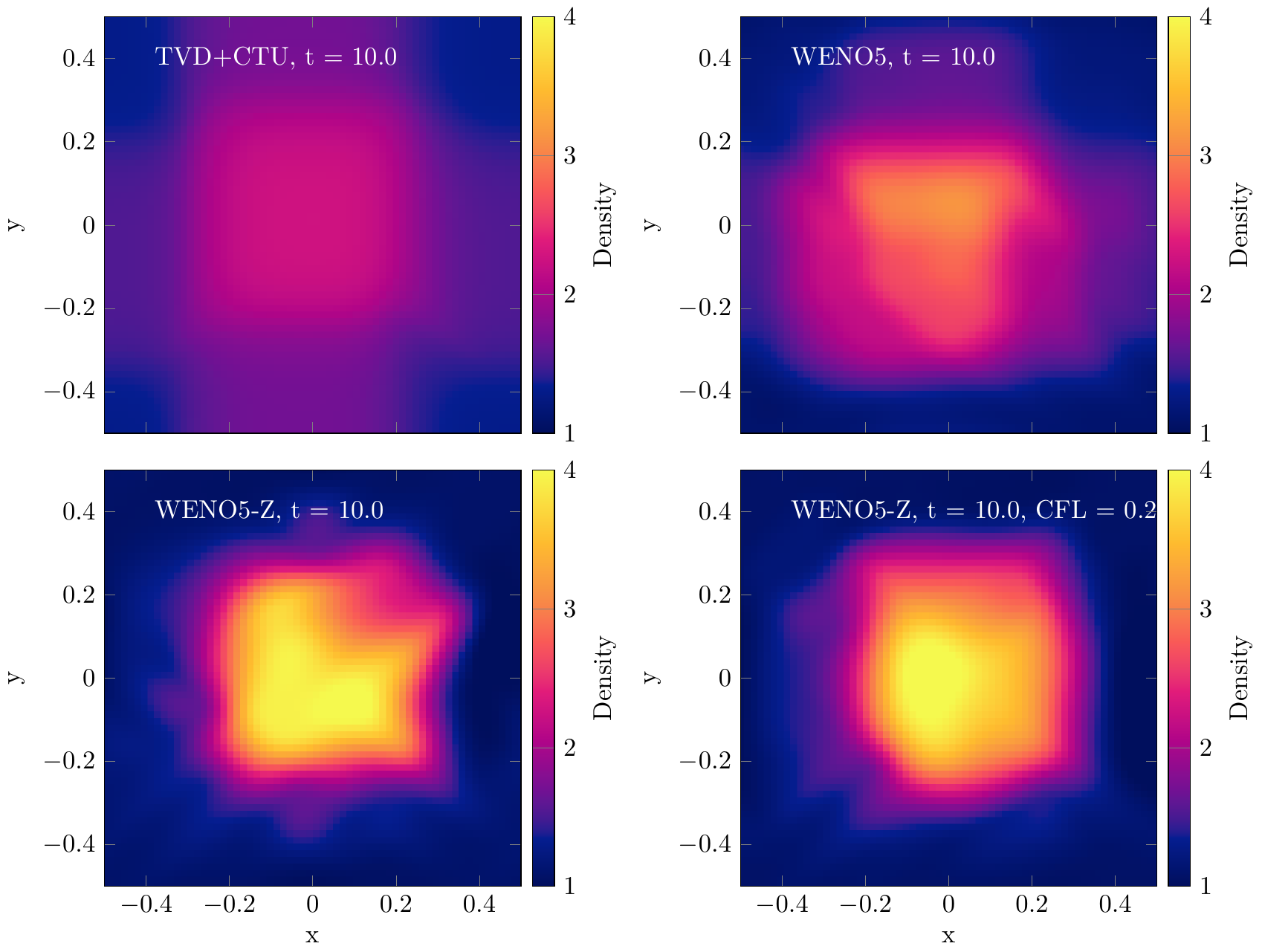}
    \caption{Density in the square advection test at $t=10$ with $64^2$ zones. Top left to bottom right: TVD+CTU, WENO5, WENO-Z, WENO-Z with CFL=0.2.}
    \label{fig.dens_adv_2d}
\end{figure*}

\begin{figure}
    \centering
    \includegraphics[width=0.45\textwidth]{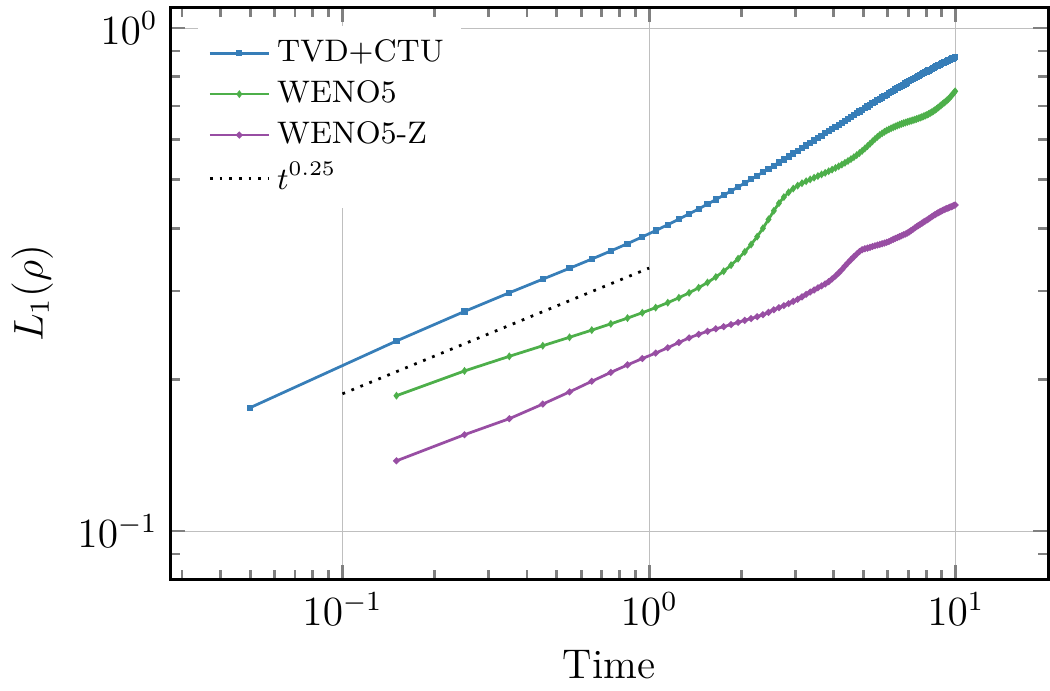}
    \caption{$L_1$ error norm over time in square advection test with $64^2$ zones. We show TVD+CTU (blue), WENO5 (green) and WENO-Z (purple).}
  \label{fig.dens_adv_conv_2d}
\end{figure}

The velocity dependent truncation error of Eulerian grid methods can be clearly exposed in the density advection test \citep{2010MNRAS.401.2463R,2015MNRAS.450...53H}. In 2 dimensions a density square in hydrodynamic equilibrium   with the surrounding medium is advected supersonically  through a unit domain: $\vec{U} = (1, 100, 50,0,0,2.4)^T$, except $\rho = 4$ when $-0.25 < x < 0.25$ and $-0.25 < y < 0.25$. Here $\gamma=5/3$, thus $c_s = 2$, so the advection is supersonic ($M > 100$) with respect to the reference frame of the mesh. While this test is trivial with Lagrangian schemes such as SPH, meshless finite volume and moving finite volume schemes, it presents a very serious challenge to Eulerian grid methods. \par
We show the result of the test run with TVD+CTU, WENO5 and WENO5-Z in figure \ref{fig.dens_adv_2d} at time $t = 10$. We also include a WENO5-Z run with $CFL=0.2$.  The convergence rate is shown in figure \ref{fig.dens_adv_conv_2d}. \par
Due to the high velocity relative to the grid, the solution is smeared out significantly by the TVD+CTU algorithm, as expected from a second order method. WENO5 performs a bit better, however still with significant dispersion. This is because the edges of the density square are critical points in the flow, where the classical WENO5 weights are only third order accurate. WENO5-Z performs significantly better, with the density square being preserved, but strongly distorted. Quantitatively WENO5-Z shows an improvement in $L_1$ error by roughly a factor of two at late times. \par
These results can be directly compared to the results shown with the discontinuous Galerkin (DG) code \texttt{\small TENET} in \citet{2015MNRAS.453.4278S}. In terms of $L_1$ error, WENO5 performs roughly like a second order DG code. WENO5-Z performs better than second order DG, but worse than third order DG. We note that increasing the convergence order in a DG scheme reduces the time step by a factor of 2. Thus these DG schemes are more computationally expensive than WENO5, where the time step does not change with order of the scheme. DG schemes also require more memory, because the full polynomial has to be stored in every zone in one way or another. We ran the WENO5-Z simulation with a reduced CFL number of 0.2 to approach the computational cost of a DG3 algorithm. The distortion of the square is reduced further, but the $L_1$ error remains roughly the same. \par
This result suggests to run strongly advection dominated problems with WENO5-Z not the classical WENO5 method. As the method is significantly less robust, this results in a trade-off between more protection fluxes and reduced advection error for complex problems. It will depend on the  problem under consideration, which approach delivers better results. 

\subsection{Advection of a Magnetic Field Loop}

\begin{figure*}
    \centering
    \includegraphics[width=0.9\textwidth]{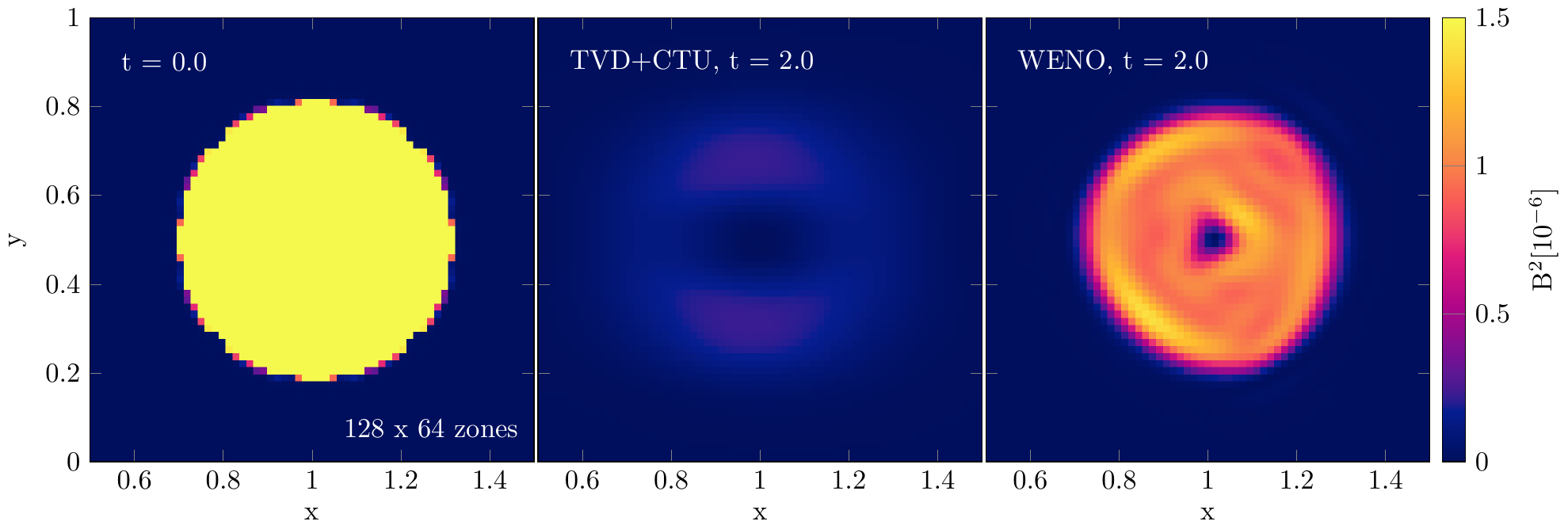}
    \caption{Magnetic field energy density $B^2$ at zone center in units of $10^{-6}$ of the magnetic field loop advection test in 2 dimensions with 128 x 64 zones. Left to right: Initial conditions, TVD+CTU at time $t=2$ and WENO5 at time $t=2$.}
    \label{fig.bloopadv_2d}
\end{figure*}

\begin{figure}
    \centering
    \includegraphics[width=0.45\textwidth]{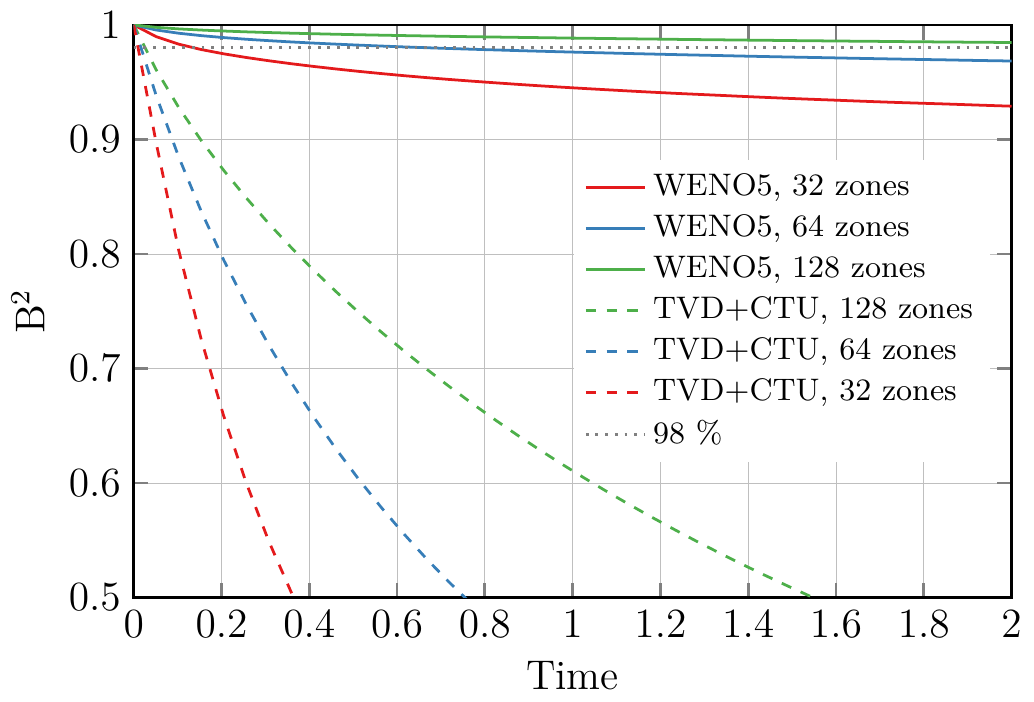}
    \caption{Total magnetic field energy density over time for the magnetic field loop test in 2 dimension. We show WENO5 at three resolutions: 64x32 zones (red), 128x64 zones (blue) and 256x128 zones (green). }
    \label{fig.bloopadvE_2d}
\end{figure}

The field loop advection test is instrumental for the development and testing of constrained transport schemes for the magnetic field evolution in Eulerian codes \citep{2008JCoPh.227.4123G}. This is a difficult test for the CT scheme of the code, as only the electric field and magnetic field are moving through the domain. In applications this will rarely be the case. \par
In 2 dimensions, we set up a periodic domain of size $L_x \times L_y = 2 \times 1$ with $\vec{U}=(1,2,1,0,B_x,B_y,0,1)^T$  and a vector potential on the interfaces with zero x and y component, but :
\begin{align}
    A_z &= A_0 \left( r_c - r \right), 
\end{align}
where $A_0 = 10^{-3}$, $r_c = 0.3$. The magnetic field on the interfaces is then obtained as the curl of the vector potential and interpolated to the grid at fourth order with (Jang et al. in prep.):
\begin{align}
    B_x &= ( 3 B_{x,i-3,j} - 25 B_{x,i-2,j} + 150 B_{x,i-1,j} \nonumber\\ 
        &+ 150 B_{x,i,j} - 25 B_{x,i+1,j} + 3 B_{x,i+2,j} ) /16
\end{align}
In figure \ref{fig.bloopadv_2d}, we show the resulting magnetic field energy density of the initial conditions (left) and the simulation at time $t=2$ with TVD+CTU (middle) and WENO5 (right) with $128 \times 64$ zones. The WENO5 scheme keeps the shape of the loop well, while it is advected through the box. The shape of the loop is comparable to second order CT schemes \citep{2008ApJS..178..137S,2013JCoPh.243..269L}. In contrast, the TVD+CTU loop is barely visible, most of the magnetic energy has dissipated into heat. To quantify this, we show the time evolution of normalized magnetic energy in three runs with 32, 64, and 128 zone base resolution in figure \ref{fig.bloopadvE_2d}. We mark 98 \% as dotted horizontal line. \par
WENO5 retains more than 90\% of the magnetic energy of the field loop, even at the lowest resolution run, while TVD+CTU loses more than half of the magnetic energy until that time, even at the highest resolution. The conservation of magnetic energy also compares favourably with the results shown in \citet{2013JCoPh.243..269L}. Even though our CT scheme is much less elaborate than theirs, the magnetic field fluxes are fifth order accurate and thus improve energy conservation. Again WENO5 roughly doubles the effective resolution of the simulation. i.e. our run $64^2$ zones conserves magnetic energy roughly as \texttt{ATHENA} with $128^2$ zones. These results are in-line with our findings from the wave convergence tests, where wave diffusion is fifth order accurate, but wave dispersion is second order accurate.  \par
\begin{figure}
    \centering
    \includegraphics[width=0.45\textwidth]{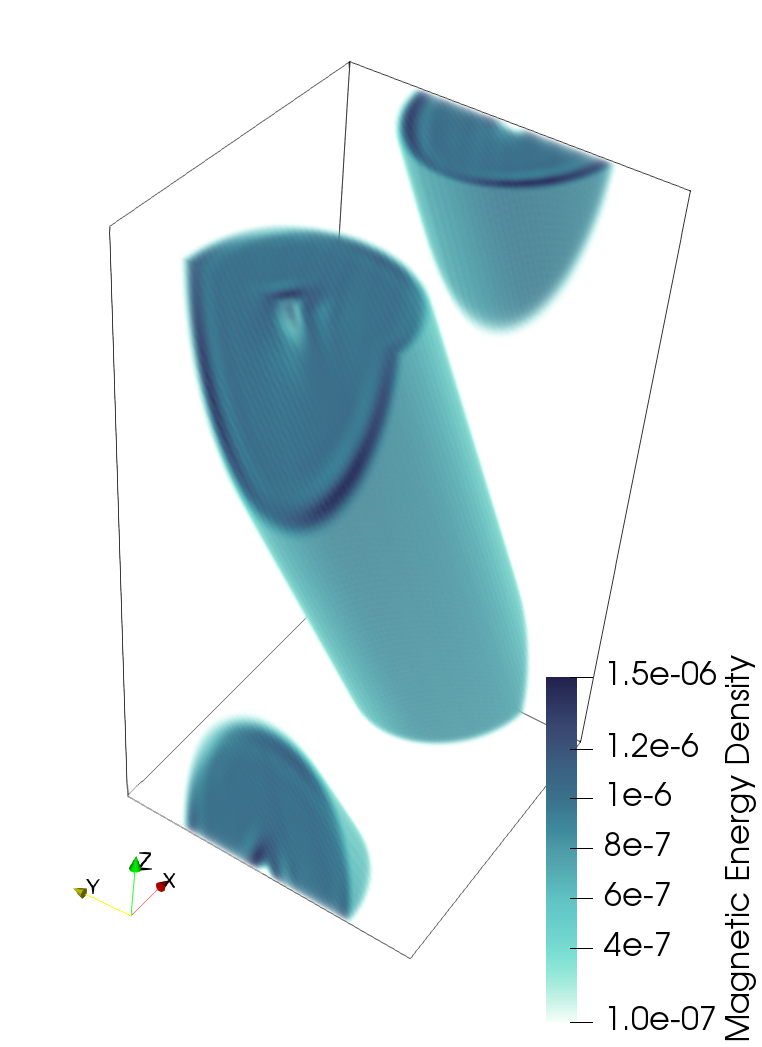}
    \caption{Magnetic field energy density $B^2$ of the magnetic field loop advection test in 3 dimensions with $128^2 \times 256$ zones at time $t=2$. A threshold was imposed at $10^{-7}$.}
    \label{fig.bloopadv_3d} 
\end{figure}
In three dimensions, the loop is rotated around the y-axis by $\phi = \arctan(0.5)$\citep{2008JCoPh.227.4123G}. We set up the magnetic vector potential as:
\begin{align}
    A_x &=  -A_0 \frac{\left(r_c -r^\star \right)}{\sqrt{5}}\\
    A_y &= 0 \\
    A_z &= -A_0 \frac{2\left(r_c -r^\star \right)}{\sqrt{5}}, 
\end{align}
where $r^\star$ is the radius obtained from the coordinates :
\begin{align}
    x^\star &= \begin{cases}
                    \frac{ (2x + z) + 2}{\sqrt{5}} & \Leftrightarrow x < -\sqrt{5}\\
                    \frac{ (2x + z) - 2}{\sqrt{5}} & \Leftrightarrow x > -\sqrt{5}
               \end{cases} \\
    y^\star &= y \\
    z^\star &= \frac{-x + 2z}{\sqrt{5}}
\end{align}
A rendering of the resulting magnetic energy at  time $t = 2$ is shown in figure \ref{fig.bloopadv_3d}, i.e. after advecting the loop twice through the grid. Following \citep{2008JCoPh.227.4123G} we impose a threshold in the projection at $B^2 = 10^{-7}$ to show the shape of the loop edge. Again the structure of the loop is comparable with the rendering shown in  \citep{2008JCoPh.227.4123G}.

\subsection{Sedov-Taylor Blast Wave}

\begin{figure*}
    \centering
    \includegraphics[height=0.25\textheight]{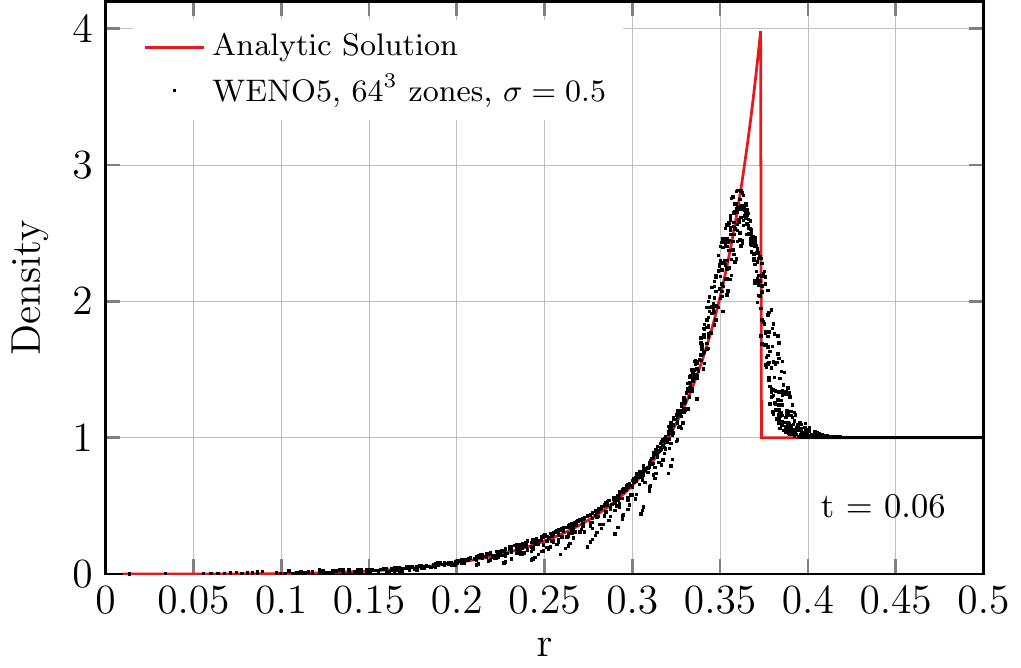}
    \includegraphics[height=0.25\textheight]{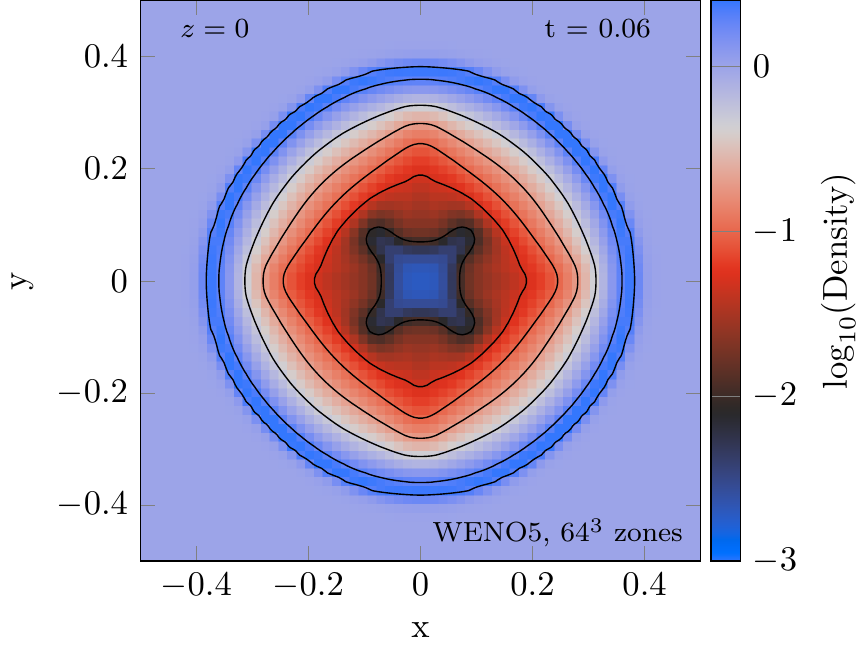}
    \caption{Left: Radial density profile of the three dimensional Sedov-Taylor test at $t=0.06$, WENO5 simulation as black dots, analytic solution as red line. Plotted is only every 20th zone. Right: Slice through the simulation at $z=0$. We include contours at  0.01, 0.05,0.1,0.2,0.5 and 2.}\label{fig.STB}
\end{figure*}

This test models the self-similar evolution of a point-like thermal detonation \citep{vonNeumann:1942:PSS,1950RSPSA.201..159T,1946Sedov}. Despite its unpleasant historical context, it also is a model for early supernova explosions and is useful to expose the effects of the regular grid on the evolution of a strong spherical shock wave propagating into a thin medium with negligible pressure. We set $\vec{U} = (1, 0,0,0, 10^{-8},0,0, 10^{-5})$ everywhere in a three dimensional computational domain with $L_x = L_y = L_z = 1$. We inject a unit energy $E_1 = 1$ in the center of the box, and distribute it using a Gaussian with FWHM $\sigma = 0.5 \, \mathrm{dx}$, where dx is the zone size. Our simulation uses $\gamma = 5/3$ and $N=64^3$ zones. The self-similar analytic solution of the problem can be found in \citet{1966hydr.book.....L}, with $\beta=1.15$.  \par
In the left of figure \ref{fig.STB}, we plot the radial profile of the shock at $t=0.06$: analytic solution as red line and every 20th zone of the simulation as black dot. On the right of figure \ref{fig.STB}, we show a slice through the simulation at $z=0, t=0.06$ with colours on a log scale from $10^{-3}$ to $2$ and contours at 0.01, 0.05, 0.1, 0.2, 0.5 and 2. The imprint of the grid on the explosion is clearly visible at radii below 0.3 and comparable to the DG3 method reported in \citet{2015MNRAS.453.4278S}, but better than the lower order WENO method presented in \citet{2004ApJ...612....1F}, likely due to our higher order time integrator. In this test, Lagrangian schemes perform much better than our Eulerian scheme, due to the adaptive sampling \citep[e.g.][]{2010MNRAS.401..791S}.

\subsection{MHD Blast Wave}

\begin{figure*}
    \centering
    \includegraphics[width=0.45\textwidth]{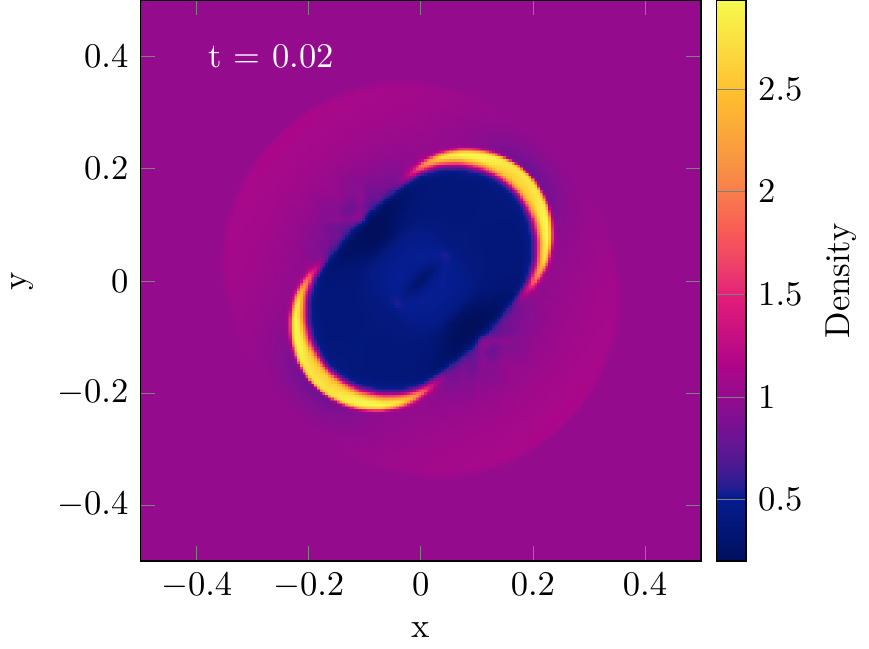}
    \includegraphics[width=0.45\textwidth]{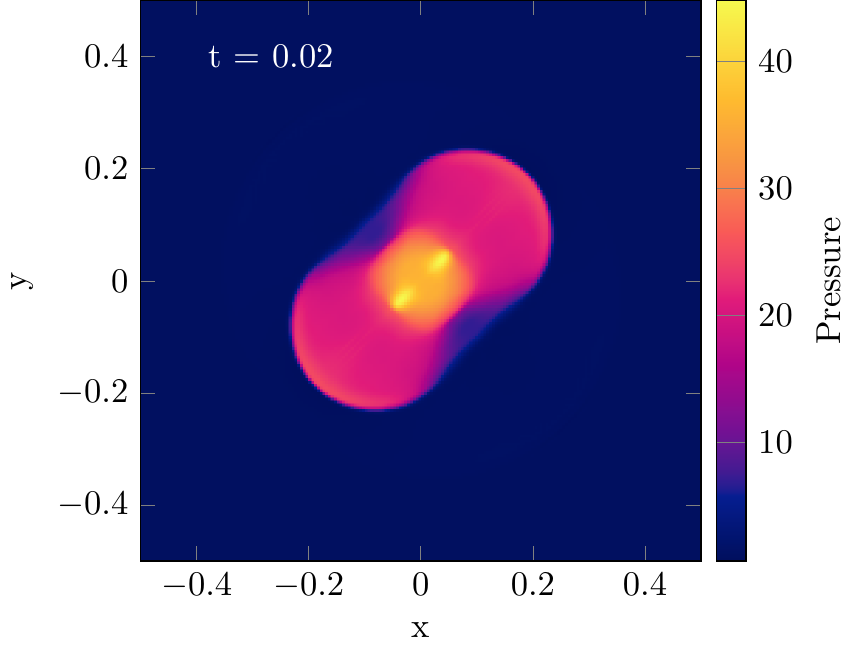}\\
    \includegraphics[width=0.45\textwidth]{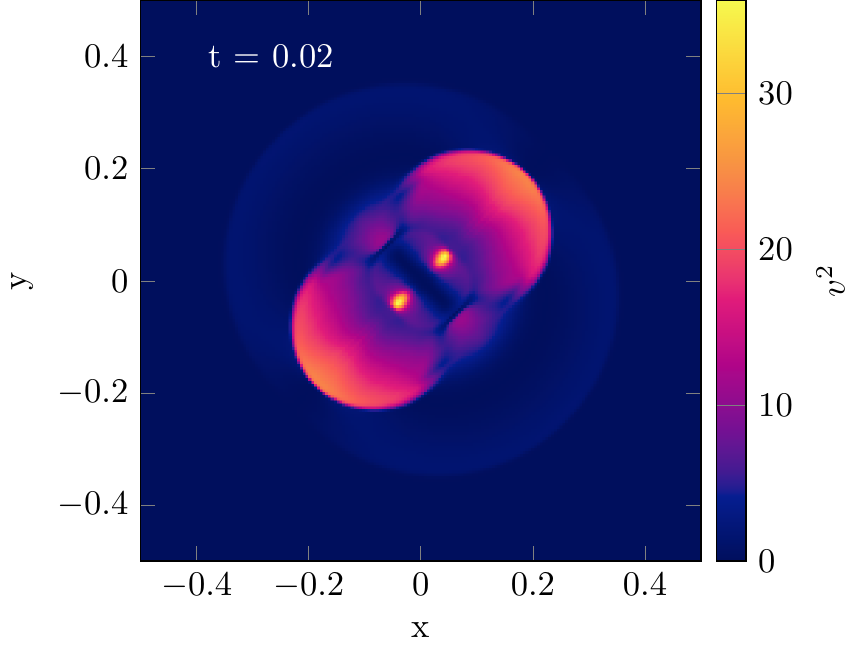}
    \includegraphics[width=0.45\textwidth]{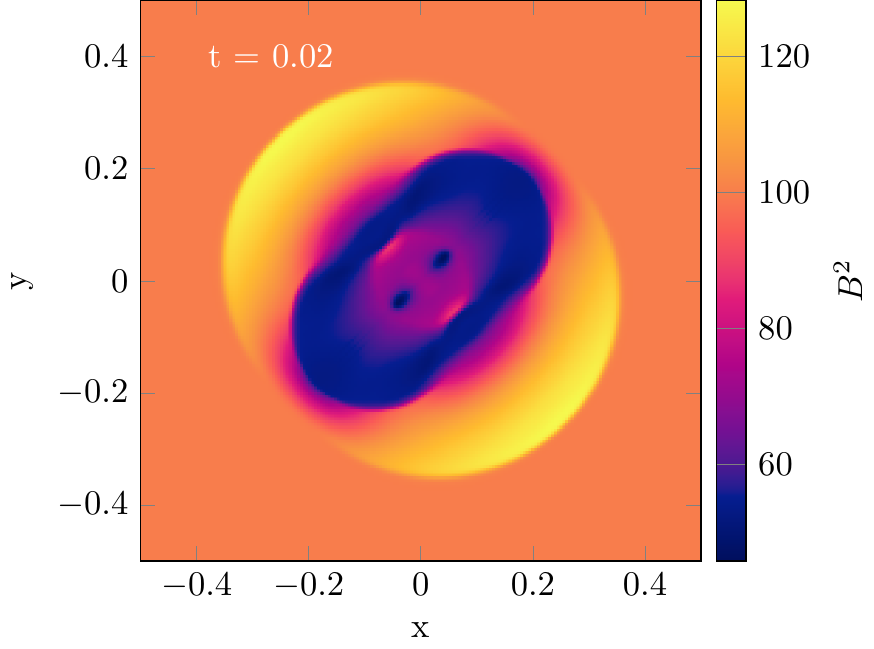}
    \caption{Slices through the MHD blast wave test at time $t=0.02$. Shown is the center of the domain with $200\times300\times200$ zones along $z=0$. Top-left to bottom right: density, pressure, $v^2$ and $B^2$.}
    \label{fig.MHDBlast_3d}
\end{figure*}

The magnetized blast wave simulation test code performance in the low-$\beta$ regime, i.e. when shocks are evolved in the presence of strong magnetic fields. Following \citet{2000ApJ...530..508L}, we set $\vec{U} = (1,0,0,0,B_0/\sqrt{2},B_0/\sqrt{2},0,P)^T$, with $P(r < r_c) = 1$, $P(r > r_c) = 100$, $B_0 = 10$ and $r_c = 0.125$. It follows that $\beta = 0.02$ in the medium ahead of the shock. The domain is $L_x \times L_y\times L_z = 1 \times 1.5 \times 1$. Slices through density, pressure,  $v^2$ and $B^2$ from a WENO5 simulation at time $t = 0.2$ and $z = 0$ with resolution $200\times300 \times 200$ zones are shown in figure \ref{fig.MHDBlast_3d}. The imprint of the magnetic field that is oriented along $(1,1,0)^T$  on the shock is clearly visible. We note that this test evaluates the robustness of the algorithm and indeed WENO5 required protection floors to complete the simulation. We leave a more elaborate implementation of protection fluxes to future work.

\section{Code Performance} \label{sect.perf}

\subsection{Cache Blocking}

\begin{figure}
    \centering
    \includegraphics[width=0.45\textwidth]{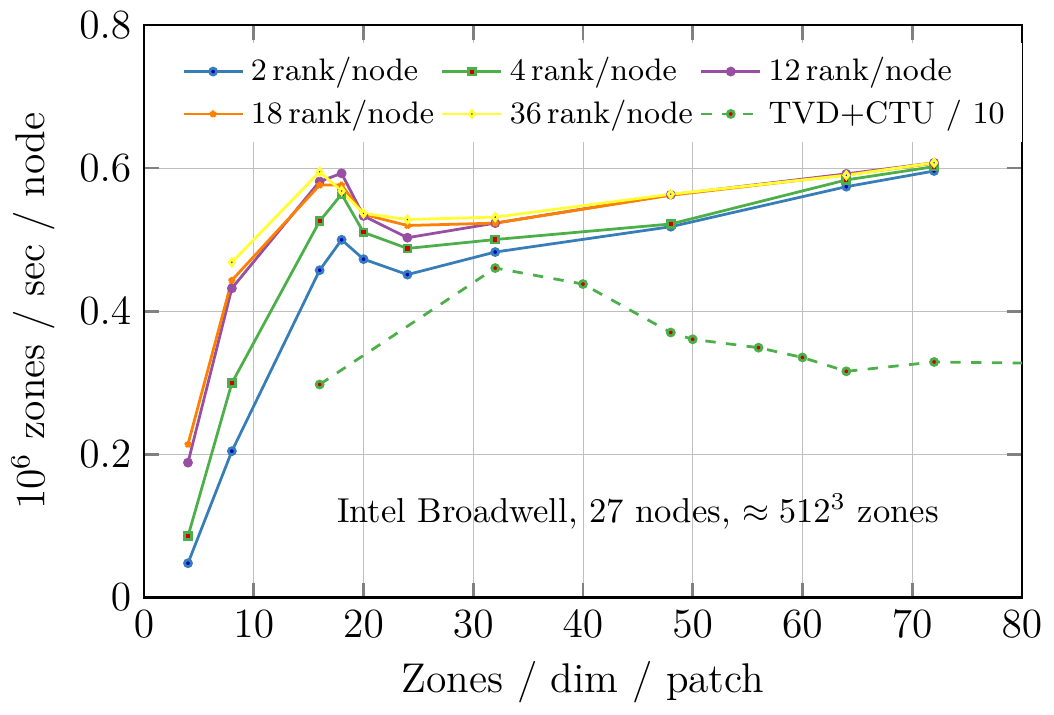}
    \caption{Compute performance in million zones per second per node over patch size (i.e. number of zones per dimension per patch) for 27 nodes on Intel Broadwell architecture. Problem size was about $512^3$ zones. Blue: 2 ranks \& 18 threads per node. Green: 4 ranks \& 9 threads per node. Purple: 12 ranks \& 3 threads per node. Orange: 18 rank \& 2 theads per node. Yellow: 36 ranks \& 1 thread per node. }\label{fig.patch}
\end{figure}

All modern HPC systems feature a cache hierarchy to increase effective memory bandwidth to the CPUs (level 1, level 2, ...), with the fastest cache usually being the smallest due to silicon cost. To achieve good performance, it is desirable to divide the computational problem into sub-problems that fit into the cache hierarchy, so that memory bandwidth is increased. This technique is called \emph{cache blocking}. The optimal size will depend on the architecture of choice and the overhead in the algorithm, thus it has to be determined empirically. In \wombat, cache blocking is mitigated by tuning the patch size for a given architecture. Small patch sizes are desirable, because they enable more fine-grained load-balancing. \par
We run a patch size optimization study with 27 nodes to saturate MPI communication on the Aries interconnect. In figure \ref{fig.patch}, we show the performance  of the WENO5 algorithm on the Broadwell architecture with 2 x 18 cores per node in Million zones per second per node as a function of number of zones per dimension per patch. The problem size is at least $512^3$ zones, but varies by about 30 \% between runs, as the number of patches must be evenly divisible by the number of threads to not induce load imbalance in the threading. Colours correspond to runs with varying number of MPI processes per node: Blue: 2 MPI ranks/12 OpenMP threads. Green: 4 ranks/9 threads. Purple: 12 ranks/3 threads. Orange: 18 ranks/2 threads. Yellow 36 ranks/1 thread.  \par
For all decompositions, the performance is maximized at $18^3$ zone patches, then drops slightly and increases again toward $70^3$ zone patches. This behaviour differs from the TVD+CTU solver that peaks at $32^3$ and shows a strong drop in performance for larger patch size (see green dashed line from \citep[][]{2017ApJS..228...23M}). The flattening of the data arrays increases the vector length of all loops in the algorithm and shifts the optimal cache block to smaller patch sizes in the WENO5 solver. It also leads to effective use of the hardware pre-fetcher at large patch sizes, so the cache blocking is effectively done in hardware. This is not possible in the TVD+CTU solver that is not flattened. With the side constraint of small patch sizes, we conclude that WENO5 performs at 0.6 Million zones per node with $18^3$ zones per patch on Broadwell, with 4 MPI rank per node. WENO5 is thus a factor 7.5 slower than the TVD+CTU solver at the same resolution, but doubles the effective resolution. It follows that the WENO5 solver is more efficient than the second order scheme, because increasing the resolution by a factor 2 per dimension increases runtime of the TVD+CTU solver by roughly a factor 16: 2x per dimension and roughly 2x because the time step halves due to the factor 2 in $\Delta x$ in equation \ref{eq.cfl} (TVD+CTU uses a 1D criterion, but with $CFL=0.4$ in 3D).

\subsection{Vectorization}

\begin{figure}
    \centering
    \includegraphics[width=0.45\textwidth]{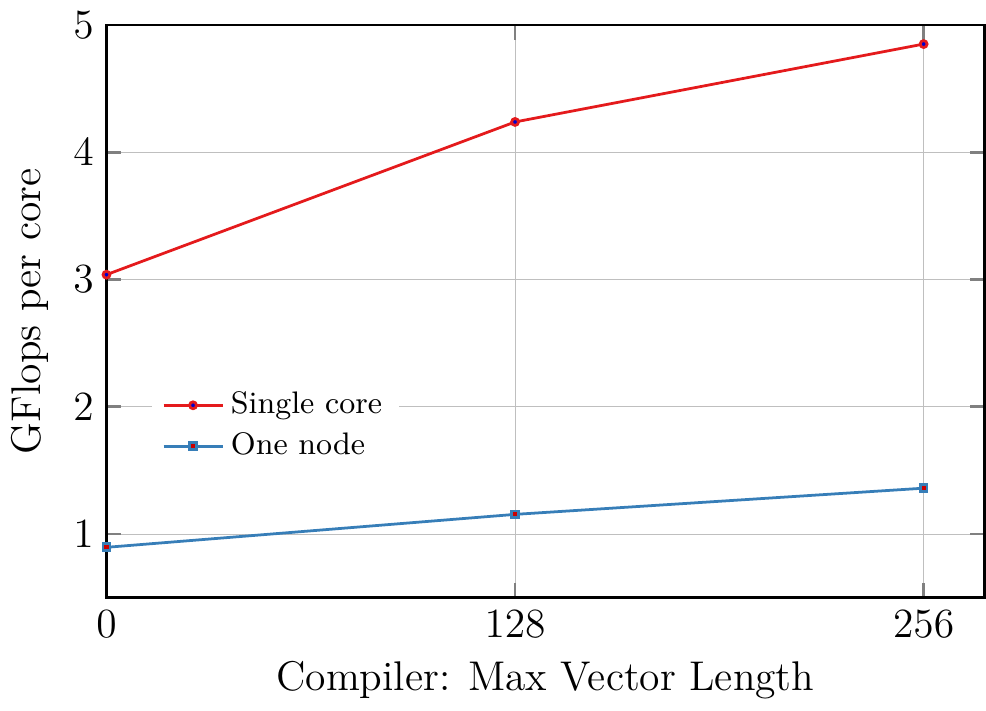}
    \caption{Performance per core in billion floating point operations per second over SIMD vector length of the WENO solver on a single Broadwell core (red) and a full Broadwell node (blue).}\label{fig.vec}
\end{figure}

Our implementation uses vector processing registers (SIMD) in modern CPUs extensively. Flattening the data arrays leads to large trip counts in the vector loops that are well cache blocked. For every memory access, the virtual address has to be translated into a physical address. In the translation look-aside buffer (TLB), the memory management unit buffers page addresses for this translation from virtual to physical memory. On a standard system with 4kb memory pages our optimized vector loops can lead to a large miss rate in the TLB, the resulting page walk reduces performance by a factor of two (TLB thrashing). Thus it is important to run \wombat with some form of \emph{huge pages} enabled. \par
To demonstrate the performance gain from vectorization, we show the performance of the WENO5 solver on a 3D problem in GFlops on the Intel Broadwell architecture in figure \ref{fig.vec}. In red a single core without threads, in blue a single core with whole node active (36 cores as 4 ranks with 9 threads.). The performance of the run was measured by Cray Perftools.  \par
\begin{figure}
    \centering
    \includegraphics[width=0.45\textwidth]{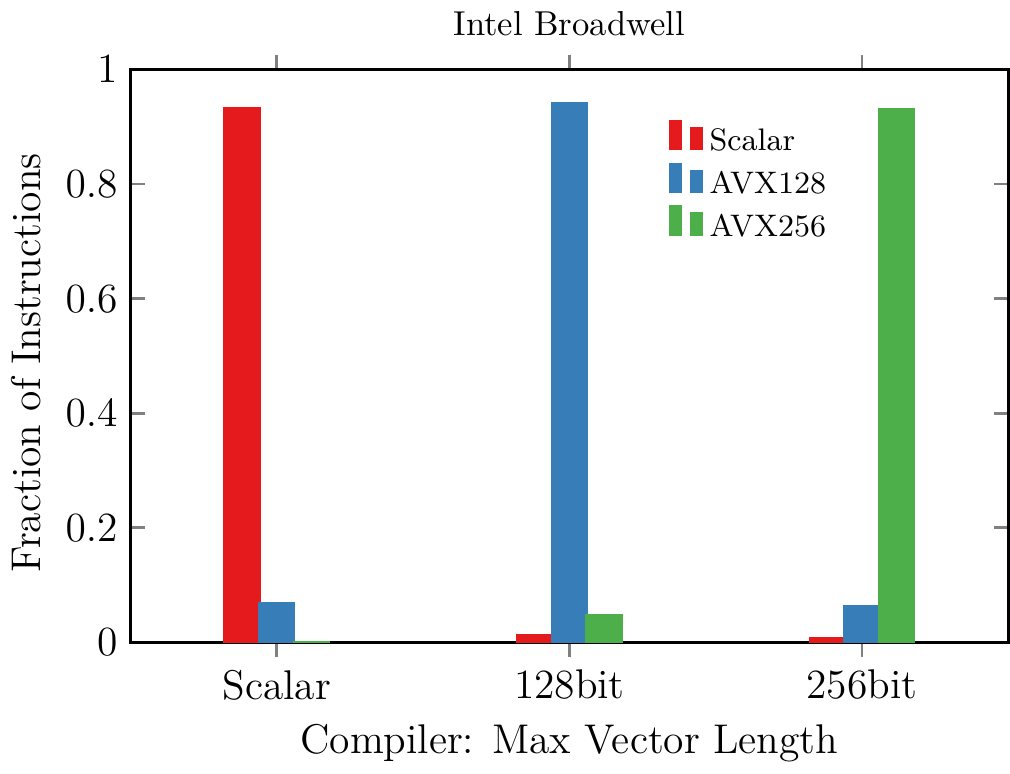}\\
    \includegraphics[width=0.45\textwidth]{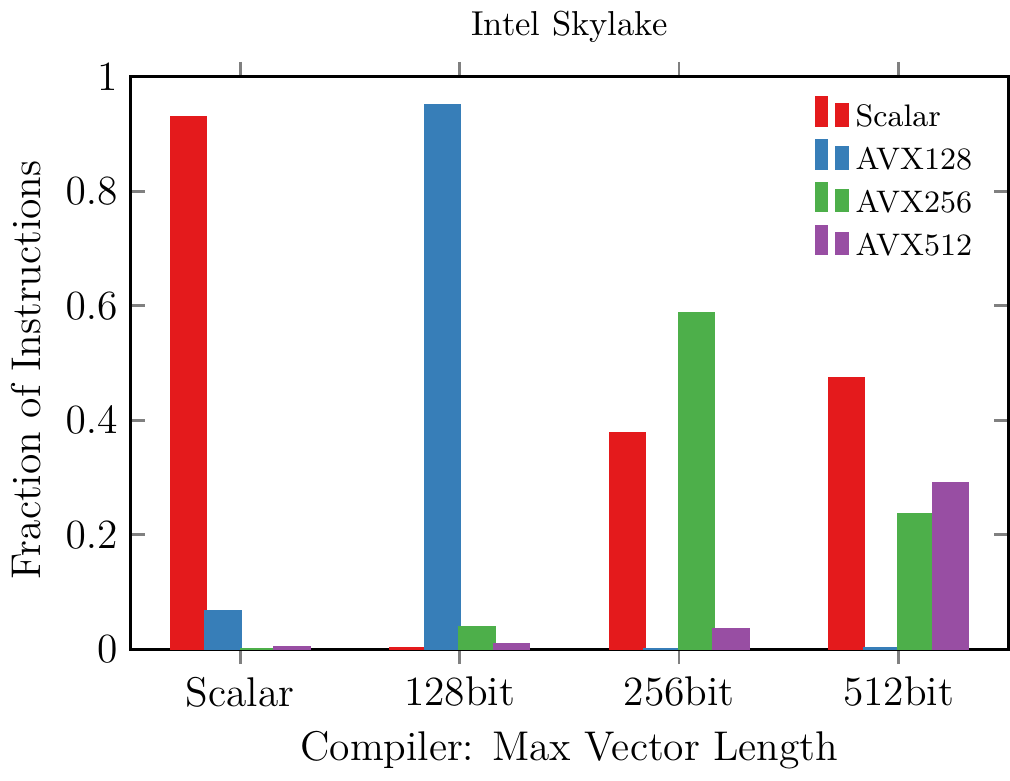}
    \caption{Fractional number of instructions generated by the Cray compiler over limit of the vector length. Top for Intel Broadwell, bottom for Intel Skylake.}\label{fig.instr}
\end{figure}
In figure \ref{fig.instr} we compare the fractional number of instructions generated by the Cray compiler for the Intel Broadwell and Intel Skylake architecture, given a largest vector length using the \texttt{-vector0} and \texttt{-preferred-vector-length} compiler flags. The top panel shows the instruction for the Intel Broadwell architecture. Clearly the compiler vectorizes virtually all of the code at the highest vector length. The bottom panel shows the same graph, but for the Intel Skylake architecture. Here the compiler only vectorizes the complete code at 128 bit vector length, but chooses a mix of vectorization and scalar instructions at broader vector lengths. In particular, at 512bit vectors, scalar instructions make up half of the program. We note that the version with the widest vectors is the fastest for both architectures, reaching about 20\% of double precision peak performance. Thus Skylake is still faster than Broadwell by about 500 MFlops.\par
This shows that on some architectures other constraints than vector length influence performance of the code. This is a good argument against using intrinsics or compiler pragmas to vectorize the whole program manually. Aside from the additional maintenance effort required to port the program to new architectures, the approach can actually increase execution time on architectures like Skylake. The better strategy is to expose instruction level parallelism to the compiler and let the auto-vectorizer choose the instructions depending on its internal metrics. Given the increasing variety of HPC architectures competing for new exa-scale systems in the next years, the compiler remains the central tool for the programmer to achieve optimal performance on a given architecture.

\subsection{Weak Scaling}

\begin{figure}
    \centering
    \includegraphics[width=0.45\textwidth]{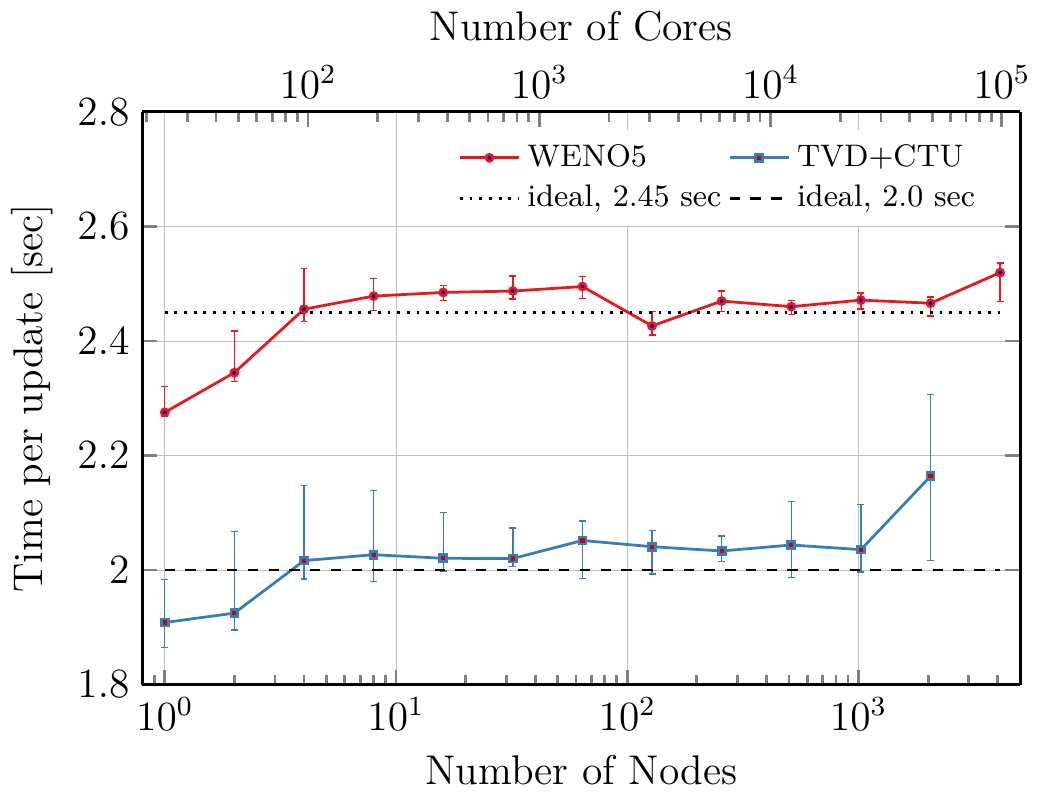}
   \caption{Time per update in seconds over number of nodes and cores (weak scaling) on HLRS Cray XC 40 ''Hazel Hen''. Red: WENO5 solver with $18^3$ patches and $3\times4\times4$ patches per rank. Blue: TVD+CTU solver with same configuration but $32^3$ patches.}\label{fig.weak}
\end{figure}

We test the weak scaling of \wombat on the Cray XC40 ''Hazel Hen'' at HLRS Stuttgart. The machine features a Cray Aries interconnect, which uses a Dragonfly network with adaptive routers that are well suited for high rate of small MPI messages in \wombat's communication pattern. 16 MB large pages are enabled by default. We note that the problem is well balanced, i.e. that the work load per node is the same on all ranks. Thus \wombat's load balancing capabilities are not tested here.  For large problems, communication is only a small fraction of the total workload, which effectively hides communication inefficiencies. Hence we choose a particularly small workload to clearly expose overhead from the MPI communication in the run. Unfortunately, this is not true for all weak scaling tests in the literature, and direct comparison with other codes is not always straight forward. We also note that for large enough machines (Millions of cores), communication overhead will eventually always be exposed. Thus the argument that enough work is available per time step to hide communication inefficiencies is just another way of saying that an implementation does not weak scale beyond a certain point. \par
Here we use $3\times4\times4$ patches per MPI rank with $18^3$ zones per rank, which results in a step time of about $2.5$ seconds and a throughput of about 500.000 zones per second per node. We evolve the 3D problem for 100 steps, which means a total runtime of about 4 minutes. Every run was a separate submission on a different set of nodes on the system. The resulting mean step time over number of nodes \& cores is shown in figure \ref{fig.weak}, from 1 node (24 cores) to 4096 nodes (98304 cores). The minimum and maximum time per step is shown as error bars. Ideal scaling at 2.45 sec per update is shown as dotted line. We also run the same test with the TVD+CTU solver, but with $32^3$ patches, which is its optimal cache blocked patch size on Broadwell. The resulting scaling is shown as a blue line. The step time is reduced by a factor of 1.2 relative to the WENO solver and the problem is a factor 5.6 larger, so the TVD+CTU solver is a factor 6.7 faster at scale than the WENO solver. But the WENO solver doubles effective resolution, which would increase the runtime of TVD+CTU by a factor of 16.  TVD+CTU runs with $CFL=0.4$, half the step size of WENO. Thus the WENO solver is more efficient (''faster'') than TVD+CTU by a factor of about $2\times2.3=4.6$ at scale.\par 
The scaling is quite flat at about 2.45 seconds per update with about 10\% scatter around the mean. Differences arise from network contention on the system, as runtime is dominated by the slowest node during the run. They are much smaller than seen previously on the Blue Waters system with the Gemini interconnect \citep{2017ApJS..228...23M}. This view is supported by the decrease in scatter in step times for the largest runs, where a significant fraction of the machine is used. The largest run uses more than half of the Hazel Hen system and performs at about 150 TFlops or about 5\% of the peak performance of the system. We note that the performance is quite competitive with recent efforts presented in \citet{2018MNRAS.477..624N}. 

\subsection{Strong Scaling}

\begin{figure}
    \centering
    \includegraphics[width=0.45\textwidth]{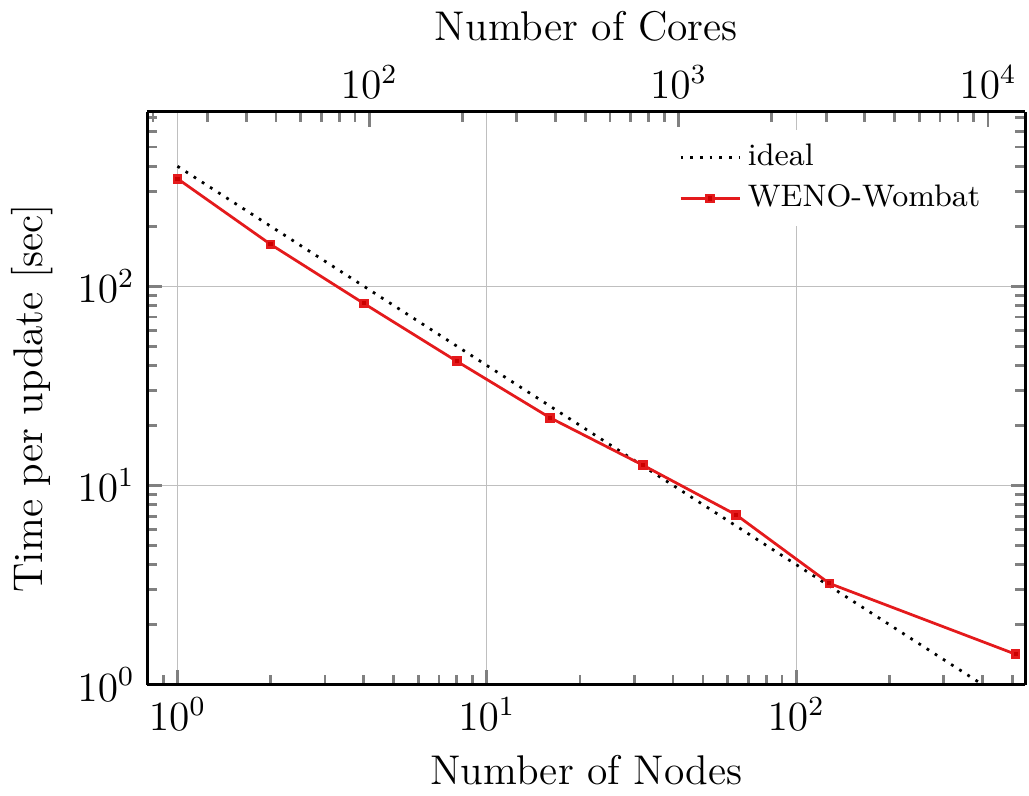}
     \caption{Strong scaling of WENO-Wombat on HLRS ''Hazel Hen'' as time per update over number of nodes/cores. }\label{fig.strong}
 \end{figure}

Strong scaling refers to the speedup of an algorithm as the problem size is kept fixed and the compute-power is increased. Ideally, execution time should half as the computing power doubles. According to Amdahl's law \citep{Amdahl:1967:VSP:1465482.1465560}, the execution time of any parallel algorithm will be dominated by the non-parallelizable part of the work eventually, which can be in the form of branching, communication or load imbalance. \par
We run a strong scaling test on the Cray XC40 ''Hazel Hen'' at HLRS Stuttgart 4 MPI ranks and 12 OpenMP threads per node. We use $18^3$ patches with an initial patch distribution of 96 x 12 x 12 patches, so the world grid has $1728 \times 216^2$ zones. For the final point at 512 nodes, we reduced the patch size to $9\times18^2$ zones. \par
The result is shown in figure \ref{fig.strong} as seconds per time step over number of nodes and cores. WENO-Wombat follows the ideal scaling closely down to 128 nodes. The point at 512 nodes deviates from the ideal scaling due to the decreased patch size. There was simply not enough work available anymore with 512 nodes, which limits \wombat's ability to further strong scale.

\section{Conclusions}\label{sect.conclusions}

High order Eulerian schemes will likely become instrumental to simulations of turbulence and magnetic field amplification in the astrophysical context \citep{2017LRCA....3....2B,2018JCoPh.375.1365F,2018arXiv180602343G}. The inevitable growth of computing power has lead to ample resources available for highly optimized numerical implementations that scale to $> 10^5$ cores. With the advent of accelerators and high bandwidth memory on the horizon, complex high-order simulations with rich sub-grid physics are within reach for the community. \par
We have presented an implementation of a fifth-order  Weighted-Essentially-Non-Oscillatory scheme in the numerical \wombat framework. We combined the classical {finite difference} scheme with a simple constrained transport scheme for the evolution of magnetic fields. We have demonstrated algorithmic correctness and fidelity on a variety of test problems in one, two and three dimensions. We argued that the CT-WENO5 scheme:
 \begin{itemize}
     \item doubles the effective resolution compared to common second order schemes like CTU+CT or TVD+CTU. 
     \item resolves instabilities better than a lower order scheme, due to its very low numerical diffusivity.
     \item is less affected by advection relative to the grid. It is competitive with Lagrangian methods in tests with moderate advection velocities.
     \item is more computationally efficient than a lower order scheme in three dimensions at similar solution quality.
     \item needs to use double precision floating point arithmetic to deliver good results. The low diffusivity of the scheme exposes noise very quickly.
 \end{itemize}
The limitations of our CT-WENO5 implementation are in the dispersion of MHD waves, where due to the simple CT scheme, the implementation roughly matches a good second order scheme.  Nonetheless, we have shown that MHD wave dissipation by numerical diffusion is accurate to fifth order.  Furthermore, the CT-WENO5 scheme does not handle advection and angular momentum conservation as well as discontinuous Galerkin schemes of third or higher order. These limitations are compromises that keep the implementation computationally cheap. E.g. WENO5 is significantly cheaper than third or higher order discontinuous Galerkin implementations, due to the larger time step in three dimensions. \par
 We showed that our implementation reaches about $20\%$ of peak double precision performance on a single Broadwell or Skylake core and $\sim 5\%$ on 98k cores on a Cray XC40. On Intel Broadwell processors we observe a throughput of about 0.5 million zones per node at twice the solution fidelity of a typical second order scheme with a Roe solver. We conclude that our implementation represents a favourable compromise of fidelity, robustness and computational efficiency, awaiting astrophysical applications.

\subsection{Outlook}

 \texttt{\small WENO-WOMBAT} already includes a treatment of cold supersonic flows using an elegant entropy scheme based on flux splitting (Jang et al. in prep). A few improvements to the implementation presented here come to mind. A high order CT scheme that matches the accuracy of the spatial interpolation would reduce magnetic field dispersion, albeit at considerable computational cost. High order CT schemes exist for finite volume approaches \citep[e.g.][]{2004JCoPh.195...17L,2019MNRAS.482..416V,2018JCoPh.375.1365F}, but need to be adapted to our finite difference algorithm. Global smoothness indicators could be used to improve the robustness of the scheme  \citep[e.g.][]{doi:10.1137/040610246}. Worth another look is likely a low-storage fourth order Runge-Kutta integrator that reduces the memory needed to store the grid by one third. A wide variety of strong-stability preserving schemes exist, usually with five of more stages and CFL numbers $> 1$ \citep{doi:10.1137/S0036142901389025}. Furthermore, a ninth order WENO implementation might become affordable once super-computers reach exa-Flop performance, further doubling the effective resolution of the simulation and reducing storage and memory requirements. An extension of our WENO5 implementation would be straight forward. On the technical side, the flattened array implementation will make it trivial to port the WENO solver to accelerators. The OpenMP 4 standard would be the natural choice, providing a portable approach for a wide variety of accelerator technologies. \par

\subsection{Reproducibility}

 It has been shown that computational fluid dynamics without code level transparency is not reproducible, even by the authors of the original publication: \citet{8012284} found that open source code, logs and data are a bare minimum to reproduce results years later. \wombat follows their reproducibility policy: sources and logs used to produce this document are available at \url{https://wombatcode.org/publications/}. Data is made available wherever possible, but has to adhere to certain space limitations. A public version of \wombat is available under MIT license, however not yet with the WENO5 solver. 

\section{Acknowledgements}\label{sect.ack}
JD thanks Louiz de Rose and the programming environment group for 2 years of hospitality and a great work environment at Cray Inc. in Minnesota. JD thanks H. Roettgering and E. Deul for support towards the end of the project. The authors thank J. Delgado for comments on the manuscript. \par
We used Julia\footnote{www.julialang.org} and PGFPlots for  post-processing and graphs shown in this document. Volume rendering was done with ParaView. Perceptually uniform colour maps were obtained from \citep{2015arXiv150903700K} and \url{colorbrewer.org}.  \par
Parts of this work were computed on the Cray development system ''Kay'', JD thanks Cray Inc. for continuous access to the system. Some code tests were performed on the MACH64 machine at IRA Bologna. JD thanks F. Bedosti for support with the system. TWJ was supported at the University of Minnesota by NSF grant AST. The work of H.J. and D.R. was supported by the National Research Foundation (NRF) of Korea through grants 2016R1A5A1013277 and 2017R1A2A1A05071429. This research  has received funding from the People Programme (Marie Sklodowska Curie Actions) of the European Union’s Eighth Framework Programme H2020 under REA grant agreement no 658912, ''Cosmo Plasmas''. Access to the 'Hazel Hen' at HLRS has been granted through PRACE preparatory access project ''PRACE 4477''. 

\bibliographystyle{apj} \bibliography{master}
\newpage
\appendix

\section{MHD Eigenvectors} \label{app.ev}
    The eigenvectors decouple the system of partial differential equations into scalar advection equations. Every component $m$ of the decoupled system, corresponds to an eigenvalue $\lambda_m$ with a left eigenvector $L_i(m)$ and right eigenvector $R^i(m)$ with $L_i R^i = \Matrix{I}$ \citep{1964nlwp.book.....J}. The eigenvalues by component $m$ are $\lambda(1,7) = v_x \mp c_\mathrm{fast}, \lambda(2,6) = v_x \mp c_\mathrm{A}, \lambda(3,5) = v_x \mp c_\mathrm{slow}, \lambda(4) = v_x$. For MHD, degeneraries occur in the case of vanishing magnetic field components \citep[][]{1988JCoPh..75..400B}, so we set:
\begin{align}
    (\beta_y, \beta_z) &= \begin{cases} 
        \frac{(B_y,B_z)}{\sqrt{B^2_x + B^2_z}}  &\, \text{if $B^2_y + B^2_z > \epsilon_B$}       \\
                            \left( \frac{1}{\sqrt{2}}, \frac{1}{\sqrt{2}}\right) &\, \text{otherwise}
                          \end{cases} \\
    \mathrm{sgn}(B_x) &= \begin{cases}
                            1 &\, \text{if $B_x \ge \epsilon_B$} \\
                            -1 &\, \text{otherwise}
                        \end{cases} \\
    (\alpha_f, \alpha_s) &= \begin{cases}
        \frac{\sqrt{c_\mathrm{s}^2 - c_\mathrm{slow}^2}, \sqrt{c^2_\mathrm{fast} - c_\mathrm{s}^2}}{\sqrt{c_\mathrm{fast}^2 - c_\mathrm{slow}^2}} &\, \text{if $ \sqrt{c_\mathrm{fast}^2 - c_\mathrm{slow}^2} \ge \epsilon_B$} \\
                                \left( 1, 1 \right) &\, \text{otherwise}
                            \end{cases} \label{eq.alpha}
\end{align}
where $\epsilon_B = 10^{-30}$. We note that in some cases, numerical noise from compiler optimizations can break the degeneracy of the MHD eigenvalues with the sound speed (but not between fast and slow mode) in the calculation of equation \ref{eq.alpha}. This can lead to noise in the calculation that can be exposed e.g. in the implosion test, section \ref{sect.implosion}. It is advisable to explicitly check for the eigenvalue degeneracy in the code and set equation \ref{eq.alpha} explicitly to zero. For similar reasons we found it practical to explicitly store the sound speed and not its square in the eigenvector calculations. \par
Our eigenvectors follow \citet[][]{1999JCoPh.150..561J}, Jang et al. in prep.  With 
\begin{align}
    \gamma_1 &= \frac{\gamma - 1}{2} &\, \gamma_2 &= \frac{\gamma-2}{\gamma-1}  &\, \tau &= \frac{\gamma-1}{c_s^2}
\end{align}
and 
\begin{align}
    \Gamma_f &= \alpha_f c_\mathrm{fast}v_x - \alpha_s c_\mathrm{slow} \mathrm{sgn}(B_x)\left( \beta_yv_y + \beta_z v_z \right)\\
    \Gamma_a &= \mathrm{sgn}(B_x) \left( \beta_z v_y - \beta_y v_z\right)\\
    \Gamma_s &= \alpha_s c_\mathrm{slow} v_x + \alpha_f c_\mathrm{fast} \mathrm{sgn}(B_x)\left( \beta_yv_y + \beta_z v_z \right),
\end{align}
the eigenvectors for the fast mode ($m=1, m=7$) are:
\begin{align}
    L_1(m) &= \frac{1}{2c_s^2} \left[ \alpha_f \left( \gamma_1 v^2 \pm \Gamma_f  \right) \right]              &\,  R^1(m) &= \alpha_f \\
    L_2(m) &= (1-\gamma) \alpha_f v_x \mp \alpha_f v_y \pm c_\mathrm{slow}\alpha_s \beta_y \mathrm{sgn}(B_x)  &\,  R^2(m) &= \alpha_f (v_x \mp c_\mathrm{fast})  \\
    L_3(m) &= (1-\gamma) \alpha_f v_y \pm c_\mathrm{slow}\alpha_s \beta_y \mathrm{sgn}(B_x)                   &\,  R^3(m) &= \alpha_f v_y \pm c_\mathrm{slow}\alpha_s\beta_y\mathrm{sgn}(B_x) \\ 
    L_4(m) &= (1-\gamma) \alpha_f v_z \pm c_\mathrm{slow}\alpha_s\beta_z\mathrm{sgn}(B_x)                     &\,  R^4(m) &= \alpha_f v_z \pm c_\mathrm{slow}\alpha_s\beta_z\mathrm{sgn}(B_x)\\
    L_5(m) &= (1-\gamma) \alpha_f B_y - \sqrt{\rho} c_s \alpha_s \beta_y                                      &\,  R^5(m) &= c_s\alpha_s\beta_y/\sqrt{\rho} \\ 
    L_6(m) &= (1-\gamma) \alpha_f B_z - \sqrt{\rho}c_s\alpha_s\beta_z                                         &\,  R^6(m) &= c_s\alpha_s\beta_z / \sqrt{\rho}\\
    L_7(m) &= (\gamma-1) \alpha_f                                                                             &\,  R^7(m) &= \alpha_f \left( \frac{1}{2} v^2 + c_\mathrm{fast}^2 - \gamma_2 c_s^2 \right) \mp \Gamma_f
\end{align}
The eigenvectors for the slow mode ($m=2, m=6$) are:
\begin{align}
    L_1(m) &= \frac{1}{2} \Gamma_a                    &\, R^1(m) &= 0 \\
    L_2(m) &= 0                                       &\, R^2(m) &= 0 \\
    L_3(m) &= -\frac{1}{2} \beta_z \mathrm{sgn}(B_x)  &\, R^3(m) &= -\beta_z \mathrm{sgn}(B_x) \\
    L_4(m) &= \frac{1}{2} \beta_y \mathrm{sgn}(B_x)   &\, R^4(m) &= \beta_y \mathrm{sgn}(B_x) \\
    L_5(m) &= \mp \frac{1}{2}\sqrt{\rho} \beta_z      &\, R^5(m) &= \mp \beta_z/\sqrt{\rho}\\
    L_6(m) &= \pm \frac{1}{2}\sqrt{\rho} \beta_y      &\, R^6(m) &= \pm \beta_y/\sqrt{\rho}\\
    L_7(m) &= 0                                       &\, R^7(m) &= -\Gamma_a
\end{align}
The eigenvectors for the Alfv\'{e}n mode ($m=3, m=5$) are:
\begin{align}
    L_1(m) &= \frac{1}{2c_s^2} \left( \gamma_1 \alpha_s v^2 \pm \Gamma_s \right)                                              &\, R^1(m) &= \alpha_s \\
    L_2(m) &= \frac{1}{2c_s^2} \left( (1-\gamma) \alpha_s v_x \mp \alpha_s c_\mathrm{slow} \right)                            &\, R^2(m) &= \alpha_s (v_x\mp c_\mathrm{slow})\\
    L_3(m) &= \frac{1}{2c_s^2} \left( (1-\gamma) \alpha_s v_y \mp c_\mathrm{fast} \alpha_f \beta_y \mathrm{sgn}(B_x) \right)  &\, R^3(m) &= \alpha_s v_y \mp c_\mathrm{fast} \alpha_f \beta_y \mathrm{sgn}(B_x)\\
    L_4(m) &= \frac{1}{2c_s^2} \left( (1-\gamma) \alpha_s v_z \mp c_\mathrm{fast}\alpha_f \beta_z \mathrm{sgn}(B_x) \right)   &\, R^4(m) &= \alpha_s v_z \mp c_\mathrm{fast} \alpha_f \beta_z \mathrm{sgn}(B_x)\\
    L_5(m) &= \frac{1}{2c_s^2} \left( (1-\gamma) \alpha_s B_y - \sqrt{\rho} c_s \alpha_f \beta_y \right)                      &\, R^5(m) &= -c_s\alpha_f\beta_y / \sqrt{\rho} \\
    L_6(m) &= \frac{1}{2c_s^2} \left( (1-\gamma) \alpha_s B_z - \sqrt{\rho} c_s \alpha_f \beta_z \right)                      &\, R^6(m) &= -c_s \alpha_f \beta_z / \sqrt{\rho} \\
    L_7(m) &= \frac{1}{2c_s^2} \left( (\gamma-1) \alpha_s \right)                                                             &\, R^7(m) &= \alpha_s \left( \frac{1}{2} v^2 +c_\mathrm{slow}^2 - \gamma_2c_s^2 \right) \mp \Gamma_s \\
\end{align}
The eigenvectors for the entropy mode ($m=4$) are:
\begin{align}
    L_1(m) &= 1 - \frac{1}{2} \tau v^2  &\, R^1(m) &= 1 \\
    L_2(m) &= \tau v_x                  &\, R^2(m) &= v_x \\
    L_3(m) &= \tau v_y                  &\, R^3(m) &= v_y \\
    L_4(m) &= \tau v_z                  &\, R^4(m) &= v_z \\
    L_5(m) &= \tau B_y                  &\, R^5(m) &= 0 \\
    L_6(m) &= \tau B_z                  &\, R^6(m) &= 0 \\
    L_7(m) &= -\tau                     &\, R^7(m) &= \frac{1}{2}
\end{align}
These components are evaluated at the boundary using simple arithmetic averaging.

\label{lastpage}
\end{document}